\documentclass[a4paper,11pt]{article}
\pdfoutput=1 

\usepackage{jheppub} 
                     \usepackage{mathrsfs} 
   \usepackage{gensymb}
\usepackage{amsfonts}
\usepackage{amsmath,amssymb}
\usepackage{
mathdots}
\usepackage{pagecolor}

\usepackage{graphicx}				

\usepackage{extarrows}
\newcommand{\calN}{\mathcal{N}}

\newcommand{\calM}{\mathcal{M}}
\newcommand{\CM}{\mathcal{M}}

\newcommand{\calZ}{\mathcal{Z}}
\newcommand{\calR}{\mathcal{R}}
\newcommand{\ft}{t}
\newcommand{\fq}{q}

\def\be{\begin{equation}}
\def\ee{\end{equation}}
\def\bea{\begin{eqnarray}}
\def\eea{\end{eqnarray}}
\def\bal{\begin{align}}
\def\eal{\end{align}}
\def\nn{\nonumber}

\def\qqq{\qquad}

\def\ii{\text{i}}
\def\ln{\text{ln}}

\def\exp{\text{exp}}

\def\cB{\mathcal{B}}
\def\cC{\mathcal{C}}

\def\cI{\mathcal{I}}
\def\cM{\mathcal{M}}
\def\cN{\mathcal{N}}

\def\fq{\mathsf{q}}

\def\ft{t}
\newcommand{\defeq}{\stackrel{\textup{\tiny def}}{=}}

\title{\center 
Trinion Conformal Blocks from Topological strings
}

\author[a,b]{Ioana Coman}
\author[a]{Elli Pomoni}
\author[a,c]{Joerg Teschner}

\affiliation[a]{DESY, Theory Group, Notkestra\ss e 85, Building 2a, 22607 Hamburg, Germany}
\affiliation[b]{Institute of Physics, University of Amsterdam,1098 XH Amsterdam, the Netherlands}
\affiliation[c]{Department of Mathematics, University of Hamburg, Bundesstrasse 55, 20146 Hamburg, Germany}

\emailAdd{i.comanlohi@uva.nl}
\emailAdd{elli.pomoni@desy.de}
\emailAdd{joerg.teschner@desy.de}
\abstract{

\bigskip

In this paper we investigate the relation between conformal blocks of Liouville CFT
and the topological string partition functions of the rank one trinion theory $T_2$.
 The partition functions exhibit jumps when passing from one chamber in the 
 parameter space to another. Such jumps can
 be attributed to a change of the integration contour in the free field representation of Liouville conformal blocks.
  We compare the partition functions of the $T_2$ theories
 representing trifundamental half hypermultiplets in $N=2$, $d=4$ field theories to the partition functions  
 associated to
  bifundamental hypermultiplets. We find that both are related to the same Liouville conformal blocks up to inessential 
  factors. 
  In order to establish this picture we combine and compare results obtained using topological vertex techniques, 
  matrix models and topological recursion. 
 We furthermore check that the partition functions obtained by gluing two $T_2$ vertices can be represented in terms of a 
 four point Liouville conformal block. Our results  indicate that the $T_2$ vertex offers a useful starting point for 
 developing an analog of the instanton calculus for SUSY gauge theories with trifundamental hypermultiplets.
}

\begin{document}

\thispagestyle{empty}
\setcounter{page}{0}
\begin{flushright}\footnotesize
\texttt{DESY 19-106}\\
\vspace{0.5cm}
\end{flushright}
\setcounter{footnote}{0}

\begin{center}
\end{center}
\maketitle


\section{Introduction}\label{sec:Intro}

The discovery of relations between $\mathcal{N}=2$, $d=4$ supersymmetric gauge theories  
and conformal field theory by
Alday, Gaiotto and Tachikawa \cite{Alday:2009aq} has stimulated a large amount of work. 
Such relations have meanwhile been proven for the linear or circular quiver gauge theories built from vector multiplets and 
hypermultiplets in the (bi-)fundamental and adjoint representation of the gauge group \cite{Alba:2010qc,Ne17}. 

In a parallel development, the $\mathcal{N}=2$, $d=4$ 
supersymmetric field theories of class $\mathcal{S}$ have been introduced in \cite{Gaiotto:2009we,Gaiotto:2009hg}. 
The theories of class
$\mathcal{S}$ are classified by punctured Riemann surfaces $C$ and Lie-algebras $\mathfrak{g}$ of ADE-type.
In the cases where $\mathfrak{g}=A_1$ 
there exists a Lagrangian representation for any pants decomposition of $C$. It involves vector multiplets for any cutting curve, and certain matter multiplets associated to any pair of pants.
By considering the S-dualities of theories in class $\mathcal{S}$ 
associated to changes of pants decomposition 
\cite{Gaiotto:2009we} one can  identify
the half-hypermultiplet in the tri-fundamental representation of $[SU(2)]^3$ as the most natural 
candidate for the field theory associated to the pairs of pants, see \cite{Ta16} for a review. 
 For the cases $\mathfrak{g}=A_N$ with $N>1$ one generically expects to find 
non-Lagrangian theories as the building blocks associated to pairs of pants.

Almost all of the available checks or proofs 
\cite{Fateev:2009aw,Mironov:2009qn,Hadasz:2010xp,Alba:2010qc,Mironov:2010qe,Mironov:2010pi,Fateev:2011hq,Kanno:2013vi,Mironov:2013oaa} 
of AGT-type relations have been performed in the linear quiver theories obtained by 
using fundamental or bi-fundamental 
hypermultiplets of $U(N)$ instead of tri-fundamental half-hypermulti\-plets of $SU(N)$.
 For these cases one can use the instanton calculus for $U(N)$-theories developed in 
\cite{MNS,Nekrasov:2002qd}. 
The instanton calculus for half-hypermultiplets in the tri-fundamental representation 
of $SU(2)$ is much less developed. Apart from the work 
\cite{Hollands:2011zc} 
we are not aware of any direct checks of AGT-type relations in class $\mathcal{S}$ theories
built from tri-fundamental half-hypermultiplets. 
The difficulties become even more severe in many cases with  
$\mathfrak{g}=A_{N-1}$, $N>2$. The theories associated to the pair of pants will then generically
be non-Lagrangian, preventing us from using any of the conventional tools based of
Lagrangian  QFT.

A possible way out is offered by the geometric engineering of supersymmetric field theories within string theory. Many such field theories can be described as certain scaling limits of string theory on $R^4\times M$, with $M$ being a toric CY manifold \cite{Katz:1996fh,Katz:1997eq}. 
In such cases one expects that the instanton partition functions coincide with the  topological string partition functions computable with the help of 
the (refined) topological vertex \cite{Aganagic:2003db,Iqbal:2007ii,Awata:2008ed}. 
Combined with the AGT-correspondence one gets relations between
topological string partition functions and conformal blocks for which more direct explanations have been proposed in 
\cite{Dijkgraaf:2009pc,Cheng:2010yw}\footnote{For related important works see \cite{Schiappa:2009cc}, as well as the review \cite{Maruyoshi:2014eja} and references therein.}. 
Dualities are expected to relate string theory on $R^4\times M$ for various toric CY $M$ to intersecting 
5-brane systems called brane webs. The toric diagrams encoding the geometry of the CY manifold $M$ 
coincide with the web diagrams describing the intersection patterns of the brane webs 
\cite{Leung:1997tw,Gorsky:1997mw}, simplifying the identification of the gauge
theories describing the relevant scaling limits.  
This line of reasoning has lead to a family of candidates for string-theoretic 
descriptions of the class $\mathcal{S}$ theories associated to the 
three-punctures spheres, which are often referred to as the $T_N$-theories and 
associated to certain web diagrams
proposed in \cite{Benini:2009gi}. 
One might hope that the refined topological vertex offers tools for the computation of partition 
functions in cases where the gauge-theoretic instanton calculi are not available or hard to use.
This line of thought has been the motivation for a series of investigations
 \cite{Bao:2013pwa,Hayashi:2013qwa,Mitev:2014isa,Isachenkov:2014eya}.

A considerable amount of evidence is available supporting the role of the $T_N$ vertex as
 a building block of class $\mathcal{S}$ theories\footnote{The web diagrams of \cite{Benini:2009gi}  
reproduce the correct  the dimensions
of Coulomb and Higgs branches, the $c$ and $a$ anomaly coefficients \cite{Gaiotto:2009gz},
the correct Seiberg Witten curves
  \cite{Benini:2009gi,Bao:2013pwa} and
 when identified with the toric diagrams of   CY manifolds, they reproduce the correct 5D superconformal index \cite{Iqbal:2012xm} (the partition function on $\mathbb{S}^4 \times \mathbb{S}^1$)
 of the 5D $T_N$ trinion theories   \cite{Bao:2013pwa,Hayashi:2013qwa}, including the $E_{6,7,8}$ symmetry enhancement \cite{Bao:2013pwa,Hayashi:2013qwa,Hayashi:2014wfa}. See \cite{Tachikawa:2015bga} for a review.}.
Analogs of the AGT-correspondence have been found relating instanton partition 
functions in five-dimensional supersymmetric
gauge theories to q-deformations of conformal blocks for the $W_N$-algbra. Using free field representations
for the q-deformed conformal blocks one can derive such relations in many cases
\cite{Aganagic:2013tta,Aganagic:2014oia}.

One should note, however, that
these results are not immediately applicable to the study of the AGT-correspondence for the 
class $\mathcal{S}$ theories.\footnote{The five-dimensional analogs of the AGT correspondence studied in the literature either concern a deformation of the linear quiver theories, or dual versions of this correspondence in the sense of fiber-base duality. It is not known how to go beyond the well-understood case of linear quiver theories in this way.}
It seems to us, in particular, that the relation between the $T_N$ theories and conformal blocks is not well-understood.
There are basic open questions even in the simplest case $N=2$.
The partition function of the $T_2$ theory has been calculated in \cite{Kozcaz:2010af} where it was
observed that the absolute value of this partition function is closely related to the three point 
function of Liouville CFT.
This gives a first piece of evidence for a relation between the $T_2$ theory and conformal blocks on 
three-punctured sphere, but it is pretty far from proving such a relation. At the moment it does not 
even seem to be known if the gluing of $T_2$ vertices reproduces the conformal blocks on 
spheres with more than three punctures.

It is interesting to compare the $T_2$ vertex with 
another composite vertex which has been discussed in the literature, often referred to as the strip \cite{Iqbal:2004ne}.  
 In the context of geometric engineering 
one may naturally associate the strip diagrams with fundamental- or bi-fundamental 
hypermultiplets. 
It is well-known that the strip partition functions represent the basic building blocks
for the instanton partition functions in linear or circular quiver gauge theories. For this case
one has a clear understanding of the relation between the instanton partition functions and conformal blocks. 

This state of affairs naturally raises two basic questions: 
\begin{itemize}
\item[(a)] Can we  demonstrate AGT-type relations
for the class $\mathcal{S}$-theories engineered from the $T_2$ vertex? 
\item[(b)] Does the AGT-correspondence 
distinguish between class $\mathcal{S}$-theories built from different types of 
vertices? 
\end{itemize}
These are the two main questions we will aim to address in this paper.
We thereby hope to
alleviate slightly the tension between the basic role of the $T_2$-vertex in class
$\mathcal{S}$-theories on the one hand, and the shortage of hard evidence for 
the AGT relation involving the $T_2$-vertex on the other hand.
We also view our investigation as a natural warm-up for investigations of the $T_N$-theories 
with $N\geq 2$. On the way we will address the following questions.

  \begin{enumerate}

\item \label{Problem:limit}
 The limits giving the partition functions of 4D gauge theories 
 from the topological string partition functions are tricky. It turns out that
 the partition functions diverge in this limit, and need to be renormalised\footnote{
 Here we use the word renormalisation in a wider sense, as a procedure to define the limit of a formally divergent quantity.} 
to get meaningful results. As usual, there can be freedom in the definition of the
renormalisation prescription. This raises the question how predictive this approach can be. This becomes particularly important in the cases $N>2$, but even for $N=2$ we did not find a discussion of this issue in 
the literature.  
 
\item Can we obtain the known instanton partition functions of non-abelian gauge theories by gluing $T_2$ partition functions? Having such a check would confirm that the $T_2$ vertex can indeed 
play the expected role as a building block for more class $\mathcal{S}$  theories.  

\item \label{Problem:map}  What is the precise map between the K\"ahler moduli of the toric CY $M$ defining the
$T_2$ vertex and the parameters labelling 
bases for the three point conformal blocks?


\end{enumerate}

On the way we take the opportunity to close some gaps in the existing literature.
The paper is organised as follows. 
We first review the derivation of the $T_2$ topological strings partition function, discussing its resummation into 
a product formula in Section \ref{sec:TopologicalPartitionFun}. This is followed in Section \ref{sec:4dLim} by an 
analysis of the 4D limit and a prescription for taking this limit in a meaningful way. Section \ref{sec:topol2matrix} 
then presents an alternative route towards deriving this limit, rewriting the partition function as a matrix integral. Here 
we also compare the integral formulation to the three point conformal block of $q$-Liouville CFT. We then proceed 
in Section \ref{sec:MM4dLim} to take the 4D limit and rewrite the matrix integral as a Selberg integral. 
As a building block, the $T_2$ needs to be compared to a better studied counterpart,  commonly referred to as the strip. 
This analysis can be found in Section \ref{sec:CompareT2-Strip}, followed by Section \ref{sec:B-modelPicture} which  
comments on this picture from the B-model point of view. 
Section \ref{sec:TNvsBlocks} then offers some words on the generalisation of this story to the higher rank case for the 
$T_N$ theories. 
Afterwards, Section \ref{sec:gluing} follows this discussion with explicitly gluing $T_2$ blocks and comparing to the 
gluing of strips. 
The appendices then gather the definitions and properties of ubiquitous special functions.
For the reader's (and our) convenience we have collected some of the key formulae in Appendix \ref{Summary}.

\begin{figure}[b]
   \centering
   \includegraphics[height=3.5cm]{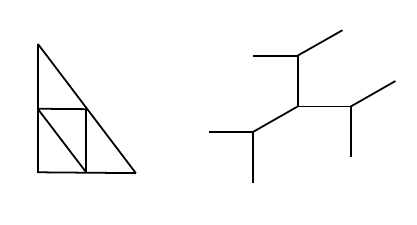} 
   \caption{\it The toric diagram (on the left) and its dual 5-brane web diagram (on the right) for the $T_2$ geometry, 
   the supersymmetric $\mathbb{C}^2/\mathbb{Z}_2 \times \mathbb{Z}_2$ orbifold.  }
   \label{fig:T2toricweb}
\end{figure}

\section{The topological vertex computation}
\label{sec:TopologicalPartitionFun}

In this section we review 
the computation of the topological sting partition function 
in  \cite{Bao:2013pwa}.

The theory of $[SU(2)]^3$ half-hypermultiplet in the tri-fundamental representation  
can be obtained from string theory in various ways. There exists a five 
dimensional uplift of this theory compactified on $\mathbb{S}^1$, which can be constructed in type IIB string theory
via the $(p,q)$ web 5-brane diagram \cite{Aharony:1997ju,Aharony:1997bh} depicted in Figure~\ref{fig:T2toricweb}, due to 
\cite{Benini:2009gi}.
Equivalently, this theory can be constructed in M-theory/type IIA string theory on Calabi-Yau threefolds via geometric 
engineering \cite{Katz:1996fh,Katz:1997eq}.
The toric diagram of the corresponding  $\mathbb{C} \times \mathbb{C}^2/\mathbb{Z}_2 \times \mathbb{Z}_2$  Calabi-Yau 
manifold has the same shape as the dual of the $T_2$ web diagram \cite{Leung:1997tw}. 
The partition function of the $T_2$  theory and any $\mathcal{N}=2$ theory  geometrically  engineered with a toric 
Calabi-Yau compactification can be obtained by computing the topological string partition function on the Calabi-Yau  
\cite{Nekrasov:2002qd,Iqbal:2003ix,Iqbal:2003zz}.
A very efficient method to compute 5D Nekrasov partition functions is the refined topological vertex formalism 
\cite{Iqbal:2007ii,Awata:2008ed,Taki:2007dh}.

\subsection{The computation}

\begin{figure}[t]
   \centering
      \includegraphics[height=4.5cm]{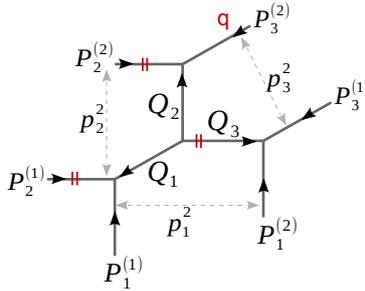} 
   \caption{\it The 5-brane web diagram for $T_2$ with the choice of the   preferred direction and the K\"ahler parameters.
     The edges along the preferred
direction are marked by two red strips.  }
   \label{fig:T2refined}
\end{figure}

The $T_2$ geometry is parametrised by three independent K\"ahler parameters $Q_1$, $Q_2$ and $Q_3$ that are related 
to the 5D gauge theory parameters 
$P_1,P_2,P_3$ in Figure \ref{fig:T2refined} as
\be
\label{eq:Kahler2Mass}
Q_1 Q_3 = \frac{P_1^{(1)}}{P_1^{(2)}} = P_1^2~, \qquad Q_1 Q_2 = \frac{P_2^{(1)}}{P_2^{(2)}} = P_2^2 ~, 
\qquad  Q_2 Q_3 = \frac{P_3^{(1)}}{P_3^{(2)}} = P_3^{2}
\ee
which are then related to  the 4D gauge theory parameters as $-R^{-1}\ln P_i$, 
with $R$ the radius of the circle of the 5th dimension. The topological partition function is also a function of the 
parameters $q,t$ which are the 5D version of Nekrasov's Omega background
\be
\label{eq:Omegatq}
q= e^{-R \epsilon_1} \, , \quad  t= e^{R \epsilon_2} \, .
\ee

For completeness we sketch the rules for reading off the refined topological string partition function from a dual toric diagram. The procedure is very similar to calculating Feynman diagrams where we associate functions of the K\"ahler parameters, the Omega background as well as Young diagrams
 to the edges and vertices of the dual toric diagram.
The  topological partition function is then obtained as the product of all the edge and vertex factors, summed over all possible partitions associated with the internal edges.

In order to compute  the refined topological partition function, using the refined topological vertex formalism we first need to pick 
a preferred direction on  the  toric diagram. This is denoted with two red lines in Figure  \ref{fig:T2refined}.
The final result for the closed topological string partition function is conjectured to not depend on the choice of the preferred 
direction \cite{Iqbal:2007ii,Awata:2009yc}. However, the closed topological string partition function corresponds to the  full 
correlation function. We are interested in its holomorphic half, the conformal block obtained  from the open topological string 
amplitude. The later carries the information of the choice of the preferred direction  \cite{Zenkevich:2016xqu,Kimura:2017auj}. 

\begin{figure}[h!]
   \centering
\includegraphics[scale=1]{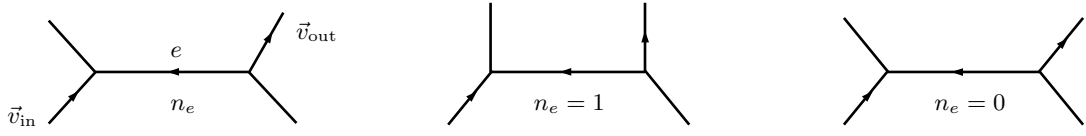}
   \put(-407,5){\footnotesize$ \vec{v}_{\text{in}}$}   
       \put(-299,37){\footnotesize$ \vec{v}_{\text{out}}$}
              \put(-346,30){\footnotesize$e$} 
     \put(-346,10){\footnotesize$n_e$} 
       \put(-210,10){\footnotesize$n_e=1$}       
  \put(-60,10){\footnotesize$n_e=0$} 
\caption{ \it The integer $n_e$ associated to the edge $e$ in the graph is defined as 
  $n_e=\det \left(\vec{v}_{\rm {in}} \,  \vec{v}_{\rm {out}}\right)$. 
  Thus for the graphs in the middle and on the right, this yields $n_e=1$ and $n_e=0$, respectively.}
\label{fig:FramingFactor}
\end{figure}

\begin{figure}[h!]
   \centering
\includegraphics[scale=1.2]{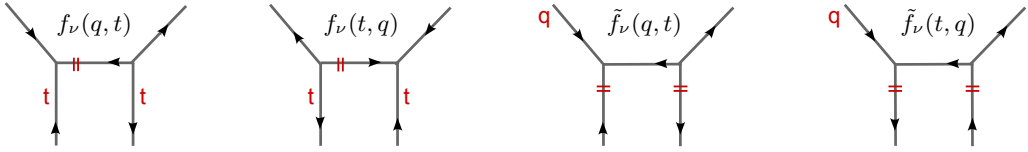}
     \put(-366,57){\footnotesize$f_\nu(q,t)$} 
     \put(-266,57){\footnotesize$f_\nu(t,q)$}      
     \put(-160,57){\footnotesize$\tilde{f}_\nu(q,t)$}       
     \put(-50,57){\footnotesize$\tilde{f}_\nu(t,q)$} 
\caption{ \it The assignment of framing factors to edges of a web diagram.}
\label{fig:FramingFactor2}
\end{figure}
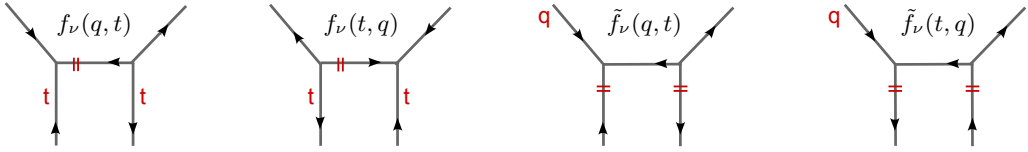

Returning to the rules for computing the refined topological partition function,  to each edge we associate a partition $\mu$. The partitions associated to external lines of the diagram  are empty.
The \textit{edge factor} is the function that we associate to each edge:
\begin{equation}
\text{edge factor}\defeq (-Q)^{|\mu|}\times \text{framing factor}.
\end{equation}
For the $T_2$ toric diagram depicted in Figure \ref{fig:T2refined} the only \textit{framing factors} which we need to use 
are trivial, equal to one. 
In general, the framing factor is given by 
\be 
f_\mu (t,q) =(-1)^{n_e |\mu|} t^{\frac{n_e ||\mu^t||^2}{2}} q^{-\frac{n_e ||\mu||^2}{2}} ~, ~
\tilde{f}_\mu (t,q) =(-1)^{n_e |\mu|} t^{\frac{n_e (||\mu^t||^2+|\mu|)}{2}} q^{-\frac{n_e (||\mu||^2+|\mu|)}{2}} ~,
\ee
with $n_e=\det \left(\vec{v}_{\rm {in}} \,  \vec{v}_{\rm {out}}\right)$ an integer defined like in Figure \ref{fig:FramingFactor} 
and framing factors assigned to edges like in Figure \ref{fig:FramingFactor2}. 

The \textit{refined topological vertex} is the function that we associate to each vertex:
\begin{equation}
\label{eq:topvertex}
C_{\mu \nu  \lambda }(\ft, q) =  q^{\frac{||\nu ||^2+||\lambda ||^2}{2}}\ft^{-\frac{||\nu^t||^2}{2}}\tilde{Z}_{\lambda }(\ft, q)
\sum_{Y}\left(\frac{ q}{\ft}\right)^{\frac{|Y|+|\mu |-|\nu |}{2}}s_{\mu^t/Y}(\ft^{-\rho} q^{-\lambda })s_{\nu /Y}
( q^{-\rho}\ft^{-\lambda^t})~,
\end{equation}
where we have used the functions $\tilde{Z}_{\lambda }(\ft, q)$ defined in equation~\eqref{eq:deftildeZ} in 
Appendix \ref{App:SpecialFunctions}.
The $s_{\mu /\nu }(x)$ are skew-Schur functions of the possibly infinite vector $x=(x_1,\ldots)$. We use the notation 
that for a partition $\nu= (\nu_1, \nu_2,\dots)$, the vector  $\ft^{-\rho} q^{-\nu}$ is given by
\begin{equation}
\ft^{-\rho} q^{-\nu}=(\ft^{\frac{1}{2}} q^{-\nu_1},\ft^{\frac{3}{2}} q^{-\nu_2},\ft^{\frac{5}{2}} q^{-\nu_3}, \ldots).
\end{equation}
\begin{figure}[t]
\centering
\includegraphics[height=2.2cm]{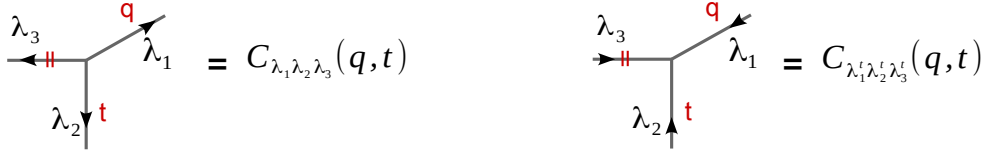}
\caption{\it The direction of the arrows determines the way partitions enter the vertex factor. Partitions are counted clockwise to the preferred direction and the variables $t, q$ enter the vertex factor in this order if $t$ is associated with the end of the first edge and $q$ with the end of the second. 
 }
\label{fig:clockwise}
\end{figure}
The topological string partition function is a sum over the partitions $\{Y_1,\cdots, Y_M\}$ of the $M$ internal edges of the toric diagram
\begin{equation}
\mathcal{Z}^{\text{top}}=\sum_{Y_1,\cdots,Y_M}\  \prod_{\text{edges}}\text{edge factor}\times \prod_{\text{vertices}}\text{vertex factor} .
\end{equation}
For the $T_2$ web depicted in Figure \ref{fig:T2toricweb} the refined topological string partition function is
\be \label{eq:T2partitionfunction1}
\mathcal{Z}_{T_2}^{\text{top}} = \mathcal{Z}_2^{\text{top}}(\boldsymbol{Q})= \sum_{\boldsymbol{R}}
\prod_{i=1}^3(-Q_i)^{\vert R_i\vert} 
C_{ R_1^t\emptyset\emptyset}(q,\ft)C_{\emptyset R_2^t \emptyset}(q,\ft) 
C_{\emptyset\emptyset R_3^t}(q,\ft) C_{R_1R_2 R_3}(\ft , q),
\ee
where $\boldsymbol{Q}=(Q_1,Q_2,Q_3)$ and $R_i$, for $i=1,2,3$, is the Young diagram decorating the edge with 
corresponding K\"ahler parameter $Q_i$. 
The $T_2$ geometry is special and following the work of \cite{Sulkowski:2006jp} for the unrefined partition function we 
can perform some of the sums and achieve a more compact way to write \eqref{eq:T2partitionfunction1}.
For that we use the following identities involving Schur functions
\begin{multline}
\label{eq:schuridentity1}
\sum_{Y}Q^{|Y|}s_{Y/R_1}(\ft^{-\rho} q^{-\nu_1})s_{Y^t/R_2}( q^{-\rho}\ft^{-\nu_2})=\\=\calR_{\nu_2\nu_1}(-Q;\ft, q)\sum_{Y}Q^{|R_1|+|R_2|-|Y|}s_{R_2^t/Y}(\ft^{-\rho} q^{-\nu_1})s_{R_1^t/Y^t}( q^{-\rho}\ft^{-\nu_2}) ,
\end{multline}
and
\begin{multline}
\label{eq:schuridentity2}
\sum_{Y}Q^{|Y|}s_{Y/R_1}(\ft^{-\rho} q^{-\nu_1})s_{Y/R_2}( q^{-\rho}\ft^{-\nu_2})=\\=\calR_{\nu_2\nu_1}(Q;\ft, q)^{-1}\sum_{Y}Q^{|R_1|+|R_2|-|Y|}s_{R_2/Y}(\ft^{-\rho} q^{-\nu_1})s_{R_1/Y}( q^{-\rho}\ft^{-\nu_2}) \,,
\end{multline}
where the new function $\calR_{\nu_2\nu_1}(Q;\ft, q)$ is defined in \eqref{eq:defcalR}. The final result is \cite{Bao:2013pwa}
\be \label{eq:T2ver1}
\mathcal{Z}_2^{\text{top}}(\boldsymbol{Q})= \calM\big(Q_1Q_2\big)
\sum_{R_3}(-Q_3)^{\vert R_3\vert} \ft^{\frac{\parallel R_3^t\parallel^2}{2}} 
q^{\frac{\parallel R_3\parallel^2}{2}}
\tilde{Z}_{R_3^t}(q,\ft)\tilde{Z}_{R_3}(\ft , q)  
\calR_{R_3^t\emptyset}(Q_2;t,q)\calR_{\emptyset R_3}(Q_1;t,q)  
\ee
and using \eqref{app:Mswap}, \eqref{app:Nekswap}, \eqref{eq:deftildeZ} and \eqref{eq:defcalR} 
can be brought to the form 
\begin{align}\label{eq:T2MLN}
&\mathcal{Z}_2^{\text{top}}(P_1,P_2,P_3;t,q)
= \\
&=\frac{\mathcal{M}(P_2^2)}{\mathcal{M}\big(\sqrt{\frac{t}{q}}
\frac{P_2 P_3}{P_1}\big)\mathcal{M}\big(\sqrt{\frac{t}{q}}\frac{P_1 P_2}{P_3}\big)}
\,
\sum_{\nu}\left( \sqrt{\frac{t}{q} }\frac{P_1 P_3}{P_2}\right)^{|\nu|}
\frac{\cN_{\nu\emptyset}\left(\sqrt{\frac{t}{q}}\frac{P_2 P_3}{P_1}\right)  
\cN_{\emptyset\nu}\left(\sqrt{\frac{t}{q}}\frac{P_1 P_2}{P_3}\right)}{\cN_{\nu\nu}(t/q)}
\, .\notag
\end{align}
\paragraph{Note:} Here we have used the relations of the three K\"ahler parameters $Q_1$, $Q_2$ and $Q_3$ to 
the 5D parameters $P_1,P_2,P_3$ \eqref{eq:Kahler2Mass} and the shorthand notations 
$\cM(Q)=\cM(Q;\ft,q)$ and $\cN_{RP}(Q)=\cN_{RP}(Q;t,q)$.

\subsection{Resummation into a product formula}
\label{sec:Resummation}

In this section we will show how the partition function of the $T_2$ theory can be brought to an infinite product formula \cite{Kozcaz:2010af}
\be \label{eq:T2partitionfunction}
 \mathcal{Z}_2^{\text{top}}(\boldsymbol{Q})=
\prod_{i,j=1}^{\infty}\frac{(1-Q_1Q_2Q_3q ^{i-\frac{1}{2}} t^{j-\frac{1}{2}})
\prod_{k=1}^3(1-Q_k q^{i-\frac{1}{2}} t^{j-\frac{1}{2}})}{
(1-Q_1Q_2 q^i t^{j-1})(1-Q_1Q_3 q^{i-1} t^{j})(1-Q_2Q_3 q^{i} t^{j-1})}.
\ee
We begin with the formula \eqref{eq:T2MLN} that was derived using the topological vertex formalism. It is useful to write it, using \eqref{def:Nek2}, as $\mathcal{Z}_2^{\text{top}}(\boldsymbol{Q})=  
\mathcal{Z}_2^{\text{prod}}(\boldsymbol{Q})\check{\mathcal{Z}}_2^{}(\boldsymbol{Q})$,
with
\be \label{eq:T2middle}
\check{\mathcal{Z}}_2^{}(\boldsymbol{Q})=    \sum_R (-Q_3)^{\vert R\vert}
q^{\frac{\parallel R\parallel^2}{2}} t^{\frac{\parallel R^t\parallel^2}{2}}\prod_{(i,j)\in R}
\frac{(1-Q_1\sqrt{\frac{t}{q}} q^{1-j} t^{i-1})(1-Q_2\sqrt{\frac{q}{t}}q^{j-1} t^{1-i}) }{
(1 - q^{R_i-j+1} t^{R^t_j-i})(1 - q^{R_i-j} t^{R^t_j-i+1})} \, ,
\ee
and we denote by
$\mathcal{Z}_2^{\text{prod}} = \mathcal{M}(P_2^2) / \mathcal{M}(P_2P_3/P_1v)\mathcal{M}(P_1P_2/P_3v)$  
the part of the partition function that is already in product form.
This sum is a formal series (for which we have not given a proof of convergence yet) and it has the form
\begin{equation}
\label{eq:formalExpansion}
 \check{\mathcal{Z}}_2^{}(\boldsymbol{Q})=
  \sum_{k=0}^\infty (-Q_3)^{k}
\mathrm{P}_k (Q_1 , Q_2)
\end{equation}
where $\mathrm{P}_k (Q_1 , Q_2)$ is a degree $k$ polynomial in $Q_1$ and $Q_2$ which can also be explicitly written as
\begin{equation}
\mathrm{P}_k (Q_1 , Q_2) = \sum_{\ell=0}^k Q_1^\ell 
 \mathrm{P}_{k \ell} (Q_2) = \sum_{\ell, m=0}^k Q_1^\ell  Q_2^m \mathrm{P}_{k \ell}  \,. 
\end{equation}
The idea of the proof is to find a function that has the same formal series expansion, but which is known 
to be convergent. 

Note that the polynomials $\mathrm{P}_k (Q_1 , Q_2) $ 
have degree $k$ in both variables $Q_1$ and $Q_2$. In order to determine 
the polynomials $\mathrm{P}_k$ uniquely, it suffices to know their values for
$(k+1)^2$ different values of $Q_1$ and $Q_2$.
We will consider the values of $Q_1$ and $Q_2$ defined by
\begin{align}
\label{eq:specialisation}
Q_1(N)=q^N 
\sqrt{\frac{q}{t}}~, 
\qquad Q_2(M) = q^M \sqrt{\frac{t}{q}}~, 
\qquad \mbox{for} ~ N,M\in\mathbb{Z} ~.
\end{align} 
For these cases we can further rewrite 
expression \eqref{eq:formalExpansion} using a specialisation of the Macdonald symmetric polynomials
\begin{align}
P_R(\ft^{N-\frac{1}{2}},\ft^{N-\frac{3}{2}},\cdots,\ft^{\frac{1}{2}}; q,\ft)
=\ft^{\frac{\parallel R^t\parallel^2}{2}}\prod_{(i,j)\in R}
\frac{(1- q^{j-1}\ft^{N+1-i})}{(1- q^{R_i-j}\ft^{R_j^t-i+1})}
\end{align}
from \cite{MacdonaldSymmetric},
equation (6.11') on page 337. Then, by further using the Cauchy formula 
\begin{equation}
\sum_R Q^{|R|} P_{R}(x;q,t)   P_{R}(y;t,q) = \prod_{i,j=1}^{\infty} \left(1 + Q x_i y_j \right)
\end{equation}
 for the Macdonald symmetric polynomials, we can rewrite
\begin{align}  \label{eq:ZT2temp}
 \check{\mathcal{Z}}_2^{}(Q_1(N), & Q_2(M),Q_3)    =   \nonumber\\
 = &\sum_{R}
(-Q_3)^{\vert R\vert}
P_{R^t}(q^{N-\frac{1}{2}}, q^{N-\frac{3}{2}},\cdots, q^{\frac{1}{2}}; t,q) 
P_R( t^{M-\frac{1}{2}}, t^{M-\frac{3}{2}},\cdots, t^{\frac{1}{2}};q, t) \nonumber\\
=   &
\prod_{i=1}^N\prod_{j=1}^M\big(1-Q_3 q^{i-\frac{1}{2}} t^{j-\frac{1}{2}}\big).
\end{align}
It is not hard to find a function $\mathcal{Y}(Q_1,Q_2,Q_3)$ such that 
the values $\mathcal{Y}(Q_1(N),Q_2(M),Q_3)$ are given by the right side of (\ref{eq:ZT2temp}),
\begin{equation}
\mathcal{Y}(Q_1,Q_2,Q_3):=\prod_{i,j=1}^\infty\frac{(1-Q_3 q^{i-\frac{1}{2}} t^{j-\frac{1}{2}})(1-Q_1Q_2Q_3 
q^{i-\frac{1}{2}} t^{j-\frac{1}{2}})}{(1-Q_1Q_3 q^{i-1} t^{j})(1-Q_2Q_3 q^{i} t^{j-1})} ~.
\end{equation}
The function $\mathcal{Y}(Q_1,Q_2,Q_3)$ is meromorphic in all three variables.
In order to see that the function $\mathcal{Y}(Q_1,Q_2,Q_3)$ is analytic in a neighbourhood of 
$Q_3=0$  one may represent it in the form
\be  \label{eq:T2partitionfunctionProduct}
 \mathcal{Y}(\boldsymbol{Q})=
 \frac{\calM(Q_1Q_3 \frac{\ft}{ q} ) \calM\big(Q_2Q_3\big)}
 {\calM\big(Q_3\sqrt{\frac{\ft}{ q}}\big)\calM\big(Q_1Q_2Q_3\sqrt{\frac{\ft}{ q}}\big)},
\ee
with the function $\calM(Q)\equiv \calM(Q;\ft, q)$ 
having a representation as an exponential function of a convergent power series, 
\be
\label{eq:defMpexp}
\calM(Q;\ft, q)=\exp\left[\sum_{m=1}^{\infty} \frac{Q^m}{m}\frac{ q^m}{(1-\ft^m)(1- q^m)}\right],
\ee
which converges for all $\ft$ and all $q$ provided that $|U|< q^{-1+\theta(| q|-1)}\ft^{\theta(|\ft|-1)}$, where 
$\theta(x)$ denotes the step function which is $\theta(x)=1$ if $x>0$ and  $\theta(x)=0$ if $x\leq 0$.

The expansion of $\mathcal{Y}(Q_1,Q_2,Q_3)$ in powers of $Q_3$
has the same form as the expansion (\ref{eq:formalExpansion}), 
$\mathcal{Y}(Q_1,Q_2,Q_3)=\sum_{k=0}^\infty Q_3^k\,\mathrm{P}_k'(Q_1,Q_2)$. 
The equations 
$ \check{\mathcal{Z}}_2^{}(Q_1(N), Q_2(M),Q_3)={\mathcal{Y}}(Q_1(N), Q_2(M),Q_3)$ imply 
that the polynomials $\mathrm{P}_k^{}(Q_1,Q_2)$ and $\mathrm{P}_k'(Q_1,Q_2)$ agree on an infinite 
set of values. We must therefore have $\mathrm{P}_k^{}(Q_1,Q_2)=\mathrm{P}_k'(Q_1,Q_2)$ for all 
$k\in\mathbb{Z}$, $k\geq 0$. But this implies that 
$\check{\mathcal{Z}}_2^{}(Q_1, Q_2,Q_3)=\mathcal{Y}(Q_1,Q_2,Q_3)$, 
which gives the result we wanted to prove.

\begin{figure}[t]
   \centering
   \includegraphics[height=8cm]{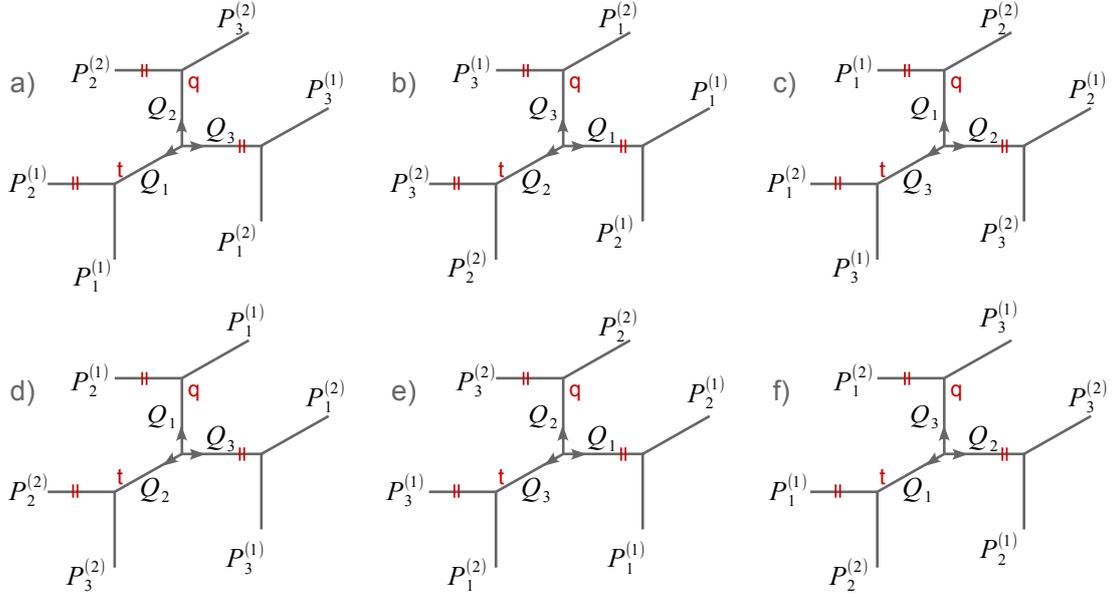} 
   \caption{\it The upper figures $a)$, $b)$ and $c)$ correspond to the three different ways to choose the the preferred direction while the three lower figures $d)$, $e)$ and $f)$ correspond to the three reflections over the middle axis labeled by the preferred direction.}
   \label{fig:PreferedDirection}
\end{figure}

\subsection{Dependence on the choice of preferred direction}

To round off the picture let us comment on the dependence on the choice of preferred direction on the web diagram. 
There are three possible ways of choosing the preferred direction and they correspond to rotating around the middle point of 
the $T_2$ web diagram the lines labeled by the preferred direction, while keeping the conventions of the topological vertex 
(Figure \ref{fig:clockwise}) fixed as well as keeping the K\"ahler parameters fixed to their position.
Equivalently, we can keep the  preferred direction and the placing of $t,q$  fixed and rotate the positions of the  K\"ahler 
parameters. This is depicted in Figure~\ref{fig:PreferedDirection}  $a)$, $b)$ and $c)$.
We may note that 
\begin{equation}
\begin{aligned}
&\calZ_{2, b}^{\text{top}}(Q_1, Q_2, Q_3)= \calZ_{2}^{\text{top}}(Q_2, Q_3, Q_1) \\ 
&\calZ_{2, c}^{\text{top}}(Q_1, Q_2, Q_3)=\calZ_{2}^{\text{top}}(Q_3, Q_1, Q_2)
\end{aligned}\qquad
\calZ_{2, a}^{\text{top}}(Q_1, Q_2, Q_3)\equiv \mathcal{Z}_2^{\text{top}}(Q_1, Q_2, Q_3).
\end{equation}

Moreover, changing the conventions and defining the vertex factor counterclockwise (instead of clockwise) leads a further 
possibility. It can be shown that
\begin{equation}
\calZ_{2, d}^{\text{top}}(Q_1, Q_2, Q_3)=  \mathcal{Z}_2^{\text{top}}(Q_2, Q_1, Q_3).
\end{equation}
The relation between $\calZ_{2, d}^{\text{top}}$ and $\mathcal{Z}_2^{\text{top}}$
may be equivalently described by 
exchanging $t\leftrightarrow q$. 
In total we get the six different options depicted in Figure \ref{fig:clockwise}.

However, inspection of the explicit formula (\ref{eq:T2partitionfunction}) reveals that
for all of these cases one finds the same functions of $Q_1,Q_2,Q_3$ in the numerator. 
The denominator factorises into three functions depending separately on only one of
the variables $P_1$, $P_2$ and $P_3$,  in the following called leg factors. 
As will be explained in Section \ref{sec:extending}, we will mostly be interested in the part of the result which 
does not factorise in this way.
It will therefore not represent an essential loss
of generality for us to
consider only  $\calZ_{2, a}^{\text{top}}(Q_1, Q_2, Q_3)=  \mathcal{Z}_2^{\text{top}}(Q_1, Q_2, Q_3)$
from now on.

\subsection{$T_2$ web with non-empty external Young tableaux}


\begin{figure}[h]
   \centering
      \includegraphics[height=3.5cm]{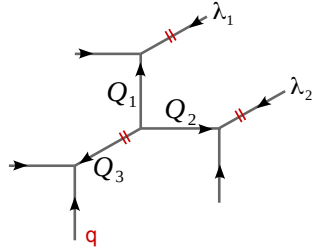} 
   \caption{\it The $T_2$ web with non-empty Young diagrams associated to the external $(1,1)$ branes.}
   \label{fig:T2-nonempty}
\end{figure}

In this paper one of our goals is to explicitly check that  by gluing two $T_2$ 
vertices we obtain the known instanton partition function of the 5D $SU(2)$ gauge theory with four flavours. 
For that we need to allow the Young diagrams associated to two of the external legs to be non-empty, like in 
Figure \ref{fig:T2-nonempty}. Then the topological strings partition function for the $T_2$ diagram is given by 
\begin{equation}\label{eq:T2-nu1}
\mathcal{Z}_{2,\vec{\lambda}}^{\text{top}}(Q_1,Q_2,Q_3;t,q)
= \sum_{\boldsymbol{\nu}} \prod_{i=1}^3(-Q_i)^{\vert \nu_i\vert} 
C_{ \emptyset\emptyset \nu_3^t}(q,\ft)C_{\emptyset \nu_2^t \lambda_2^t}(q,\ft) 
C_{\nu_1^t \emptyset\lambda_1^t }(q,\ft) C_{\nu_1\nu_2 \nu_3}(\ft , q).
\end{equation}
By applying the formulas  \eqref{eq:schuridentity1}-\eqref{eq:schuridentity2} repeatedly, 
this expression can be simplified to
\bea\label{eq:T2-nu2-6} 
\mathcal{Z}_{2,\vec{\lambda}}^{\text{top}}(Q_1,Q_2,Q_3;t,q)&=& 
\ft^{\frac{||\lambda_1^t||^2+||\lambda_2^t||^2}{2}} \tilde{Z}_{\lambda_1^t}(q,\ft) \tilde{Z}_{\lambda_2^t}(q,\ft) 
\frac{\cM(Q_1 Q_2)}{\cM(Q_1\sqrt{\frac{\ft}{q}}) \cM(Q_2\sqrt{\frac{\ft}{q}})} \nn\\
&~& \sum_\nu \left( Q_3 \sqrt{\frac{q}{\ft}} \right)^{|\nu|}
\frac{\cN_{\lambda_1 \nu}(Q_1\sqrt{\frac{\ft}{q}}) \cN_{\nu \lambda_2}(Q_2\sqrt{\frac{\ft}{q}})}{
\cN_{\nu\nu}(1) \cN_{\lambda_1 \lambda_2}(Q_1 Q_2)} ~.
\eea
At this point it is currently not known how to perform the summation over $\nu$.


\section{The 4D limit}\label{sec:4dLim}

The previous section has reviewed the derivation of the topological strings partition function for the $T_2$ brane web. 
Recall that this partition function is for the 5D uplift on $\mathbb{S}^1$ of the theory describing the  $2\times 2$ free 
hypermultiplet. We are now going to analyse its 4D limit, with $\mathbb{S}^1$ radius $R\rightarrow 0$. 
In order to define this limit precisely we find it useful to parametrise the variables 
 $q$, $t$, $P_1$, $P_2$ and $P_3$, as 
\be \label{eq:mapping3-1}
q=e^{-\epsilon_1 R}\,, \quad t=q^{\beta^2}\,,\quad
P_1^2 =\frac{ t^{\nu}}{t}  \, , \quad P_2^2 = {t^\mu \over q} \, , \quad P_3^2 = {t^\lambda \over q} \, .
\ee
The limit of our interest is $R\rightarrow 0$, keeping $\epsilon_1,\beta,\lambda,\mu,\nu$ finite.
Introducing the notation 
\be
s=\frac{1}{2}(\lambda+\mu-\nu-\beta^{-2}) ~
\ee
allows us  to rewrite the expression 
for $\mathcal{Z}_2^{\rm top}\equiv\mathcal{Z}_2^{\rm top}(\lambda,\mu,\nu;\epsilon_1,\beta,R)$ 
in the following useful form
\begin{equation}
\mathcal{Z}_2^{\rm top}=\frac{1}{\CM(t^{1+s}/q)}
\frac{\CM(t^{\lambda}/q)\CM(t^{\mu}/q)\CM(t^{\nu}/q)}
{\CM(t^{\lambda-s}/q)\CM(t^{\mu-s}/q)\CM(t^{\nu+s}/q)}~.
\end{equation}
It turns out that each of the functions $\CM$ 
becomes singular in the limit defined above, see Appendix \ref{App:phiMq21limit}
for a discussion. It follows that the function $\mathcal{Z}_2^{\rm top}(\lambda,\mu,\nu,\epsilon_1,\epsilon_2,R)$ does not
have a finite limit when $R\rightarrow 0$. In the rest of this section we will discuss how a meaningful limit 
can be defined nevertheless.

\subsection{A useful factorisation}

In the following it will be demonstrated that  $\mathcal{Z}_2^{\rm top}$ can be factorised in a singular and a regular factor\begin{equation}\label{factorsing}
\mathcal{Z}_{2}^{\rm top}(\lambda,\mu,\nu;\epsilon_1,\beta,R)=
\mathcal{Z}_{2,\rm sing}^{\rm top}(s,\epsilon_1,\beta,R)
\mathcal{Z}_{2,\rm reg}^{\rm top}(\lambda,\mu,\nu,\epsilon_1,\beta,R),
\end{equation}
where $\mathcal{Z}_{2,\rm reg}^{\rm top}(\lambda,\mu,\nu,\epsilon_1,\beta,R)$ stays finite 
in the limit $R\rightarrow 0$ and
the singular part $\mathcal{Z}_{2,\rm sing}^{\rm top}$ is explicitly 
given as 
\begin{equation}\label{Zsing-a}
\mathcal{Z}_{2,\rm sing}^{\rm top}(s,\epsilon_1,\beta,R)=\frac{1}{\CM(t/q)} 
\frac{(1-e^{-\epsilon_1R})^{\beta^2s}}{(1-e^{-\epsilon_1\beta^2 R})^s}~.
\end{equation}
In order to prove our factorisation (\ref{factorsing}) one may start by rewriting $\mathcal{Z}_2^{\rm top}$ as
\begin{equation}\label{balancfactor}
\mathcal{Z}_2^{\rm top}=\frac{1}{\CM(t/q)}\;\frac{(t^{s+1};t)_\infty (q;q)_\infty}{ (t;t)_\infty(q^{\beta^2s+1};q)_\infty}\;
\frac{\CM(t^{\lambda}/q)\CM(t^{\mu}/q)\CM(t^{\nu}/q)\CM(1)}
{\CM(t^{\lambda-s}/q)\CM(t^{\mu-s}/q)\CM(t^{\nu+s}/q)\CM(t^s)}~.
\end{equation}
The equality between the two expressions for $\mathcal{Z}_2^{\rm top}$ above is easily verified using the 
functional equations
\begin{equation}
\CM(ut)=(uq,q)_\infty\CM(u)~,\qquad
\CM(uq)=(uq,t)_\infty\CM(u)~.
\end{equation} 
The middle factor in (\ref{balancfactor}) 
may be represented   in terms of the $q$-Gamma function
\begin{equation}
\Gamma_q(x)=(1-q)^{1-x}\frac{(q;q)_\infty}{(q^x;q)_\infty}\,,
\end{equation}
as
\begin{equation}
\frac{(t^{s+1};t)_\infty (q;q)_\infty}{ (t;t)_\infty(q^{\beta^2s+1};q)_\infty}
=\frac{(1-q)^{\beta^2s}}{(1-t)^s}\frac{\Gamma_q(1+\beta^2 s)}{\Gamma_t(1+s)}.
\end{equation}
The function $\Gamma_q(x)$ is known to have the ordinary Gamma-function $\Gamma(x)$
as its limit $q\rightarrow 1$, so that \eqref{Zsing-a} displays a factorisation into a simple
singular and a finite part for $R\rightarrow 0$.

We thereby arrive at the factorisation (\ref{balancfactor}) with
 \begin{equation}
 \mathcal{Z}_{2,{\rm reg}}^{\rm top}=\frac{\Gamma_q(1+\beta^2 s)}{\Gamma_t(1+s)}
  \mathcal{Z}_{2,{\rm bal}}^{\rm top},\quad \mathcal{Z}_{2,{\rm bal}}^{\rm top}=
\frac{\CM(t^{\lambda}/q)\CM(t^{\mu}/q)\CM(t^{\nu}/q)\CM(1)}
{\CM(t^{\lambda-s}/q)\CM(t^{\mu-s}/q)\CM(t^{\nu+s}/q)\CM(t^s)}
\end{equation}
which will be shown to have a finite limit when $R\rightarrow 0$.

\subsection{The limit of the regular part}

In order to show that  $\mathcal{Z}_{2,{\rm reg}}^{\rm top}$ has a finite limit when $R\rightarrow 0$
it is useful to observe that it is a ratio of products of functions 
that is ``perfectly balanced'' in the following sense. For any given function $F(x)$ 
we may call ratios of the form
\begin{equation}\label{Rdef}
R(\lambda,\mu,\nu,\delta)=
\frac{F(\delta)}{F(\delta-s)}\frac{F(\nu)}{F(\nu-s)}
\frac{F(\mu)}{F(\mu+s)}\frac{F(\lambda)}{F(\lambda+s)}
\end{equation} 
perfectly balanced if $(\lambda,\mu,\nu,\delta)$ satisfy $\mu+\lambda-\nu-\delta=2s$. If this is the
case one easily finds that the function $\tilde{R}(\lambda,\mu,\nu,\delta)$ obtained by 
replacing the function $F(x)$ in (\ref{Rdef}) by $\tilde{F}(x)=e^{\alpha x^2+\beta x}F(x)$ is 
identically equal to 
the function $R (\lambda,\mu,\nu,\delta)$. It is easily checked that $ \mathcal{Z}_{2,{\rm bal}}^{\rm top}\equiv
\mathcal{Z}_{2,{\rm bal}}^{\rm top}(\lambda,\mu,\nu,\beta^{-2})$ is perfectly balanced in this sense. 
It can be represented as the infinite product
\be  
\mathfrak{T}_s (\delta, \nu , \mu, \lambda)
=
\prod_{i,j=0} \mathfrak{t}_{ij}(\delta, \nu , \mu, \lambda)~,
\ee
where $\delta=\beta^{-2}$ and
\be
\mathfrak{t}_{ij}(\delta, \nu , \mu, \lambda)=
 \vartheta_{ij}^{+}(\delta,s) 
 \vartheta_{ij}^{+}(\nu,s) 
 \vartheta_{ij}^{-}(\mu,s) 
 \vartheta_{ij}^{-}(\lambda,s) 
  \,,\quad
\vartheta^\pm_{ij}(\mu,s)=\frac{1-t^{\mu\pm s+i}q^j}{1-t^{\mu+i}q^j}.
\ee
In the limit $R\rightarrow 0$ one finds that 
\be
\mathfrak{T}_s (\delta , \nu , \mu, \lambda) \rightarrow  \mathfrak{R}_s (\delta , \nu , \mu, \lambda)=
\prod_{i,j=0} \mathfrak{r}_{ij}(\delta, \nu , \mu, \lambda)~,
\ee
where 
\be
\mathfrak{r}_{ij}(\delta, \nu , \mu, \lambda)=
 \varrho_{ij}^{+}(\delta,s) 
 \varrho_{ij}^{+}(\nu,s) 
 \varrho_{ij}^{-}(\mu,s) 
 \varrho_{ij}^{-}(\lambda,s) 
  \,,\quad
\varrho^\pm_{ij}(\mu,s)=\frac{j+ \beta^2 (\mu \pm s +i)}{j+ \beta^2 (\mu  +i)}\,.
\ee
The crucial point to observe is that the infinite product that defines
$\mathfrak{R}_s (\delta , \nu , \mu, \lambda)$ is still absolutely convergent thanks to the 
fact that it is the product of perfectly balanced factors. In order to see this, let us introduce
\be\label{eq:Gammabeta}
\Gamma_{\beta}(\beta x) = {\Gamma_2(x|1,\beta^{-2})}~, 
\ee
with $\Gamma_2(x|\epsilon_1,\epsilon_2)$ defined by the absolutely convergent infinite product
\be
\label{defGamma2}
 \Gamma_2(x|\epsilon_1,\epsilon_2)=\frac{e^{-\alpha x +\frac{\beta x^2}{2}}}{x} \prod'_{n_1,n_2\geq 0}\frac{e^{\frac{x}{\epsilon_1n_1+\epsilon_2n_2}-\frac{x^2}{2(\epsilon_1n_1+\epsilon_2n_2)^2}}}{1+\frac{x}{\epsilon_1n_1+\epsilon_2n_2}}  \, , \quad \epsilon_1,\epsilon_2>0 \, .
\ee
It is then easy to see that
\begin{equation}\label{R-Gamma}
\mathfrak{R}_s (\beta^{-2} , \nu , \mu, \lambda)=
\frac{\Gamma_\beta(\beta^{-1})}{\Gamma_\beta(\beta^{-1}+\beta s)}
\frac{\Gamma_\beta(\beta\nu)}{\Gamma_\beta(\beta(\nu+s))}
\frac{\Gamma_\beta(\beta\mu)}{\Gamma_\beta(\beta(\mu-s))}
\frac{\Gamma_\beta(\beta\lambda)}{\Gamma_\beta(\beta(\lambda-s))}.
\end{equation}
Indeed, each factor in the infinite product over $i,j$ obtained by inserting (\ref{defGamma2}) into 
(\ref{R-Gamma}) is perfectly balanced, making it easy to see that all 
exponential factors cancel each other, factor by factor. Thus the infinite 
product defining $\mathfrak{R}_s (\delta , \nu , \mu, \lambda)$ is 
absolutely convergent since the infinite products defining the function $\Gamma_{\beta}(x)$ also have this property.

\subsection{Renormalising the singular part}
\label{sec:RenormalizationAmbiguity}


One may now be tempted to simply define $\mathcal{Z}_{2}^{\rm 4d}$ to be 
$\mathcal{Z}_{2,reg}^{\rm top}(\lambda,\mu,\nu,\epsilon_1,\beta,R)$.
However, it is clear that the factorisation (\ref{factorsing}) is ambiguous. 
One could 
modify $\mathcal{Z}_{2,sing}^{\rm top}$ by multiplying 
it with an arbitrary function  while dividing $\mathcal{Z}_{2,reg}^{\rm top}$
by the same function. 
Additional requirements have to be imposed in order to 
arrive at an unambiguous definition for $\mathcal{Z}_{2}^{\rm 4d}$. 

In our case it seems natural to require that the key analytic properties of the function 
$\mathcal{Z}_{2}^{\rm top}$ are preserved in the 
limit.
In this regard let us note that
the factorisation (\ref{factorsing}) has some special features distinguishing it
from other possible factorisations. The singular piece,
here recalled for convenience
\begin{equation}\label{Zsing}
\mathcal{Z}_{2,\rm sing}^{\rm top}(s,\epsilon_1,\beta,R)=\frac{1}{\CM(t/q)} 
\frac{(1-e^{-\epsilon_1R})^{\beta^2s}}{(1-e^{-\epsilon_1\beta^2 R})^s}~,
\end{equation}
depends (i) 
on the variables $\lambda,\mu,\nu$ only through the combination $s=\frac{1}{2}(\lambda+\mu-\nu-\beta^{-2})$
and (ii) depends on the variable $s$ in a very simple way: the dependence of 
$\mathcal{Z}_{2,\rm sing}^{\rm top}$ on the variable $s$ 
is  entire analytic, $\mathcal{Z}_{2,\rm sing}^{\rm top}$ is nowhere vanishing
as function of $s$, and $\mathcal{Z}_{2,\rm sing}^{\rm top}$ has at most exponential 
growth. This means that $\log\mathcal{Z}_{2,\rm sing}^{\rm top}$ is a linear 
function. 

Imposing the requirement that these features are preserved in the limit $R\rightarrow 0$
eliminates most of the ambiguities in the renormalisation of $\mathcal{Z}_{2}^{\rm top}$.
The factor $\frac{1}{\CM(t/q)} $ does not depend on $s$ at all, while 
\begin{equation}\label{Zsing}
\frac{(1-e^{-\epsilon_1R})^{\beta^2s}}{(1-e^{-\epsilon_1\beta^2 R})^s}\sim 
R^{s(\beta^2-1)}\,\beta^{-2s}\,.
\end{equation}
We conclude that the most general renormalised limit $R\rightarrow 0$ satisfying the requirements
formulated above is 
\begin{equation}
\lim_{R\rightarrow 0} \eta(\rho R^{\beta^2})^{-s} \CM(t/q) \mathcal{Z}_{2}^{\rm top}(\lambda,\mu,\nu;\epsilon_1,\beta,R).
\end{equation}
The factors $\eta$ and $\rho^{-s}$ represent the ambiguity in the definition of the limit that cannot be removed by the
requirements above.

Collecting our findings above, introducing the notations
\begin{equation}\label{eq:mapping3-2}
\beta \lambda =2a_3+2\beta^{-1}-\beta ,\qquad\beta \mu = - 2a_2+\beta,
\qquad \beta \nu = 2a_1 + \beta^{-1},
\end{equation}
and using the identity
\begin{equation}
\frac{\Gamma_{\beta}(\beta^{-1})}{\Gamma_\beta(\beta^{-1}+\beta s)}=\beta^{-s(1+\beta^2)}
\frac{\Gamma(1+s)}{\Gamma(1+\beta^2s)}\frac{\Gamma_{\beta}(\beta)}{\Gamma_\beta(\beta+\beta s)}
\end{equation}
we arrive at the statement that
\begin{align}
& \lim_{R\rightarrow 0}\eta (\rho R^{\beta^2})^{-s} \CM(t/q)  \mathcal{Z}_{2}^{\rm top}
 =\eta {\big(\beta^{1+\beta^2}\rho\big)^{-s}}
\frac{\Gamma_\beta(\beta)}{\Gamma_\beta(\beta(1+ s))}\\
& \qquad \qquad \qquad \qquad \times 
\frac{\Gamma_\beta(\beta^{-1}+2a_1)}{\Gamma_\beta(\beta^{-1}+2a_1+ s \beta )}
\frac{\Gamma_\beta(\beta - 2a_2)}{\Gamma_\beta((1- s) \beta -2a_2 )}
\frac{\Gamma_\beta(2\beta^{-1}-\beta + 2a_3)}{\Gamma_\beta(2\beta^{-1}- (s+1)\beta + 2a_3)} \, .
\notag\end{align} 
With $a_3=a_1+a_2+s \beta$, the above equation becomes
\begin{align}
& \lim_{R\rightarrow 0}\eta (\rho R^{\beta^2})^{-s} \CM(t/q)  \mathcal{Z}_{2}^{\rm top}
 =\eta {\big(\beta^{1+\beta^2}\rho\big)^{-s}}
\frac{\Gamma_\beta(\beta)}{\Gamma_\beta(\beta+ a_3-a_1-a_2))}\\
& \qquad \qquad   \times 
\frac{\Gamma_\beta(\beta^{-1}+2a_1)}{\Gamma_\beta(\beta^{-1}+a_1+a_3-a_2)}
\frac{\Gamma_\beta(\beta -2a_2)}{\Gamma_\beta(\beta+a_1-a_2-a_3)}
\frac{\Gamma_\beta(2\beta^{-1}-\beta + 2a_3)}{\Gamma_\beta(2\beta^{-1}-\beta +a_3-a_1-a_2)} \, .
\notag\end{align} 
Note that the arguments 
of the double gamma functions can {\it not} all be positive when $a_2>0$.
However, iff $a_2<0$ there exists a regime where the 
arguments of the double gamma functions are all positive.
This should be compared with the expression for the topological string partition function in terms of 
K\"ahler parameters $Q_i$, $i=1,2,3$. It is manifest in formula \eqref{eq:T2partitionfunctionProduct} that the 
combinations of 
K\"ahler parameters appearing in this expression are all positive. 

Recall that the calculation above is performed for the case where
$q=e^{-\epsilon_1 R}<1$ and $t=q^{\beta^2}<1$.
In the regime where $|t|>1$ it is more natural to parameterise $t=q^{- b^2}>1$. In this case we obtain
\begin{align}
\lim_{R\rightarrow 0}\eta(\rho R^{\beta^2})^{-s}   \CM(t/q)  \mathcal{Z}_{2}^{\rm top}=
& \eta  (b^{1-b^2} \rho )^{-s}
\left(\frac{b^{1+b^2}}{2\pi\mathrm{i}}\Gamma(-b^2)\right)^{-s}
\frac{\Gamma_b(-s b
)}{\Gamma_b(0)}\\
&\frac{\Gamma_b(Q+2 \alpha_1-sb)\Gamma_b(-2 \alpha_2+sb)\Gamma_b(2\alpha_3+2Q+sb)}
{\Gamma_b(Q+2 \alpha_1)\Gamma_b(-2 \alpha_2)\Gamma_b(2Q+2\alpha_3)}.
\notag\end{align}
Comparing to the Liouville three point function we see that 
the parameter $\rho$ parametrising the ambiguity in the definition of the limit $R\rightarrow 0$
gets related to the parameter called `cosmological constant' in the  Liouville CFT literature.

\subsection{Extending the domain of definition of the $T_2$ partition functions}\label{sec:extending}

It will be 
useful to parameterise the distances between the pairs of legs emanating from the $T_2$-vertex vertically, horizontally 
and diagonally by the variables $p_1$, $p_2$ and $p_3$, respectively, which is 
equivalent to the parameterisation
\begin{equation}
P_1^2=q^{2\beta p_1}, \qquad P_2^2=q^{2\beta p_2},\qquad P_3^2=q^{2\beta p_3} ~.
\end{equation}
From the point of view of the world-sheet sigma model it is not unexpected that the definition of the partition function
is somewhat ambiguous. As we are dealing with non-compact target spaces having infinite ends indicated by the pairs 
of parallel external lines in the toric diagram, we expect that naive definitions of the partition functions will be divergent,
and need to be regularised. As the cut-off defining the regularisation for a pair of external lines could depend on the 
parameter describing the asymptotic geometry of the corresponding infinite end, one expects that changes of the
cut-offs will change the partition functions by multiplicative factors depending on one of the variables $p_i$, $i=1,2,3$, only.
Such factors will be called leg factors in the following. Contributions to the partition functions with mixed dependence on 
$p_1,p_2,p_3$, on the other hand, are naturally interpreted as contributions coming from a compact region in the target 
space containing the region where the three legs meet. We'd therefore expect that such contributions are meaningful.

\begin{figure}[t]
   \centering
      \includegraphics[height=9.5cm]{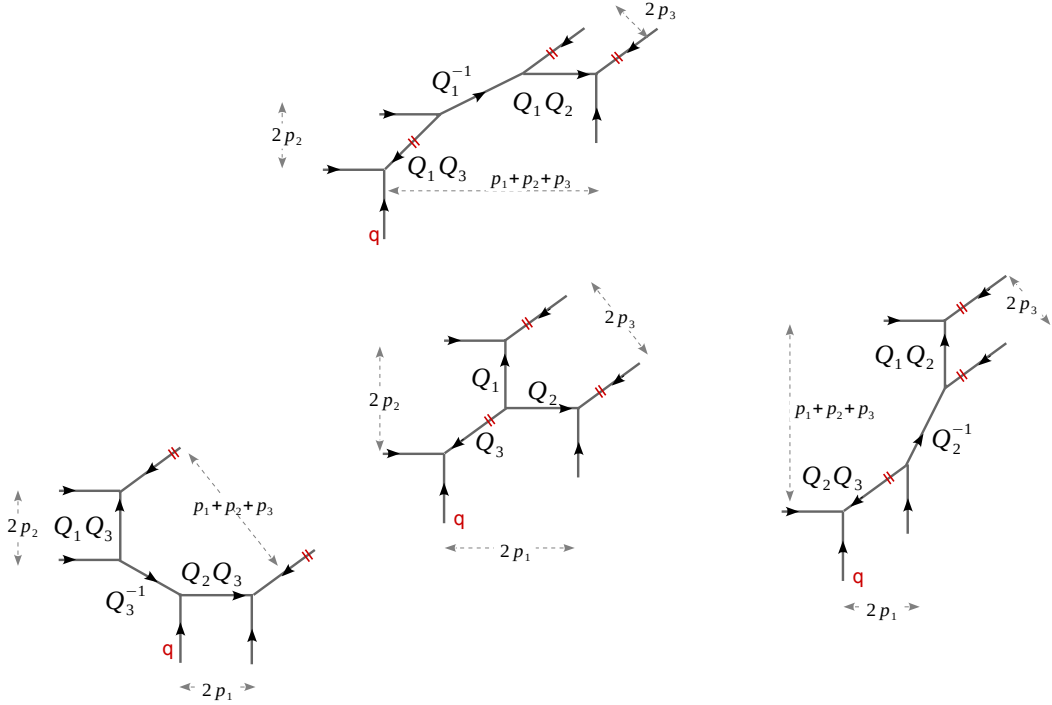} 
   \caption{\it Different $T_2$ diagrams related by flops. }
   \label{fig:T2flops}
\end{figure}

The result for the $T_2$ vertex can then be represented as
\begin{equation}\label{ZT2simple}
{\mathcal{Z}}^{T_2}(p_1,p_2,p_3)=\frac{[\text{Leg factors}]}{G_\beta(p_1+p_2+p_3)
\prod_{i=1}^3 G_\beta(p_1+p_2+p_3-2p_i)},
\end{equation}
using the notation 
\be 
G_\beta(x)=\Gamma_\beta\bigg(\frac{1}{2}(\beta+\beta^{-1})+x\bigg) ~.
\ee 
This result has been derived under the condition that the K\"ahler parameters in the $T_2$ diagram are
all positive, which is equivalent to the inequalities
\begin{equation}
p_1+p_2>p_3,\qquad p_2+p_3>p_1,\qquad p_3+p_1>p_2.
\end{equation}
Toric CY having sets of parameters $p_i$, $i=1,2,3$ 
violating any of these inequalities are related to the toric CY described by the $T_2$ 
by flop transitions. The corresponding toric diagrams are depicted in Figure
\ref{fig:T2flops}.

The subspace $\mathfrak{P}_{\mathbb{R}}$ of the space of 
parameters of the $T_2$ diagrams with real and positive K\"ahler parameters
can be parameterised by the variables $p_i$, $i=1,2,3$. This space is covered exactly once if $p_i>0$ for $i=1,2,3$.
It breaks up into four chambers $\mathfrak{P}_{\mathbb{R}}^{(i)}$, $i=1,2,3$ and $\mathfrak{P}_{\mathbb{R}}^{(s)}$, 
defined by the positivity of all
K\"ahler parameters. These chambers are
in one-to-one correspondence with the toric diagrams in Figure \ref{fig:T2flops}, with 
$T_2^{(s)}\equiv T_2$ being the most symmetric one in the middle.
Note that the parameter
$p_i$ describing the width of the corresponding region can grow arbitrarily large in the 
chamber $\mathfrak{P}_{\mathbb{R}}^{(i)}$, $i=1,2,3$. One may easily check that the arguments of the 
functions $G_\beta$ appearing in the expressions for $\mathcal{Z}^{T_2^{(i)}}$
are always positive within the respective chambers $\mathfrak{P}_{\mathbb{R}}^{(i)}$ for $i=1,2,3,s$.

Each of the new diagrams also has three asymptotic regions bounded by two parallel
edges. The distances between these parallel legs are $\sum_{i=1}^3 p_i$, 
$2p_{i+1}$ and $2p_{i+2}$ respectively for the diagram $T_2^{(i)}$, using the notations $p_{i+3}=p_i$. 
There are known simple rules describing the ratios of the topological 
string partition functions associated to toric CY related by flop transitions \cite{Konishi:2006ev,Mitev:2014jza,Coman:2018uwk}.
A flop transition describing the continuation from positive to negative values of 
a K\"ahler parameter $\log Q_i$ will be described 
by substituting the factor $\cM(Q_i)$ in $\mathcal{Z}^{T_2^{(i)}}$ 
by $\cM(Q_i^{-1})$.
 The corresponding partition functions are thereby found to be  
\begin{equation} \label{eq:floppedT2s}
\mathcal{Z}^{T_2^{(i)}}=\frac{[\text{Leg factors}]}{\prod_{\epsilon,\epsilon'}G_\beta(p_i+\epsilon p_{i+1}+\epsilon'p_{i+2})} ~.
\end{equation} 
The dependence of the denominator on the variable $p_i$ associated to the distinguished asymptotic region is 
somewhat special in the sense that it is not invariant under $p_i\rightarrow -p_i$, as is manifestly the case for $j\neq i$.



\section{Integral representation of the $T_2$ partition function}
\label{sec:topol2matrix}

%
%
%
%
%
%

In order to determine the map between the K\"ahler parameters for the $T_2$ vertex and the variables 
parameterising chiral vertex operators in Liouville conformal field theory, we will now derive an integral representation 
for the $T_2$ partition function $\mathcal{Z}_2^{\text{top}}$ \eqref{eq:T2MLN} specialised to a discrete family of 
two-dimensional subspaces inside the three-dimensional parameter space. 
The specialisation of parameters is known in the CFT literature as the screening condition. 
We will follow the approach of \cite{Aganagic:2013tta, Aganagic:2014oia}, being interested in the 
precise form of the normalisation factors which had not been determined before. 
These normalisation factors will turn out to be important for us later.

\subsection{Imposing the specialisation condition to the topological strings}

We begin with the topological string partition function for $T_2$ derived in Section \ref{sec:TopologicalPartitionFun} and written in equation \eqref{eq:T2MLN}, which we rewrite here for the convenience of the reader:
\begin{align} \label{TopolT_2-B}
&\mathcal{Z}_2^{\text{top}}(P_1,P_2,P_3;t,q)
=\\
&=\frac{\mathcal{M}(P_2^2)}{\mathcal{M}\big(\sqrt{\frac{t}{q}}\frac{P_2 P_3}{P_1}\big)
\mathcal{M}\big(\sqrt{\frac{t}{q}}\frac{P_1 P_2}{P_3}\big)}
\,
\sum_{\nu}\left( \sqrt{\frac{t}{q} }\frac{P_1 P_3}{P_2}\right)^{|\nu|}
\frac{\cN_{\nu\emptyset}\left(\sqrt{\frac{t}{q}}\frac{P_2 P_3}{P_1}\right)  
\cN_{\emptyset\nu}\left(\sqrt{\frac{t}{q}}\frac{P_1 P_2}{P_3}\right)}{\cN_{\nu\nu}(t/q)}
\, .
\notag\end{align}
Following \cite{Aganagic:2013tta,Aganagic:2014oia}, an important first step is to observe that if we impose
 \be
 \label{eq:screening}
\sqrt{\frac{t}{q}}\frac{P_2 P_3}{P_1} = \frac{t}{q} \, t^s ~,
\ee
with $s$ an integer and the length of the partition $\nu$ denoted $N_\nu$ that satisfies $N_{\nu}>s$, then the 
partition function vanishes. Therefore, after substituting \eqref{eq:screening} in \eqref{TopolT_2-B} we can safely 
replace  $N_{\nu}$ by $s$ for the products which exist inside the expansion of the functions $\cN_{\mu\nu}$. 
Then, using equations \eqref{def:Nek1} and 
\eqref{def:Nek2} we can recast the topological string partition function of $T_2$ as
\bea
\nonumber
\mathcal{Z}_2^{\text{top}}
&= &
\frac{\prod_{i=1}^s \varphi(P_2^2qt^{-i})}{ \mathcal{M}\big(v^{-2}\big)   \prod_{i=1}^s \varphi(t^{i})}
\,
\sum_{\nu}\left(\frac{P_1 P_3}{v P_2}\right)^{|\nu|}
\\
&& \times 
\prod^{s}_{i=1} \frac{\varphi(P_2^2 v^2 /y_{\nu,i})}{\varphi(P_2^2 v^2 /y_{\emptyset,i})}
\prod^{s}_{i=1} \frac{\varphi( 1 /y_{\emptyset,i})}{\varphi(1 /y_{\nu,i})}
  \prod_{i,j=1}^{s} 
\frac{\varphi(q^{\nu_i-\nu_j} t^{j-i})}{\varphi(q^{\nu_i-\nu_j} t^{j-i+1})}
\frac{\varphi(t^{j-i+1})}{\varphi( t^{j-i})}
\label{eq:specialisationStep}
\eea
where the variables $y$ are 
\be 
y_{\nu,i} = q^{\nu_i} t^{s-i} ~ \,.
\ee
Further massaging the equation \eqref{eq:specialisationStep} we can rewrite the topological 
strings partition function as
\be \label{TopolT_2-C}
\mathcal{Z}_2^{\text{top}}
=
\frac{\prod_{i=1}^s \varphi(P_2^2qt^{-i})}{ \mathcal{M}\big(v^{-2}\big)   \prod_{i=1}^s \varphi(t^{i})}
\,
\sum_{\nu}\left(\frac{P_1 P_3}{v P_2}\right)^{|\nu|}
 ~
\frac{\cI_m(y_\nu)}{\cI_m(y_\emptyset)}
\frac{\cI_{1,1} (y_{\nu})}{\cI_{1,1} (y_{\emptyset })} ~ ,
\ee 
having defined \cite{Aganagic:2013tta,Aganagic:2014oia} 
\be \label{eq:integrantsMain}
\cI_m(y_\nu) =  \prod_{i=1}^{s} 
\frac{\varphi(P_2^2 v^2/y_{\nu,i})}{\varphi(1/ y_{\nu,i})} \, .
\, \, 
\cI_{1,1} (y_\nu) = \prod_{i\neq j=1}^{s} 
\frac{\varphi(q^{\nu_i-\nu_j} t^{j-i})}{\varphi(q^{\nu_i-\nu_j} t^{j-i+1})}= \prod_{i\neq j =1}^{s} 
\frac{\varphi(y_{\nu,i}/y_{\nu,j})}{\varphi(ty_{\nu,i}/y_{\nu,j})} \, 
\ee
For the intermediate steps to reproduce this calculation we note that when $s$ is the number 
of rows of the partition $\nu$, the expansion \eqref{def:Nek4} of the Nekrasov function $\cN_{\nu\nu}(t/q)$ 
into quantum dilogarithms allows to obtain $\cI_{1,1}(y)$ from
\bea 
\cN_{\nu\nu}(t/q)^{-1} &=& 
\cN_{\nu\emptyset}\left(t^{1+s}/q \right)^{-1}
\cN_{\emptyset \nu} \left(t^{1-s}/q \right)^{-1}
\frac{\cI_{1,1} (y_{\nu})}{\cI_{1,1} (y_{\emptyset })} 
\, .
\eea
Then, the functions 
$\cN_{\nu\emptyset}$ combine to give $\cI_m(y)$
\be 
\frac{
\cN_{\emptyset \nu}(P_2^2t^{-s})}{
\cN_{\emptyset \nu} \left(t^{1-s}/q \right)} = 
\frac{\cI_m(y_\nu)}{\cI_m(y_\emptyset)}
\, .
\ee

\subsection{The matrix integral as a sum of residues}
\label{sec:Residues}

The summation inside the topological string partition function \eqref{TopolT_2-C} is related to the matrix integral
\be \label{q-MatrixModel}
\mathcal{I}_2 = \int \frac{d'_qy_1}{y_1} \cdots \frac{d'_qy_s}{y_s} \,  \prod_{i =1}^s y^{\zeta+1}_i  \,  \mathcal{I}_{1,1}(y)  \,  \mathcal{I}_{m}(y) \, ,
\ee
where the parameter $\zeta$ is defined by 
\be\label{eq:zata-a1}
q^{\zeta+1} = \sqrt{\frac{t}{q} } \frac{P_1 P_3}{P_2}  \, ,
\ee
and
the integrals $\int d'_qy_1 \cdots d'_qy_s \prod_{i=1}^s y_i^{-1} $ are variants of the Jackson integral 
defined for meromorphic functions $M(y)$ of $s$ variables $y=(y_1,\dots,y_s)$ as  a sum over residues
\begin{equation}
\label{eq:varJack}
\int \frac{d'_qy_1}{y_1} \cdots \frac{d'_qy_s}{y_s} 
\; M(y):=(2\pi \mathrm{i})^s \sum_{\substack{R_1,\dots,R_s\in\mathbb{N}\\R_1>R_2>\dots R_s}}
\mathop{\rm Res}_{y=y_R} M(y)  \, .
\end{equation}

\paragraph{Summation over residues:} 

To understand the relation between the integral \eqref{q-MatrixModel} and its expression as a sum over 
residues precisely, we discuss the pole structure of the integrand, noting that it parallels to a large extent 
the analysis of \cite{Aganagic:2013tta,Aganagic:2014oia}. We nevertheless review this here for clarity and 
completeness. Assuming a radial ordering of the poles $|y_i|<|y_{i+1}|$, these originate from:
\begin{itemize}
\item $\mathcal{I}_{m}(y)$: outermost pole $y_s= q^{\nu_s}$ 
\item $\mathcal{I}_{1,1}(y)$: poles organised by a partition $\nu$, with $ y_i = q^{\nu_i} t^{s-i}~, ~1\leq i<s$.
\end{itemize}
The reasoning behind this statement is as follows. The 
poles of the function $\mathcal{I}_{m}(y)$ are located 
at $y_i = q^m,~m\in\mathbb{N}$ while, in the regime where 
$|q|,|t|<1$, those of the function $\mathcal{I}_{1,1}(y)$ 
satisfy $y_i/y_{i+1} = q^n t$. The outermost pole $y_s$ 
therefore has to belong to $\mathcal{I}_{m}(y)$, since it 
would otherwise be inconsistent with the radial ordering. 
Having established this fact, no other poles can 
originate from $\mathcal{I}_{m}(y)$ because any such singularities would be 
cancelled by zeros of $\mathcal{I}_{1,1}(y)$. Consequently, all of the remaining poles belong to the 
function $\mathcal{I}_{1,1}(y)$ and 
$y_{s-1}/y_s = q^n t$ implies $y_{s-1}=q^{\nu_{s-1}}t$, with $\nu_{s-1}>\nu_{s}$. Iterating this logic, we find 
the remaining poles $ y_i = q^{\nu_i} t^{s-i}$, the set 
of which is therefore labelled by a Young tableau $\nu$   \be
  y_\nu = \left( q^{\nu_1} t^{s-1} , \dots , q^{\nu_{s-1}} t, q^{\nu_s}  \right)  \, .
\ee
The matrix integral \eqref{q-MatrixModel} is therefore defined through the sum
\be
\label{Integral2Res}
\mathcal{I}_2 =  (2\pi i)^s  \left( \mbox{Res}_\emptyset \right)  \sum_{\nu}   q^{(\zeta+1)  |\nu|} \frac{\mathcal{I}_{1,1}(y_\nu)}{\mathcal{I}_{1,1}(y_\emptyset)}
\frac{\mathcal{I}_{m}(y_\nu)}{\mathcal{I}_{m}(y_\emptyset)}  ~, 
\ee
where the coefficient
\be
 \mbox{Res}_\emptyset   =  t^{\frac{1}{2}s(s-1) (\zeta+1) } \left(\frac{ \varphi(t)}{\varphi(q)}\right)^{s} \prod_{i =1}^s \frac{\varphi(P_2^2 q t^{i-s-1})}{\varphi(t^i)} \, 
\ee
is obtained by evaluating the integral \eqref{q-MatrixModel} with $\nu$ empty. One thus finds the same summation 
over Young diagrams on the right hand side of both equations \eqref{TopolT_2-C} for the $T_2$ 
topological string partition function and \eqref{Integral2Res} for the integral formulation. 


%

\subsection{The free field representation for $q$-Liouville}

In this section we briefly discuss
the relation between  the matrix integral \eqref{q-MatrixModel}
and mathematical objects 
called $q$-deformed Virasoro conformal blocks which represent a
deformation of the integrals representing conformal blocks in conformal field theory.
The $q$-deformation of 
a three point conformal block 
on $\mathbb{CP}^1 \setminus \left\{0,1,\infty\right\}$ 
with primary fields  $V_{a_1}$, $V_{a_2}$ and $V_{a_3}$   inserted at the locations of the punctures
and with $\beta s=a_3-a_2-a_1$, for example,  can be written as
 \be 
\cB_{q-\text{Liouv}}  
\equiv 
   \oint d y  ~ \cI_{q-\text{Liouv}} 
\ee
with the  integrand 
\be 
\cI_{q-\text{Liouv}} =  ~
\prod_{i=1}^s y_i^{2\beta a_1}
 \langle S (y_i) V_{a_2} (1)  \rangle 
\prod_{j>i}^s 
 \langle S (y_j) S (y_i)  \rangle ~,
\ee
where without any loss of generality we send $z_3\to \infty$.
The two point function of $q$-Liouville 
between a primary field and a screening current is   \cite{Feigin:1995sf,1997FrenkelReshetikhin}
\be
\label{eq:2ptVS}
\langle S (y_i) V_{a_2} (z)  \rangle =  y_i^{2\beta a_2} \frac{\varphi(q^{-2\beta a_2}/y_i)}{\varphi(1/y_i)}  ~, 
\ee
while that between two screening currents  is   \cite{Feigin:1995sf,1997FrenkelReshetikhin}
\be 
\label{eq:2ptSS}
\langle  S (y_j)   S (y_i) \rangle 
= (y_j)^{2\beta^2} ~ \frac{\varphi(y_i/y_j)\varphi(qt^{-1} y_i/y_j)}{\varphi(q y_i/y_j)\varphi(t y_i/y_j)} ~, \quad 
|y_i|<|y_j| ~,~ i<j  ~ .
\ee
One should note that a different definition of the two point function (\ref{eq:2ptSS})
has been used in the references  \cite{Aganagic:2013tta,Aganagic:2014oia}.

We may already observe that the factor $\cI_m$ in \eqref{eq:integrantsMain} is proportional to 
the two-point function (\ref{eq:2ptVS}) provided 
the parameter $P_2$ is related to $a_2$ by $P_2^2 = \frac{t}{q} q^{-2 \beta a_2}$.
To also reach agreement between the ratio of quantum dilogarithms in the two point function 
\eqref{eq:2ptSS} and the function $\cI_{1,1}$ inside \eqref{eq:integrantsMain}, it is necessary to rewrite 
\eqref{eq:2ptSS} using the following function
\be
\label{eq:vartheta1}
\vartheta_q \left(\frac{y_{i}}{y_{j}},1-\beta^2 \right)  
=
\left( \frac{y_{j}}{y_{i}} \right)^{\beta^2}
\frac{\varphi(qt^{-1}y_{i}/y_{j})}{\varphi(q y_{i}/y_{j})} 
\frac{\varphi(t y_{j}/y_{i})}{\varphi(y_{j}/y_{i})}
\, .
\ee
The function $\vartheta_q(x, \kappa)$  is a quasi-constant, meaning that it does not depend on multiplicative shifts of the argument by $q$,  
\be
\vartheta_q(x, \kappa)= \vartheta_q(qx, \kappa) \, .
\ee
If we now introduce the short-hand grouping all quasi-constants
\be 
\vartheta_q(y,s)=
 \prod_{j>i}^{s} 
\vartheta_q \left(\frac{y_i}{y_j},1-\beta^2 \right) ~,
\ee
gather the contributions of all   factors $y_i$ into $y_i^\zeta$ with
\be     
\zeta = 2\beta(a_1 + a_2 ) + \beta^2(s-1)  ~, 
\ee
then the three point conformal block  $\cB_{q-\text{Liouv}}$ in the free field representation with 
$z_1=0,z_2=1$  and $z_3 = \infty$ can be written proportional to the matrix model integral \eqref{q-MatrixModel} 
 \begin{align}
\label{def:qLiouvInt-secT2}
\cB_{q-\text{Liouv}}  
 &= 
    \int d'_qy_1 \cdots d'_qy_s  ~\prod_{i=1}^s y_i^\zeta ~ \cI_m (y) \cI_{1,1} (y) \vartheta_q(y,s)\\
   & =
    \, \vartheta_q(y,s) \,  \cI_2 (a_1,a_2,s;\beta)  \, .
  \nn \end{align}
The quasi-constant factors, through their invariance with respect to the multiplicative factor $q$, can all 
be pulled outside of the integral in equation \eqref{def:qLiouvInt-secT2}. 
Furthermore, the power $\zeta$ appearing in this equation, together with the identity \eqref{eq:zata-a1} and 
the specialisation \eqref{eq:screening}, fixes the remaining entries of the dictionary 
between the topological string parameters $P_1$ and $P_3$ and the conformal field theory momenta $a_1$ and $a_2$
to be
\be
\label{eq:ParameterIdentification}
P_1^2 = \frac{q}{t} q^{2 \beta a_1 }
\, , \quad 
P_2^2 = \frac{t}{q} q^{-2 \beta a_2} \, , 
\quad P_3^2 = \frac{q}{t} q^{ 2 \beta a_3}
\, , 
\quad 
t= q^{\beta^2}
\, ,
\ee
where $a_3=a_1+a_2 + s\beta$.\footnote{ 
We remark that the dictionary \eqref{eq:ParameterIdentification} fully agrees with equations 
\eqref{eq:mapping3-1} and \eqref{eq:mapping3-2} in Section \ref{sec:4dLim}.} 

Given the relations (\ref{eq:ParameterIdentification}) between the relevant parameters 
the relation between the $T_2$ topological strings partition function \eqref{TopolT_2-C} 
and the matrix model integral \eqref{q-MatrixModel} is
\be
\label{relationTopMM}
\quad
\mathcal{I}_2 (a_1,a_2,s;\beta) = (2\pi i)^s t^{\frac{1}{2}s (s-1) (\zeta +1) } \left( \frac{\varphi(t)}{\varphi(q)} \right)^s \mathcal{M}\big(t/q\big)  \mathcal{Z}_2^{\text{top}}(P_1,P_2,P_3 ;t,q)
\quad
~.
\ee 
We will next see that the integral $\mathcal{I}_2 (a_1,a_2,s;\beta)$ has a well-defined limit $q\rightarrow 1$.

\section{The $q\to 1$ limit of the Matrix integral}
\label{sec:MM4dLim}
 
 Having established the relation between the topological string partition function and the matrix integral, we wish to 
 take the  $q\to 1$ limit of the latter. 
Looking at definitions \eqref{eq:integrantsMain}, \eqref{q-MatrixModel}, as well as \eqref{eq:qdilogDiv}, one will notice that the 
function $\varphi(z)$ representing the main building block of the integrand 
diverges for $q\rightarrow 1$, and that the summation over residues 
defining $\mathcal{I}_2$ does not have an obvious limit.

Our strategy so far to resolve this issue has been to first rewrite the sum \eqref{Integral2Res} as a variant of the 
Jackson integral, then recast this in terms of combinations of functions $\varphi(z)$ which are known to have a 
well-defined limit when $q\rightarrow 1$. 
%
%
Now we will rewrite the integrand using the quasi-constants $\vartheta_q(x,s)$ defined in equation \eqref{eq:vartheta1} 
such that all the singularities are contained inside a product of quasi-constant terms. 
As a result we will discover the relation to the usual Selberg integral.


\subsection{A simple example}

Let us first consider a simple example. 
We are ultimately interested in evaluating  the $q\rightarrow 1^{-}$  limit  of integrals of the form
\begin{equation}\label{intdef}
\cI_q=\int_\cC dx\; \mathcal{R}(x;s,t),\qquad 
\mathcal{R}(x;s,t)=x^{t+s-2}\frac{\varphi(q^{1-s}/x)}{\varphi(1/x)} \, .
\end{equation}
where $\cC$ is a contour starting and ending at $0$, encircling the poles of the 
integrand in the interval $(0,1)$ in the counterclockwise direction.
The integrand \eqref{intdef} has poles at $x=q^n$, $n\in\mathbb{N}$. The integral $\cI_q$ can therefore be evaluated as a sum of the residues 
\be
\cI_q=2\pi \mathrm{i}\sum_{n=0}^\infty\mathcal{R}_n(s,t), \qquad
\mathcal{R}_n(s,t):=\mathop{\mathrm{Res}}_{x=q^n}\mathcal{R}(x;s,t)   \, .
\ee
For what follows we find it useful to introduce   a variant of the
Jacobi triple product function
\be
\vartheta_q(z):=\varphi(z)\varphi(q/z)\varphi(q) =(1-z)\prod_{n=1}(1-zq^n)(1-z^{-1}q^n)(1-q^n)~,
\ee
which we may use to rewrite the integrand  $\mathcal{R}(x;s,t)$ of $\cI_q$ as 
\be
\frac{\varphi(q^{1-s}/x)}{\varphi(1/x)}=\frac{\varphi(qx)}{\varphi(q^s x)}\frac{\vartheta_q(q^sx)}{\vartheta_q(qx)}  \, .
\ee
The function $\vartheta_q(z)$ allows us to represent $\cI_q$ as
\be
\cI_q=2\pi \mathrm{i}\,\sum_{n=0}^\infty \rho(s) \mathcal{R}_n'(s,t),\qquad 
\mathcal{R}_n'(s,t):=\bigg[x^{t-1}\frac{\varphi(qx)}{\varphi(q^s x)}\bigg]_{x=q^n}~
\ee
and where $\rho(s)$ is given by 
\be
\label{eq:Res1}
\rho(s)=\mathop{\mathrm{Res}}_{x=q^n} x^{s-1} \frac{\vartheta_q(q^sx)}{\vartheta_q(qx)}=
q^n \frac{\vartheta_q(q^s)}{(q;q)_{\infty}^3} ~.
\ee
It follows that the integral \eqref{intdef} can be rewritten in terms of an integral $\cI'_q$ 
\be \label{contour-Jackson}
\cI_q=\frac{2\pi\mathrm{i}}{1-q}\,\frac{\vartheta_q(q^s)}{(q;q)_{\infty}^3}\,\cI'_q,\qquad
\cI'_q=\int_0^1d_q x \;x^{t-1}\frac{(qx;q)_{\infty}}{(q^s x;q)_\infty},
\ee
where $\cI'_q$ can be evaluated as  Jacksons integral
\be
\int_0^1 \frac{d_qx}{x} f(x) = (1-q)\sum_{k=0}^\infty f(q^k) \, .
\ee

\subsection{The matrix integral}

We now return to the specific case in which we are interested, equation \eqref{q-MatrixModel}. 
Identity \eqref{eq:vartheta1} allows us to rewrite the integral \eqref{q-MatrixModel} as
\begin{align} \label{eq:integralI2-1}
\cI_2  =
\int d'_qy_1 \cdots d'_qy_s & \prod_{i=1}^s  y_i^\zeta 
\frac{\varphi(q y_i)}{\varphi(t y_i/ P_2^2 )}
\vartheta_q \left( y_i , 1+2\beta a_2 \right) 
\left( y_i \right)^{-2\beta a_2} \\
&\prod_{j>i}^s \frac{\varphi(y_i/y_j)}{\varphi(t y_i/y_j)}
\frac{\varphi(v^2 y_i/y_j)}{\varphi(q y_i/y_j)} 
\vartheta_q \left( \frac{ y_i }{ t y_j} , 1+\beta^2 \right) 
\left( \frac{ y_i }{ t y_j} \right)^{-\beta^2}
  \nn  \, ,
\end{align}
where we recall that the integrals $\int d'_qy_1 \cdots d'_qy_s \prod_{i=1}^s y_i^{-1}$ are variants of the Jackson integral \eqref{eq:varJack}. 
We also recall that as explained in Section \ref{sec:Residues} we have assumed  the radial ordering of poles  $|y_i|<|y_{i+1}|$, which are
labelled by a partition $\nu$  
\be \label{eq:poles}
y_\nu = \{y_1 , y_2 , \ldots , y_{s-1} , y_s\}_\nu = 
\{ t^{s-1} q^{\nu_1} ,   t^{s-2} q^{\nu_2} ,
\ldots , t q^{\nu_{s-1}} , q^{\nu_s}  \} ~ .
\ee 
Since quasi-constants are independent of  multiplicative shifts of the argument by $q$,  
they can be factored out of the integrand and $\cI_2$ in equation \eqref{eq:integralI2-1} becomes 
\bea \label{eq:integralI2-3}
\cI_2 &=& t^{\frac{s(s-1)}{2} \beta^2} 
\left( \frac{2\pi\ii}{1-q} \right)^{s}
\text{Eval}_{y=y_{\nu_\emptyset}} \left[\prod_{i=1}^{s-1} 
\vartheta_q \left( y_i , 1+2\beta a_2 \right)
\prod_{j-i\geq 2}^s 
\vartheta_q \left( \frac{ y_i }{ t y_j} , 1+\beta^2 \right) \right]
\nn \\
&~& \qqq\qqq\qqq
\text{Res}_{y_s=1} ~
\vartheta_q \left( y_s , 1+2\beta a_2 \right)
\prod_{i=1}^{s-1} \text{Res}_{y_i=t y_{i+1}} 
\vartheta_q \left( \frac{ y_i }{ t y_{i+1}} , 1+\beta^2 \right)
\nn\\
&~& \qqq
\int d'_qy_1 \cdots d'_qy_s
\prod_{i=1}^s  y_i^{2\beta^2(i-1)+2\beta a_1} 
\frac{\varphi(q y_i)}{\varphi(t y_i/ P_2^2 )}
\prod_{i<j}^s \frac{\varphi(y_i/y_j)}{\varphi(t y_i/y_j)}
\frac{\varphi(v^2 y_i/y_j)}{\varphi(q y_i/y_j)} ~.
\eea
 The residues can be evaluated, as we show in Appendix \ref{ap:JacksonIntegrals}, and in particular using the definition \eqref{eq:vartheta1}, and we obtain
\be 
\text{Res}_{y_i=t y_{i+1}} 
\vartheta_q \left( \frac{ y_i }{ t y_{i+1}} , 1+\beta^2 \right) = \frac{\vartheta_q (q^{-\beta^2})}{\varphi(q)^3}
\ee
and 
\be 
\text{Res}_{y_s=1} ~
\vartheta_q \left( y_s , 1 + 2\beta a_2 \right) = 
\frac{\vartheta_q (q^{-2\beta a_2})}{\varphi(q)^3}
\, .
\ee
Such ratios have a well defined  $q\to 1$ limit 
\be
\lim_{q\to 1}\frac{\varphi(q^{\alpha_1} x)}{\varphi(q^{\alpha_2} x)} 
 = (1-x)^{\alpha_2-\alpha_1} ~,
\quad
\frac{\vartheta_q (q^s)}{\vartheta_q (q^t)} 
 \xrightarrow{q\to 1}  
\frac{\sin (\pi s)}{\sin (\pi t)} ~, \quad
\frac{2\pi\ii}{1-q} \frac{\vartheta_q (q^s)}{\varphi(q)^3} 
 \xrightarrow{q\to 1}  
  2\ii \sin(\pi s) ~ ,
\ee
see Appendix \ref{App:phiMq21limit} and Appendix \ref{ap:JacksonIntegrals} for details. Thus 
the integral \eqref{eq:integralI2-3} becomes  
\bea \label{eq:integralI2-5} 
\lim_{q\to 1} \cI_2 \longrightarrow   &&
\prod_{k=0}^{s-1}  2\ii \sin(\pi \beta(2a_2 + k\beta))
\nn\\
&&
\int_0^1 dy_s \int_0^{y_s} dy_{s-1} \ldots \int_0^{y_2} dy_1 
\prod_{i=1}^s y_i^{2\beta a_1} (1-y_i)^{2\beta a_2} 
\prod_{i<j} (y_j-y_i)^{2\beta^2} ~.
\eea 
Obtaining the $\prod_{k=0}^{s-1}  \sin(\pi \beta(2a_2 + k\beta))$ prefactor in front of the Selberg integral is an 
important point of our paper as it reflects a particular choice of basis in the space of conformal blocks. This 
choice is not particularly spectacular for the 
space of \emph{Virasoro} three point conformal blocks, since this is one dimensional, completely fixed via Ward identities. 
The real importance comes from the implication for the higher rank $T_N$ theories. 
In this context the $T_N$ topological strings partition function is expected to correspond to a particular choice of basis 
for the space of $W_N$ three point conformal blocks, which is infinite dimensional for $N\geq 3$.

\subsection{The matrix integral as a Selberg integral}

We can finally express the limit of $\cI_2$ in 
terms of $\Gamma$-functions through the Selberg integral
\bea 
\label{eq:MMI2Selberg}
\quad
\lim_{q\to 1}\cI_2 = 
 \prod_{k=0}^{s-1}  2\ii \sin(\pi \beta(2a_2 + k\beta))
 \, 
\cI_{\text{Sel}} (1+2\beta a_1, 1+2\beta a_2, \beta^2) ~,
\quad
\eea  
using the definition
\bea \label{Selberg}
&& \cI_{\text{Sel}} (1+2\beta a_1, 1+2\beta a_2, \beta^2) 
= \nn\\
&& \qquad \qquad 
=\int_0^1 dy_s \int_0^{y_s} dy_{s-1} \ldots \int_0^{y_2} dy_1 
\prod_{i=1}^s y_i^{2\beta a_1} (1-y_i)^{2\beta a_2} 
\prod_{i<j} (y_j-y_i)^{2\beta^2} 
\nn\\
&& \qquad \qquad    =
\prod_{j=0}^{s-1} \frac{\Gamma (1+2\beta a_1 + j\beta^2)
\Gamma(1+2\beta a_2+j\beta^2)\Gamma((j+1)\beta^2)}{
\Gamma(2+2\beta(a_1+a_2)+(s-1+j)\beta^2)\Gamma(\beta^2)} . ~ \qquad 
\eea
Note that this integral is convergent for $a_1$, $a_2$ and $\beta$ real and positive. 
Furthermore, using the shift identities 
of the $\Gamma$ function to suppress the products in  equation \eqref{eq:MMI2Selberg} by rewriting this in terms of the double gamma function 
$\Gamma_\beta$:
\be 
\frac{\pi}{\sin (\pi x)} = \Gamma(1-x)\Gamma(x)
\ee
and 
\be 
\label{eq:bGammaFunctionalRel}
\frac{\Gamma_\beta(x+\beta)}{\Gamma_{\beta}(x)} = \sqrt{2\pi}
\beta^{\frac{1}{2}-\beta x} \Gamma^{-1}(\beta x) ~,
\ee
this becomes 
\begin{align}
\label{eq:Selberg2GammasA}
&\cI_{\text{Sel}} (1+2\beta a_1, 1+2\beta a_2, \beta^2)
=\left(\frac{2\pi\ii}{\beta^{\beta^2-1}\Gamma(\beta^2)}\right)^{s} 
\frac{\Gamma_\beta(\beta)}{\Gamma_\beta(2\beta^{-1}-\beta+a_1+a_2+a_3)}
\nonumber
\\ \nonumber
& \qqq\qqq\qqq\qqq
\frac{\Gamma_\beta(\beta^{-1}+2a_1)\Gamma_\beta(\beta^{-1}+2a_2)
\Gamma_\beta(2\beta^{-1}-\beta+2a_3)}
{\Gamma_\beta(\beta^{-1}+a_3+a_1-a_2)\Gamma_\beta(\beta^{-1}+a_3+a_2-a_1)
\Gamma_\beta(\beta+a_3-a_1-a_2)} ~, \\
& 
\end{align}
recalling the condition $a_3=a_1+a_2 + s\beta$. 
We can further make the following observations.
\begin{itemize}
\item  We can extend the domain of definition from integer $s$ to $s\in\mathbb{C}$.
\item The arguments of the double gamma functions are all positive if 
$a_i>0$ for $i=1,2,3$ and $s>0$. 
\end{itemize}
Taking into account the factor $\prod_{k=0}^{s-1}  \sin(\pi \beta(2a_2 + k\beta))$ 
in front of the Selberg integral 
\eqref{eq:MMI2Selberg}, this equation becomes
%
%
%
%
%
%
\begin{align} 
\label{eq:Selberg2Gammas}
&\lim_{q\to1}\cI_2
= \left(\frac{2\pi\ii}{\beta^{\beta^2-1}\Gamma(\beta^2)}\right)^{s} \frac{\Gamma_\beta(\beta)}{\Gamma_\beta((s+1)\beta)} 
 \nonumber
\\ \nonumber
&\qquad\quad\quad ~~
\frac{\Gamma_\beta(\beta^{-1}+2a_1)}{\Gamma_\beta(\beta^{-1}+a_1-a_2+a_3)}
\frac{\Gamma_\beta(\beta - 2a_2)}{\Gamma_\beta( a_1 - a_2 - a_3+ \beta)}
\frac{\Gamma_\beta(2\beta^{-1}-\beta+2a_3)}{\Gamma_\beta(2\beta^{-1}-\beta+ a_1 + a_2 + a_3)}
~. 
\\ &
\end{align}

\paragraph{Note:} 
So far we have been considering the regime where $|q|<1$, $|t|<1$ corresponding 
to CFT with central charge $c=1-6(\beta-\beta^{-1})^2$ 
if $t = q^{\beta^2}$. Another interesting regime is 
$|t|>1$  and $|q|<1$.
Similar arguments as used above would lead to 
\begin{align}\label{Liou-N}
\mathcal{Z}_{\rm Liou}^{}(\alpha_1,\alpha_2,\alpha_3)=
&\left(\frac{b^{1+b^2}}{2\pi\mathrm{i}}\Gamma(-b^2)\right)^{-s}
\frac{\Gamma_b(-s b
)}{\Gamma_b(0)} \notag \\
&\frac{\Gamma_b(Q+2 \alpha_1-sb)\Gamma_b(-2 \alpha_2+sb)\Gamma_b(2\alpha_3+2Q+sb)}
{\Gamma_b(Q+2 \alpha_1)\Gamma_b(-2 \alpha_2)\Gamma_b(2Q+2\alpha_3)}
\, .
\end{align}
The regime $|q|<1$, $|t|<1$ is related to the CFT called Generalised Minimal Models
in \cite{Zamolodchikov:2005fy}, see also \cite{Ribault:2015sxa}.
The parameters of the two regimes are 
related by analytic continuation, 
\begin{equation}
b= - \ii \beta \, , \quad \alpha = \ii a  \, ,  \quad Q = \ii q  \, ,
\end{equation}
provided that $b$ is the parameter giving the central charge of Liouville field theory as
$c_{\text{Liouv}}=1+6(b+b^{-1})^2$.

\section{Comparison with the strip vertex}
\label{sec:CompareT2-Strip}

\begin{figure}[t]
   \centering
      \includegraphics[height=4.3cm]{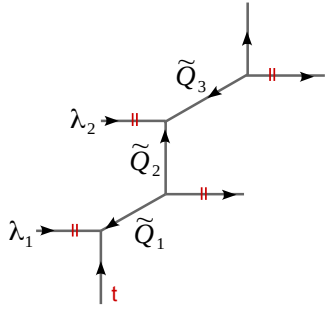} 
   \caption{\it The strip geometry, with the assignment of K\"ahler parameters, the choice of preferred direction and 
   a pair of Young tableaux decorating on external legs.}
   \label{fig:Strip1}
\end{figure}

We have so far focused on the $T_2$ vertex. 
There exists however another important building block which has been well studied in literature with which the $T_2$ 
should be compared. This is usually referred to as the strip geometry \cite{Iqbal:2004ne}, which we review below.  
One version of the strip is depicted in Figure \ref{fig:Strip1}, other variants of the strip being related to this by flop
transitions.
From the point of view of gauge theory, its partition function is well known to give the partition function of $SU(2)\times SU(2)$ 
bifundamental hypermultiplets. 

We will begin by comparing the  results of the topological vertex computations for the respective regions
in the parameter space. It will turn out, however, that the strip diagrams do not allow
us to cover the full parameter space of the $T_2$ vertex and its relatives obtained 
by flop transitions. For the regions of the parameter
space for which both the $T_2$ and the strip vertex can be used we will find agreement. 

This agreement is not unexpected.
The toric CY associated to the strip and $T_2$ have the same mirror manifolds, indicating that 
the partition functions associated to these two diagrams should be related. 
In the second half of this section we will briefly discuss two more direct ways for understanding the relation
between the $T_2$ vertex and strip from the point of view of the B-model, using the
relation between geometric transitions and matrix models on the one hand, and 
recent results on the topological recursion for the case of our interest on the other hand.

\subsection{Topological vertex computation}

The topological strings partition function for the strip geometry has the property that all the sums over Young diagrams can be performed and it can be written in product form. 
Starting with 
\be \label{eq:strip-nu1}
\mathcal{Z}_{\text{strip},\vec{\lambda}}^{\text{top}}(\tilde{Q_1},\tilde{Q}_2,\tilde{Q}_3;\ft,q)
=\sum_{\boldsymbol{\nu}} \prod_{i=1}^3(-\tilde{Q}_i)^{\vert \nu_i\vert} 
C_{\nu_1^t\emptyset\lambda_1^t}(q,\ft)C_{\nu_1\nu_2\emptyset}(\ft ,q) 
C_{\nu_3^t \nu_2^t\lambda_2^t }(q,\ft) C_{\nu_3\emptyset\emptyset}(\ft , q)~,
\ee
keeping two of the Young diagrams decorating the external legs non-empty, this equation can be brought to the form
\bea \label{eq:strip-nu2}
&&\mathcal{Z}_{\text{strip},\vec{\lambda}}^{\text{top}}(\tilde{Q}_1,\tilde{Q}_2,\tilde{Q}_3;t,q)
= 
\ft^{\frac{||\lambda_1^t||^2+||\lambda_2^t||^2}{2}} \tilde{Z}_{\lambda_1^t}(q,\ft) \tilde{Z}_{\lambda_2^t}(q,\ft) \nn\\
&~& \qquad\qquad\qquad\qquad
\frac{\cM(\tilde{Q}_1 \tilde{Q}_2;\ft , q)\cM(\tilde{Q}_2 \tilde{Q}_3\frac{\ft}{q};\ft , q)}{\cM(\tilde{Q}_1\sqrt{\frac{\ft}{q}};\ft , q) 
\cM(\tilde{Q}_2\sqrt{\frac{\ft}{q}};\ft , q)\cM(\tilde{Q}_3\sqrt{\frac{\ft}{q}};\ft , q) 
\cM(\tilde{Q}_1\tilde{Q}_2\tilde{Q}_3\sqrt{\frac{\ft}{q}};\ft , q)}
\nn\\
&~& \qquad\qquad\quad
\frac{\cN_{\lambda_1\emptyset}(\tilde{Q}_1\sqrt{\frac{\ft}{q}};\ft , q)
\cN_{\emptyset\lambda_2}(\tilde{Q}_2\sqrt{\frac{\ft}{q}};\ft , q)\cN_{\lambda_2\emptyset}(\tilde{Q}_3\sqrt{\frac{\ft}{q}};\ft , q)
\cN_{\lambda_1\emptyset}(\tilde{Q}_1\tilde{Q}_2\tilde{Q}_3\sqrt{\frac{\ft}{q}};\ft , q) }{
\cN_{\lambda_1\lambda_2}(\tilde{Q}_1\tilde{Q}_2 ;\ft , q)}. \nn\\
&~&
\eea
Setting then all of the external legs' Young tableaux to $\emptyset$, see for example also \cite{Kozcaz:2010af}, 
this partition function reduces to 
\be \label{eq:pfstripempty}
 \mathcal{Z}_\text{strip}^{\text{top}}(\tilde{Q}_1,\tilde{Q}_2,\tilde{Q}_3;\ft,q)=\frac{\cM\big(\tilde{Q}_1 \tilde{Q}_2\big) 
 \cM\big(\tilde{Q}_2 \tilde{Q}_3 \frac{\ft}{q}\big)  }{
 \calM\big(\tilde{Q}_1\sqrt{\frac{\ft}{ q}}\big)\calM\big(\tilde{Q}_2\sqrt{\frac{\ft}{ q}}\big)\calM\big(\tilde{Q}_3\sqrt{\frac{\ft}{ q}}\big)
 \calM\big(\tilde{Q}_1\tilde{Q}_2\tilde{Q}_3\sqrt{\frac{\ft}{ q}}\big)} ~.
\ee
We may compare this equation to its $T_2$ counterpart, whose topological strings partition function was given in 
product form in equation \eqref{eq:T2partitionfunctionProduct} as 
\be  
\label{eq:T2empty}
 \mathcal{Z}_2^{\text{top}}(Q_1,Q_2,Q_3;\ft,q)=
 \frac{\calM\big(Q_1Q_2\big) \calM(Q_1Q_3 \frac{\ft}{ q} ) \calM\big(Q_2Q_3\big)}
 {\calM\big(Q_1\sqrt{\frac{\ft}{ q}}\big)\calM\big(Q_2\sqrt{\frac{\ft}{ q}}\big)\calM\big(Q_3\sqrt{\frac{\ft}{ q}}\big)\calM\big(Q_1Q_2Q_3\sqrt{\frac{\ft}{ q}}\big)}~. 
\ee

%
%
%
%

At the level of the topological strings partition functions, when naively 
setting $Q_i=\tilde{Q}_i$, for all $i=1,2,3$, the 
expressions $ \mathcal{Z}_2^{\text{top}}$ \eqref{eq:T2empty} and $ \mathcal{Z}_\text{strip}^{\text{top}}$ 
\eqref{eq:pfstripempty} are identified with the partition function of $SU(2)\times SU(2)$ free 5D hypermultiplets on 
$\mathbb{R}^4\times \mathbb{S}^1$
\begin{eqnarray}
\label{eq:emptyFreeTrinion}
\mathcal{Z}_{hypers}^{\mathbb{R}^4\times \mathbb{S}^1} 
&=& \frac{\mathcal{Z}^{\text{top}}_{T_2} (Q_1,Q_2,Q_3;t,q)}{\mathcal{M}(Q_1 Q_2)\mathcal{M}(Q_2 Q_3)
\mathcal{M}(Q_1 Q_3 \frac{t}{q})} 
=
\frac{\mathcal{Z}^{\text{top}}_{\text{strip}}(Q_1,Q_2,Q_3;t,q)}{\mathcal{M}(Q_1 Q_2)\mathcal{M}(Q_2 Q_3 \frac{t}{q})}
\\ \nn
 &=& 
\frac{1}
 {\calM\big(Q_1\sqrt{\frac{\ft}{ q}}\big)\calM\big(Q_2\sqrt{\frac{\ft}{ q}}\big)\calM\big(Q_3\sqrt{\frac{\ft}{ q}}\big)
 \calM\big(Q_1Q_2Q_3\sqrt{\frac{\ft}{ q}}\big)}
\end{eqnarray}
up to products of $\cM$ functions which are interpreted as the non-full spin content \cite{Bao:2013pwa}.  
The topological string partition function contains extra degrees of freedom from strings stretching between the external 
parallel legs which do not transform properly under the 5D Lorenz group and thus have to be removed in order to obtain 
the 5D partition function \cite{Bao:2013pwa,Mitev:2014isa}. 
%
%
%
%
%
%
%
%
%
%
%
%
%
%
%

Below we give a systematic presentation comparing the $T_2$ and strip geometries, taking into account the 
various possible ways to flop internal edges of the web diagrams for each of these. 

\subsection{Comparison}

\begin{figure}[t]
   \centering
      \includegraphics[height=4.5cm]{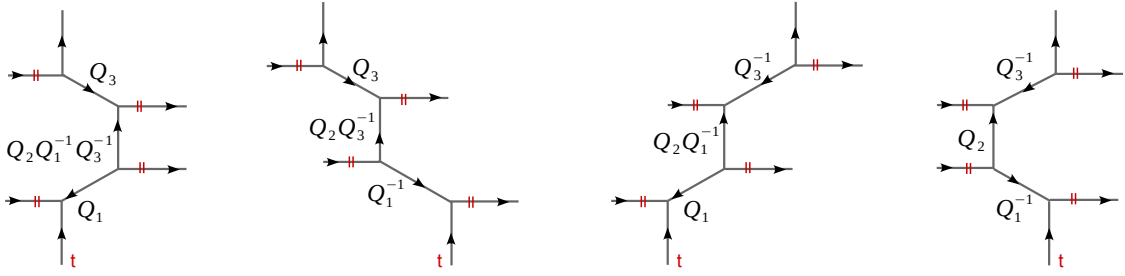} 
   \caption{\it Different strip diagrams related by flops. }
   \label{fig:Strips}
\end{figure}

\begin{figure}[b]
   \centering
      \includegraphics[height=4.0cm]{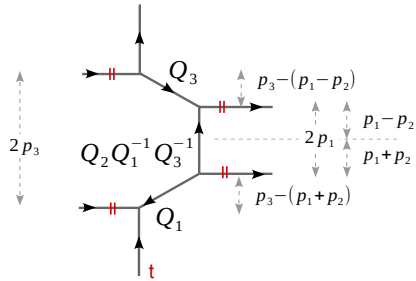} 
   \caption{\it Parametrisation of distances between pairs of external parallel legs of a strip. }
   \label{fig:Strips2P}
\end{figure}

We now want to compare the $T_2$ partition functions \eqref{ZT2simple} and \eqref{eq:floppedT2s} with 
the result for the strip vertex. The computations can be found in 
\cite{Iqbal:2003ix,Iqbal:2003zz,Eguchi:2003sj,Iqbal:2004ne}, and are reviewed in \cite{Coman:2018uwk}.  
The result for the diagrams on the left and on the right of Figure \ref{fig:Strips} is found to be
%
%
%
%
\begin{equation}\label{Strip}
 Z^{\rm strip}_{l}  
   =\frac{\mathcal{M}\big(Q_2\big)\mathcal{M}(Q_2  ( Q_1 Q_3 )^{-1} )}
   {\prod_{i=1,3}\mathcal{M}\big(Q_{i}\big)\mathcal{M}\big(Q_2  Q_{i}^{-1} \big)}, \qquad
   Z^{\rm strip}_{r}  
   =\frac{\mathcal{M}\big(Q_2\big)\mathcal{M}( Q_2  ( Q_1 Q_3 )^{-1} )}
   {\prod_{i=1,3}\mathcal{M}\big(Q_{i}^{-1}\big)\mathcal{M}\big(Q_2 Q_{i}^{-1}\big)}.
\end{equation}
Let us start with the leftmost diagram in Figure \ref{fig:Strips}. We may parameterise 
\begin{equation}
Q_1=q^{\beta(p_3-p_1-p_2)}, \qquad Q_3=q^{\beta(p_3+p_2-p_1)},\qquad
Q_2=q^{2\beta p_3},
\end{equation}
where $p_i$, $i=1,2,3$, are the distances between the pairs of parallel lines containing the external edges of the 
strip diagram depicted in Figure \ref{fig:Strips2P}. Similar parameterisations can be introduced for the remaining cases.
The resulting formulae for the limit $q\rightarrow 1$ can be written in the form
 \begin{equation}
\mathcal{Z}^{\rm strip}_{l}=
\frac{[\text{Leg factors}]}{\prod_{\epsilon,\epsilon'}G_\beta(p_3+\epsilon p_{1}+\epsilon'p_{2})},\qquad
\mathcal{Z}^{\rm strip}_{\rm r}=
\frac{[\text{Leg factors}]}{\prod_{\epsilon,\epsilon'}G_\beta(p_1+\epsilon p_{3}+\epsilon'p_{2})}.
\end{equation}
The result for the second strip diagram takes the form 
\begin{equation}\label{ZT2simple-b}
\mathcal{Z}^{strip}_{m}=\frac{[\text{Leg factors}]}{G_\beta(p_1+p_2+p_3)
\prod_{i=1}^3 G_\beta(p_1+p_2+p_3-2p_i)} ~,
\end{equation}
while for the third diagram, $\mathcal{Z}^{strip}_{m_2}$ differs from this by the 
replacement $p_2\rightarrow -p_2$. 

Let us now compare these results to the ones found for $T_2$. First one may note that 
chamber $\mathfrak{P}_{\mathbb{R}}^{(2)}$ is not covered by the strip diagrams.
One may otherwise observe the following relations 
\begin{equation} \label{comparison}
\mathcal{Z}^{\rm T_2^{(3)}} \simeq \mathcal{Z}^{\rm strip}_l,\qquad
\mathcal{Z}^{\rm T_2^{(s)}} \simeq \mathcal{Z}^{\rm strip}_m,\qquad
\mathcal{Z}^{\rm T_2^{(1)}} \simeq \mathcal{Z}^{\rm strip}_r,
\end{equation}
where $\simeq$ means equality up to leg factors. We see that the results
calculated using $T_2$ vertex and strip agree within each chamber up to leg factors.


\section{B-model picture}\label{sec:B-modelPicture}

We'd now like to shed some light on our findings using the tools offered by the B-model approach to topological string theory.

\subsection{Relation between mirror curves of strip and $T_2$}

In the above we had observed simple relations between $T_2$ vertices and strips. This is not surprising in view 
of the fact that the toric CY described by $T_2$ vertex and the strip have mirror manifolds related by a coordinate 
change. 

\begin{figure}[t]
   \centering
      \includegraphics[height=4.2cm]{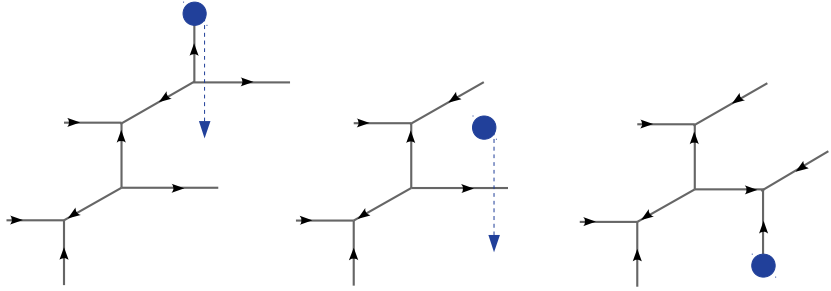} 
   \caption{\it Starting with the strip geometry, the $T_2$ geometry can be obtained via two successive Hanany-Witten moves. The blue dot represents the 7-brane on which the 5-branes end.}
   \label{fig:HananyWitten}
\end{figure}

Indeed, the mirror of the $T_2$ toric diagram is known \cite{Bao:2013pwa} to be to the curve in $\mathbb{C}^*\times\mathbb{C}^*$ 
defined by the equation
\begin{equation}\label{T2curve}
W^2-(P_2+P_2^{-1})W+(P_3+P_3^{-1})WT-(P_1+P_1^{-1})T+T^2+1=0 ~.
\end{equation}
It is easy to check that the external legs of the $T_2$ toric diagram describe the infinite ends of the curve \eqref{T2curve}. 
Changing coordinates $T=U(W-P_2)$ one gets
\begin{equation}\label{Stripcurve}
W(U+P_3)(U+P_3^{-1})=P_2(U+P_1 P_2^{-1})(U+P_1^{-1}P_2^{-1})   \, .
\end{equation}
This is recognised as the mirror of the toric diagram for the strip.  
A dual interpretation of this relation in terms of the intersecting brane systems one
is provided by the two successive Hanany-Witten moves depicted in Figure \ref{fig:HananyWitten}.

The fact that the local CY mirror to the toric geometries associated to the strip and $T_2$ diagrams
are related by a change of coordinates suggests that the corresponding topological string partition 
functions should be related.
This gives us a simple way to predict relations between the partition functions $\mathcal{Z}^{\rm T_2}$ 
and $\mathcal{Z}^{\rm strip}$, as observed in Section \ref{sec:CompareT2-Strip} above. 
To make this argument precise one needs to have a more direct way to compute the topological string partition 
functions within the B-model description. This is provided by the matrix model representation predicted
by the Gopakumar-Vafa geometric transitions, as will be discussed next.

\subsection{Geometric transitions and matrix models}\label{geomtrans}

It is well-known that 
the matrix model representation of topological string partition function reflects
a duality between closed and open topological string referred to as geometric transition
\cite{Dijkgraaf:2002vw,Dijkgraaf:2009pc}.
We will briefly discuss the implications of these dualities for the case of our interest. 
This can be done directly on the level of the four-dimensional partition functions obtained
in the limit $q\rightarrow 1$ we are mostly interested in.

In the B-model one considers the local CY defined by the equation
\begin{equation}\label{C03curve}
x^2(x-1)^2\,y^2=p_3^2x^2-(p_3^2+p_1^2-p_2^2)x+p_1^2 ~,
\end{equation}
which can be written as 
\begin{equation}\label{defdstrip}
\left(y-\frac{p_1}{x}-\frac{p_2}{x-1}\right)\left(y+\frac{p_1}{x}+\frac{p_2}{x-1}\right)=
\frac{(p_3-p_1-p_2)(p_3+p_2+p_1)}{x(x-1)},
\end{equation}
 Defining 
\begin{equation}
W(x)=p_1\log(x)+p_2\log(x-1) ~,
\end{equation}
we may write  equation (\ref{defdstrip}) in the form
\begin{equation}\label{Scurve}
y^2-(W'(x))^2=\frac{p_3^2-(p_1+p_2)^2}{x(x-1)}~.
\end{equation}
In this form we may recognise the curve (\ref{C03curve}) as a resolution of the singular curve 
$y^2-(W'(x))^2=0$.

For integer values of $s=p_3-p_2-p_1$ one expects to have a dual description of the 
closed topological string in terms of an open topological string on a deformation of the singular 
curves obtained by setting $p_3=p_2+p_1$ and wrapping $s$  D-branes on the resulting $S^3$. 
The arguments of \cite{Dijkgraaf:2002vw,Dijkgraaf:2009pc} then lead to the prediction that
\begin{equation}\label{matmodrep}
\mathcal{Z}^{\rm cl}(p_1,p_2,p_3)\Big|_{p_3-p_2-p_1=s}
=\mathcal{Z}^{\rm op}(p_1,p_2;s) ~,
\end{equation}
with $\mathcal{Z}^{\rm op}(p_1,p_2;s)$ having a representation as a multiple integral
\begin{equation}
\mathcal{Z}^{\rm op}(p_1,p_2;s)=\int_{\mathcal{C}_s} dx_1\dots dx_s\;\,\prod_{k<l}(x_k-x_l)^2
\prod_{k=1}^s e^{\frac{1}{\lambda}W(x_k)} ~,
\end{equation}
for a certain choice of the contour $\mathcal{C}_s$. The results for 
$\mathcal{Z}^{\rm op}(p_1,p_2;s)$ will of course depend sensitively on 
$\mathcal{C}_s$. As stressed in \cite{Cheng:2010yw} one should consider
the choice of $\mathcal{C}_s$ as an important part of the non-perturbative
definition of topological string theory.
 A natural candidate for the contour $\mathcal{C}_s$ has been identified in \cite{Cheng:2010yw}. It amounts to 
defining $\mathcal{Z}^{\rm op}(p_1,p_2;s)$ as the Selberg integral \eqref{Selberg}.

We may now observe that this proposal is perfectly consistent with the results 
of the topological vertex computations. The comparison is possible in the chamber 
of the parameter space where the real part of $p_3-p_2-p_1$ is positive. 
In this chamber one may use the toric diagram depicted in Figure \ref{fig:Strips2P}. 
In Section \ref{sec:CompareT2-Strip}
we had already observed that $\mathcal{Z}_l^{\rm strip}$ is given by the Selberg integral, 
in perfect agreement with the prediction (\ref{matmodrep}).

In the cases where $p_3<p_2+p_1$ one can no longer apply this reasoning.  
However, there is a second possibility for representing the curve (\ref{C03curve}) in the
form  (\ref{Scurve}), given by the formula
\begin{equation}\label{defdT2}
\left(y-\frac{p_1}{x}+\frac{p_2}{x-1}\right)\left(y+\frac{p_1}{x}-\frac{p_2}{x-1}\right)=
\frac{(p_3-p_1+p_2)(p_3-p_2+p_1)}{x(x-1)} ~.
\end{equation}
The same arguments as had been used to arrive at (\ref{matmodrep}) can now be applied, for example
in the case that $p_3-p_1+p_2=s$, with $s$ being a positive integer,
leading to the prediction that  
\begin{equation}\label{matmodrep-2}
\mathcal{Z}^{\rm cl}(p_1,p_2,p_3)\Big|_{p_3-p_1+p_2=s}=\tilde{\mathcal{Z}}^{\rm op}(p_1,p_2;s)
\end{equation}
with $\mathcal{Z}^{\rm op}(p_1,p_2;s)$ having a representation as a multiple integral
\begin{equation}
\tilde{\mathcal{Z}}^{\rm op}(p_1,p_2;s)=\int_{\tilde{\mathcal{C}}_s} dx_1\dots dx_s\;\,\prod_{k<l}(x_k-x_l)^2
\prod_{k=1}^s e^{\frac{1}{\lambda}\tilde{W}(x_k)} ~,
\end{equation}
for a certain choice of the contour $\tilde{\mathcal{C}}_s$, and
\begin{equation}
\tilde{W}(x)=p_1\log(x)-p_2\log(x-1) ~.
\end{equation}
One should observe, however, that the contour $\mathcal{C}_s$ proposed in \cite{Cheng:2010yw} is not suitable 
for the second case, as it would yield a divergent integral, in general. At the upper limit of integration one 
encounters a singularity of the form $(1-x)^{-p_2}$ with $p_2$ positive.
In this case it is natural to replace 
the contour by a half-open multi-contour $\mathcal{C}_s'$ starting at $x_k=0$, encircling $x_k=1$,
and returning to $x_k$ for $k=1,\dots,s$, assuming that the contour of integration for $x_k$ is in the interior of the 
disc surrounded by the contour for $x_l$ if $k<l$.
By deforming the contours\footnote{One may first deform the innermost contour in 
the sum of two segments infinitesimally below and above the real axis plus a small circle around $1$.
The integral is thereby seen to be proportional to $2\ii \sin(2\pi \beta p_2)$ times an integral having 
the interval $[0,1]$ as the contour
of integration over $x_1$. The contour of integration for $x_2$ may then be deformed in a similar way, 
and so on. We will assume that the definition of the contour $\mathcal{C}_s'$ involves 
an ordering ensuring  that the result of the  
contour deformation above is an integral with
integration variables $x_1,\dots,x_s$ ordered along the interval $[0,1]$ 
as $0<x_s<x_{s-1}<\dots<x_1<1$.} 
it is possible to show that 
\begin{equation}
\tilde{\mathcal{Z}}^{\rm op}(p_1,p_2;s)= \mathcal{Z}^{\rm op}(p_1,-p_2;s)
\prod_{k=0}^{s-1}  (-2\ii) \sin(\pi \beta(-2p_2 + k\beta)) ~.
\end{equation}
This prediction 
can be compared with the results we had obtained above.
It can be applied, on the one hand, to the case of the strip 
in the case where $p_2>0$ and $p_1-p_2<p_3<p_2+p_1$, represented by second diagram from the left in Figure 
\ref{fig:Strips}. It can, on the other hand, be applied in the case of the $T_2$ vertex. 
We find that our previous results are perfectly consistent with (\ref{matmodrep-2}).
It is interesting to observe that the factor relating the partition functions associated to the 
two chambers is directly related to the monodromy factors reflecting the change of the contours.

Reconsidering the identification between the Liouville parameters $a_1,a_2,a_3$ and the geometric
parameters $p_1,p_2,p_3$ it is important to note that the partition functions  associated to 
the toric diagrams on the left half of Figure \ref{fig:Strips2P} are related to the counterparts on the right
half found by a reflection along an axis in the middle by $p_2\rightarrow -p_2$. Consideration of the 
strip diagrams can therefore fix the relation between $a_2$ and $p_2$ only up to a sign. 
The situation is better in the case of the $T_2$ diagram. Formula (\ref{defdT2}) suggests to 
identify
\begin{equation}
\tilde{W}(x)=p_1\log(x)-p_2\log(x-1)=a_1\log(x)+a_2\log(x-1) ~,
\end{equation}
leading to the relation $p_2=-a_2$ already adopted in the above.



\subsection{Topological recursion} \label{sec:B-model}

Another  technique for the reconstruction of the topological string partition function from the 
curve $\Sigma$ is the topological recursion  \cite{Eynard:2007kz,Eynard:2014zxa}. 
This method  has recently been applied to the case at hand in 
\cite{2018arXiv180510945I,2018arXiv181002946I}.
The result can be represented in the form
\begin{equation}\label{IKT}
\big[\log\mathcal{Z}(p_3,p_2,p_1)\big]^{}_{\rm f}=
\left[\log
\frac{\prod_{\epsilon,\epsilon'=\pm}G(1+p_3+\epsilon p_2+\epsilon'p_1)}{G(1+2p_3)G(1+2p_2)G(1+2p_1)G(1)}\right]_{\rm f},
\end{equation}
where $p_i=\frac{1}{\lambda}\nu_i$ and 
$[\log G(1+w)]^{}_{\rm f}$ is the following formal power series
\begin{equation}\label{Barnesexp}
[\log G(1+w)]^{}_{\rm f}=\frac{1}{2}\bigg(w^2-\frac{1}{6}\bigg)\log(w)-\frac{3}{4}w^2-w\zeta'(0)+\zeta'(-1)
-\sum_{g=2}\frac{B_{2g}}{2g(2g-2)}w^{2-2g}.
\end{equation}
It is known that this formal series represents the aymptotic expansion of the 
Barnes $G$-function (see e.g. \cite{2003math......8086A}).

The formal series on the right side of (\ref{Barnesexp}) is Borel summable, as follows 
from the Binet type integral representation for the function $G(x)$  \cite{2003math......8086A}
\begin{equation}\label{Binet}
\begin{aligned}
\log G(1+w)=w\log\Gamma(w)&+\frac{w^2}{4}-\frac{B_2(w)}{2}\log w-\log A-\\
&-\int_0^\infty\frac{dt}{t^2}\;e^{-tw}\left(\frac{1}{1-e^{-t}}-\frac{1}{t}-\frac{1}{2}-\frac{t}{12}\right).
\end{aligned}
\end{equation}
Indeed, the integral on the right of equation (\ref{Binet}) represents the Laplace transform
of a function which is easily found to be the Borel-transform of the asymptotic series in (\ref{Barnesexp}).

However, the Borel summation displays a
Stokes phenomenon. The result depends on the chamber in the parameter space one is in. 
The integral in (\ref{Binet}) converges for $\mathrm{Re}(w)>0$. When $\mathrm{Re}(w)<0$
one may notice that $G(1-w)$ has almost the same asymptotic expansion as 
$G(1+w)$ has. There are two differences, though. The first is due to the factor $\log(-w)$, 
the second to the term proportional to $\zeta'(0)$.  
There is, of course, an inevitable 
ambiguity in the choice of the branch of the logarithm in (\ref{Barnesexp}). 
Replacing $\log(w)$ by $\log(-w)$ appears to be a natural way
to fix this ambiguity. This motivates us to define the piecewise holomorphic 
function $\widehat{G}(w)$ by the equation
\begin{equation}\label{hatGdef}
\log \widehat{G}(1+w)=\left\{\begin{aligned} &\log G(1+w) \;\, & \text{for}\;\,\mathrm{Re}(w)>0,\\
&\log G(1-w)-2w\zeta'(0) \;\,& \text{for}\;\,\mathrm{Re}(w)<0,
\end{aligned}\right. 
\end{equation}
With this definition one may represent the result in the form 
\begin{equation}
\log\mathcal{Z}(p_3,p_2,p_1)=
\log
\frac{\prod_{\epsilon,\epsilon'=\pm}\widehat{G}(1+p_3+\epsilon p_2+\epsilon'p_1)}
{\widehat{G}(1+2p_3)\widehat{G}(1+2p_2)\widehat{G}(1+2p_1)G(1)}~,
\end{equation}
assuming that all variables $p_i$ are real and positive. Noting that the term proportional to $\zeta'(0)$ 
only affects the leg factors one recognises the same form as 
was found in  Section \ref{sec:extending}.

Let  us note that the partition functions represent analytic functions
within their respective domains of definition. Requiring that this is the case 
fixes the signs in front of the combination $p_3+\epsilon p_2+\epsilon'p_1$ 
appearing in the arguments of the Barnes functions\footnote{Note here that the Barnes G-function $G(1+x)$ is analytic 
if the real part of $x$ is positive, but has poles on the negative real axis.}. 
This observation is naturally explained by 
the observations above concerning the Borel summation of the topological 
recursion.

 \section{$T_N$ versus conformal blocks}
 \label{sec:TNvsBlocks}

We are now going to formulate more precisely what an AGT-type correspondence between the $T_N$ theories 
and Toda CFT conformal blocks would mean. It should relate a) the partition functions of $T_N$ theories 
computed using the topological vertex, and
b) conformal blocks of $A_{N-1}$ conformal Toda field theories. To this aim we will again 
compare with the 
case of the strip for which the relation to the 
AGT-correspondence is fairly well-understood. 

The basic observation is the following:
Let $\mathcal{Z}^{\rm strip}(\vec{\mu},\vec{\nu};\mathbf{p},\mathbf{p}',w;q,t)$ be the partition 
function defined by the strip diagram with horizontal legs decorated by
$N$-tuples of Young diagrams $\vec{\mu}=(\mu_1,\dots,\mu_N)$, $\vec{\nu}=(\nu_1,\dots,\nu_N)$ (see for example Figure 
\ref{fig:Strip1} for the case $N=2$). 
The variables $\mathbf{p}$ and $\mathbf{p}'$ are $N$-tuples of complex numbers assigned to the two
sets of parallel external horizontal lines appearing on the left and on the right of the strip diagram, respectively.  
The variable $w$ is associated to the horizontal shift between the vertical lines emanating at the top and on the bottom
of the strip diagram. The K\"ahler parameters 
associated to the internal edges can be easily expressed in terms of the variables $\mathbf{p}$, $\mathbf{p}'$ and 
$w$. The parameters $q=e^{-R\epsilon_1}$ and $t=e^{R\epsilon_2}$ can be expressed in terms of the topological 
string coupling $\lambda^2=\epsilon_1\epsilon_2$ and the additional refinement parameter $\epsilon_1+\epsilon_2$
introduced in \cite{Iqbal:2007ii}.

One may then consider the function $\mathcal{Z}^{\rm bif}$ defined by the  limit
\begin{equation}
\mathcal{Z}^{\rm bif}_{\vec{\mu},\vec{\nu}}(\mathbf{p},\mathbf{p}',w;b):=
\lim_{q\rightarrow 1}
\frac{\mathcal{Z}^{\rm strip}_{\vec{\mu},\vec{\nu}}(\mathbf{p},\mathbf{p}',w;q,t)}
{\mathcal{Z}^{\rm strip}_{\vec{\emptyset},\vec{\emptyset}}(\mathbf{p},\mathbf{p}',w;q,t)} ~.
\end{equation}
The function $\mathcal{Z}^{\rm bif}$ is on the one hand known to be an important  building block of the instanton 
partition functions of linear quiver theories, and it was on the other hand shown to be the matrix representing a particular 
type of intertwining operator between representations $\mathcal{R}_{\mathbf{p}}$ and $\mathcal{R}_{\mathbf{p}'}$ 
of the $W_N$-algebra in a particular basis for the $W_N$-representations called  AFLT-basis
after Alba-Fateev-Litvinov-Tarnopolsky \cite{Alba:2010qc,Fateev:2011hq},
\begin{equation}\label{AFLT}
\mathcal{Z}^{\rm bif}_{\vec{\mu},\vec{\nu}}(\mathbf{p},\mathbf{p}',w;b)=
{\phantom{\big|}}^{}_{\mathbf{p}';b}\big\langle\, \vec{\nu}\,|\,\mathrm{V}_{\mathbf{p}',\mathbf{p}}^{}(w)\,|\,\vec{\mu}\,
\big\rangle^{}_{\mathbf{p};b} ~.
\end{equation} 
The AFLT basis diagonalises a natural abelian sub-algebra of the product of the $W_N$ algebra with a free boson 
algebra within a Fock-space representation of this algebra. Identity (\ref{AFLT}) is the main ingredient in the proof
of the AGT correspondence given in \cite{Alba:2010qc,Fateev:2011hq}.

One may, of course,  simply define an operator $\mathcal{V}_{\mathbf{p}',\mathbf{p}}^{}(w)$ on a vector space 
$\mathcal{F}$ having a basis formed by vectors $|\,\vec{\mu}\,\big\rangle$ labelled by $N$-tuples of Young 
diagrams $\vec{\mu}$ such that
\begin{equation}\label{stripVO}
{\mathcal{Z}^{\rm strip}_{\vec{\mu},\vec{\nu}}(\mathbf{p},\mathbf{p}',w;q,t)}
            =\!
\big\langle\, \vec{\nu}\,|\,\mathcal{V}_{\mathbf{p}',\mathbf{p}}^{}(w)\,|\,\vec{\mu}\,\big\rangle^{} ~.
\end{equation}
The following results from \cite{FOS,Ne16} and references therein clarify in which sense the operator 
$\mathcal{V}_{\mathbf{p}',\mathbf{p}}^{}(w)$ is a 
q-deformation of the $W_N$-vertex operators  $\mathrm{V}_{\mathbf{p}',\mathbf{p}}^{}(w)$. It is known that 
\begin{itemize}
\item The spaces $\mathcal{F}_{\mathbf{p}}$ are modules of a Hopf algebra called  Ding-Iohara-Miki-algebra 
(DIM algebra). There exists a Fock-space realisation $\mathcal{F}_{\mathbf{p}}$ of the DIM algebra in which 
the states $|\,\vec{\mu}\,\big\rangle$ can be realised in terms of generalised Macdonald functions, 
$|\,\vec{\mu}\,\big\rangle\equiv|\,\vec{\mu}\,\big\rangle^{}_{\mathbf{p};q,t}$. The basis  $|\,\vec{\mu}\,\big\rangle$ 
diagonalises a large abelian sub-algebra in the DIM algebra with eigenvalues being functions of an N-tuple of  
parameters $\mathbf{p}=(p_1,\dots,p_N)$. 
\item
The q-deformation $q$-$W_N$ of the W-algebra $W_N$ can be embedded into the DIM algebra, making the spaces   
$\mathcal{F}_{\mathbf p}$ highest weight representations of the $q$-$W_N$-algebra. The eigenvalues of the zero 
modes of the $q$-$W_N$-algebra determining the $q$-$W_N$-module can be expressed in terms of the parameters 
$\mathbf{p}$. For $q\rightarrow 1$ one finds that 
\begin{equation}
\lim_{R\rightarrow 0}|\,\vec{\mu}\,\big\rangle^{}_{\mathbf{p};q,t}=|\,\vec{\mu}\,\big\rangle^{}_{\mathbf{p};b} ~,
\end{equation}
where the vectors $|\,\vec{\mu}\,\big\rangle^{}_{\mathbf{p};b}$ form the AFLT-basis.

\item The operators $\mathcal{V}_{\mathbf{p},\mathbf{p}'}(w)$ defined from $\mathcal{Z}^{\rm strip}$ in this way 
are intertwining operators between representations $\mathcal{F}_{\mathbf{p}}$ and $\mathcal{F}_{\mathbf{p}'}$ 
of the DIM algebra, characterised by simple commutation relations with the 
generators of the DIM algebra \cite{FOS} or with the ge\-nerators of the $q$-$W_N$-algebra \cite{Ne16}.
\end{itemize}

These observations explain how the symmetry algebra $W_N$ in the AGT correspondence can be understood 
as a limit of similar structures of topological string theory. They explain, in particular, why the partition functions 
associated to toric diagrams obtained by gluing two strip diagrams  get related to the conformal blocks associated 
to spheres with four punctures in the limit $q\rightarrow 1$. Such conformal blocks can be represented as matrix 
elements of compositions of the vertex operators $\mathrm{V}_{\mathbf{p}',\mathbf{p}}^{}(w)$, like, for example
\begin{equation}\label{compVO}
{\phantom{\big|}}^{}_{\mathbf{p}'';b}\big\langle\, \vec{\emptyset}\,|\,\mathrm{V}_{\mathbf{p}'',\mathbf{p}'}^{}(w_2)
\mathrm{V}_{\mathbf{p}',\mathbf{p}}^{}(w_1)\,|\,\vec{\emptyset}\,\big\rangle^{}_{\mathbf{p};b} ~.
\end{equation}
Inserting a complete set of states from the representation $\mathcal{F}_{\mathbf{p}'}$ between the vertex operators 
$\mathrm{V}_{\mathbf{p}'',\mathbf{p}'}^{}(w_2)$ and $\mathrm{V}_{\mathbf{p}',\mathbf{p}}^{}(w_1)$ yields the familiar 
power series expansions of the conformal blocks with coefficients given by products of the matrix elements
related to $\mathcal{Z}^{\rm bif}$ via (\ref{AFLT}).

If a generalisation of the AGT-correspondence holds for all class
$\mathcal{S}$ theories, as expected,  it would imply relations between the partition functions of the $T_N$ theories
and $W_N$ conformal blocks  that we'll now formulate a bit more precisely. 
The diagrams representing the toric CY used to engineer  the $T_N$ theories within string theory
now have three very similar legs consisting of $N$ parallel lines. Associating $N$-tuples of partitions to 
each of the three legs allows one to define partition functions
\begin{equation}
\mathcal{Z}^{T_N}_{\vec{\nu}_1,\vec{\nu}_2,\vec{\nu}_3}(\mathbf{p}_1,\mathbf{p}_2,\mathbf{p}_3;\mathbf{a};t,q)
\end{equation}
depending on tuples of partitions $\vec{\nu}_1$, $\vec{\nu}_2$, $\vec{\nu}_3$ associated to the three legs.
The K\"ahler parameters can be naturally parameterised through three $N$-tuples of parameters
$\mathbf{p}_1$, $\mathbf{p}_2$, $\mathbf{p}_3$ describing the asymptotic  geometry of the legs, 
supplemented by $d=\frac{1}{2}(N-2)(N-1)$ parameters $\mathbf{a}$ parameterising the widths of the
internal faces in the toric diagram. 

It is tempting to define vertex operators from $\mathcal{Z}^{T_N}$ by a construction 
similar to the one outlined above for the case of the strip. A natural analog of 
(\ref{stripVO}) could be, for example, 
\begin{equation}\label{T2VO}
{\mathcal{Z}^{T_N}_{\vec{\mu},\vec{\emptyset},\vec{\nu}}(\mathbf{p}_1,\mathbf{p}_2,\mathbf{p}_3,\mathbf{a};q,t)}
        =\!
\big\langle\, \vec{\nu}\,|\,\mathcal{V}_{\mathbf{p}_3,\mathbf{p}_1}^{\mathbf{p}_2,\mathbf{a}}(1)\,|\,\vec{\mu}\,\big\rangle^{} ~.
\end{equation}
It is not clear at the moment if the vertex operator $\mathcal{V}_{\mathbf{p}_3,\mathbf{p}_1}^{\mathbf{p}_2,\mathbf{a}}(1)$
defined through (\ref{T2VO}) has a limit for $q\rightarrow 1$ which can be identified with a vertex operator 
$\mathrm{V}_{\mathbf{p}_3,\mathbf{p}_1}^{\mathbf{p}_2,\mathbf{a}}(1)=\mathrm{V}_{\mathbf{p}_3,\mathbf{p}_1}^{\mathbf{p}_2,\mathbf{a}}(w)\big|^{}_{w=1}$
intertwining between representations of the $W_N$-algebra. This does not seem to be known 
even for $N=2$, where 
one might expect to recover the familiar vertex operators of the Virasoro algebra in this way.
The matrix elements of such vertex operators are determined by simple commutation
relations with the generators of the Virasoro algebra. It is not clear if the limit $q\rightarrow 1$ of 
the objects in (\ref{T2VO}) will reproduce these matrix elements.
A necessary condition for this to be the case would be a relation between 
matrix elements of compositions of the vertex operators 
$\mathrm{V}_{\mathbf{p}_3,\mathbf{p}_1}^{\mathbf{p}_2}(w)$ similar to (\ref{compVO}) and 
the limit $q\rightarrow 1$ of glued $T_N$-vertices.  We will check this condition below.

For $N>2$ one defines a family of vertex operators through (\ref{T2VO})
having $d$ parameters $\mathbf{a}$. This is indeed as expected. For generic triples of 
representations $\mathcal{R}_{\mathbf{p}_1}$, $\mathcal{R}_{\mathbf{p}_2}$, $\mathcal{R}_{\mathbf{p}_3}$
one expects to find an infinite dimensional vector space of chiral vertex operators
which has a basis spanned by chiral vertex operators 
$\mathrm{W}_{\mathbf{p}_3,\mathbf{p}_1}^{\mathbf{p}_2,\mathbf{s}}(w)$
labelled by $d$ parameters $\mathbf{s}$, see \cite{Coman:2015lna,Coman:2017qgv} for 
discussions of this issue. It is not clear at the moment if 
the operators
$\mathcal{V}_{\mathbf{p}_3,\mathbf{p}_1}^{\mathbf{p}_2,\mathbf{a}}(1)$ defined through 
(\ref{T2VO}) have a limit $q\rightarrow 1$ defining operators 
$\mathrm{V}_{\mathbf{p}_3,\mathbf{p}_1}^{\mathbf{p}_2,\mathbf{a}}(w)$
forming a basis for the space of chiral vertex operators. Even if this was the case it is 
not clear if such a basis would have a more direct  description in 
conformal field theory. The vertex
operators $\mathrm{V}_{\mathbf{p}_3,\mathbf{p}_1}^{\mathbf{p}_2,\mathbf{a}}(w)$ 
might, for example,  turn out to be linear combinations of  vertex operators 
 $\mathrm{W}_{\mathbf{p}_3,\mathbf{p}_1}^{\mathbf{p}_2,\mathbf{s}}(w)$  constructed
 with the help of the free field representation of the $W_N$-algbra in \cite{Coman:2017qgv}.

These issues simplify considerably when $N=2$, but do not become trivial. 
Even if the the space of chiral vertex operators is generically one-dimensional in 
this case, there still is non-trivial information needed to fix a basis for this 
one-dimensional vector space. This information is provided by the normalisation factor 
\begin{equation}
N(\mathbf{p}_3,\mathbf{p}_2,\mathbf{p}_1):=
{\phantom{\big|}}^{}_{\mathbf{p}_3,b}
\big\langle\, \vec{\emptyset}\,|\,\mathrm{V}_{\mathbf{p}_3,\mathbf{p}_1}^{\mathbf{p}_2}(1)\,
|\,\vec{\emptyset}\,\big\rangle^{}_{\mathbf{p}_1,b} ~.
\end{equation}
The results obtained in the previous section determine a preferred choice of this normalisation
factor for each region in the parameter space described in Section \ref{sec:extending}.
The relation between topological string partition functions and three-point conformal blocks
obtained in this way can be seen as a variant of the AGT-correspondence.


\section{Gluing $T_2$ building blocks vs. strips}
\label{sec:gluing}

\begin{figure}[h!]
   \centering
      \includegraphics[height=5cm]{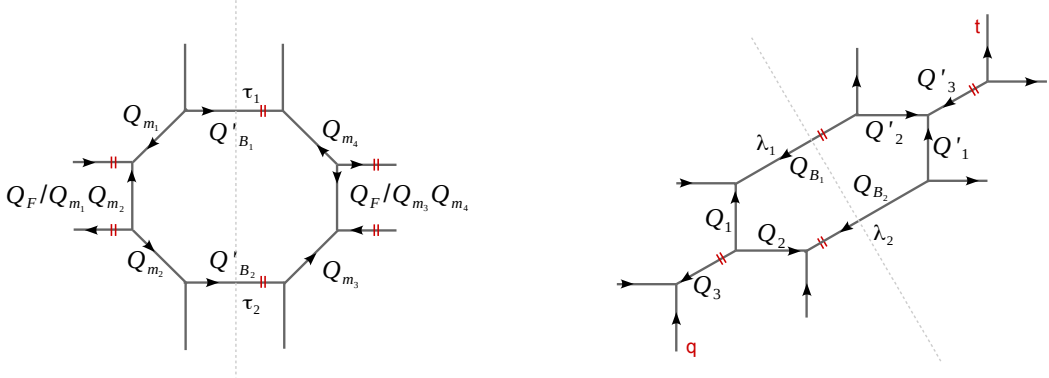} 
   \caption{\it Brane diagrams corresponding to the $\cN=1$, $SU(2)$ four-flavour theory in 5D, constructed by gluing 
   strips {\emph (left)} and two $T_2$ webs {\emph (right)}.
   Both the preferred direction and K\"ahler parameters are also indicated.}
   \label{fig:SU2Nf4-2T2}
\end{figure}

With the aim to better understand if the $T_2$ vertex can be used as a building block, we will glue two $T_2$ vertices 
and compute the 
partition function of the resulting brane web using the topological vertex formalism, comparing this result to the gluing of 
strip blocks.  
The two geometries produced by the gluing process are depicted in Figure \ref{fig:SU2Nf4-2T2} and they are both used 
to compute 
the partition function of the 5D $\cN=1$, $SU(2)$ four-flavour theory on $\mathbb{R}^4_{q,t} \times \mathbb{S}^1$, up 
to what is referred to in the literature as the \emph{non-full spin content} \cite{Bao:2013pwa}. 
This represents the 
degrees of freedom coming from the external parallel legs. 
The webs depicted in Figure \ref{fig:SU2Nf4-2T2} are related by two flops, applied to the edges of the octagon  
decorated by the K\"ahler parameters $Q_{m_2}$ and $Q_{m_4}$, or equivalently to the edges with K\"ahler parameters 
$Q_3$ and $Q_3'$. 


%

The partition function for the $SU(2)$, $N_f=4$ theory in 5D, calculated with the refined topological vertex for 
the brane diagram on the left of Figure \ref{fig:SU2Nf4-2T2}, can be expressed in terms of the partition functions 
$ \mathcal{Z}_{\rm strip}$ of the strips glued to create the octagonal web 
\be \label{stripglue}
\mathcal{Z}_{\text{oct}} ({\bf Q_{m}},{\bf Q_{m'}}, {\bf Q'}_B;\ft,q) = \sum_{\tau_1,\tau_2} 
 (-Q'_{B_1})^{|\tau_1|} (-Q'_{B_2})^{|\tau_2|} 
 \mathcal{Z}^{\rm strip}_{\tau_1,\tau_2} ({\bf Q_{m}};\ft,q) 
 \mathcal{Z}^{\rm strip}_{\tau_2^t,\tau_1^t} ({\bf Q_{m'}};q,\ft) ~.
\ee 
The dependance on the K\"ahler parameter $Q_F$ is kept implicit in this equation. 
We will see below that, in order to recast this summation in the form of an instanton expansion where 
$\mathcal{Z}_{\rm oct}  =  \mathcal{Z}^{\rm pert}_{\rm oct}  \mathcal{Z}^{\rm inst}_{\rm oct} $, the K\"ahler parameters 
${\bf Q'}_B$ are first set to 
\be \label{eq:mapQBstrip}
Q_{B_1}' =  u \frac{Q_F}{Q_{m_1} Q_{m_4}} ~, \qquad Q_{B_2}' =  u \frac{Q_F}{Q_{m_2} Q_{m_3}}  ~,
\ee
with $u$ the instanton counting parameter. It is possible to express two independent parameters $Q_{B_i}'$, $i=1,2$, 
in terms of a single parameter $u$ because the K\"ahler parameters associated to the edges of a face of a brane 
diagram satisfy face constraints, here $Q_{m_1} Q_{B_1}' Q_{m_4} = Q_{m_2} Q_{B_2}'  Q_{m_3}$.
Equation \eqref{eq:mapT2toStripGluing2} allows to represent the instanton part of the partition function in the form
\be \label{instexp-strip}
\mathcal{Z}_{\text{oct}}^{\text{inst}} ({\bf Q_m},{\bf Q_{m'}};\ft,q) = \sum_{k=0}^{\infty} \,
 u^k \, \mathcal{Z}_{k}^{\rm strip} ({\bf Q_m},{\bf Q_{m'}};\ft,q)~ ,
 \ee
which is known to reproduce the expansion of Virasoro four point conformal blocks in powers of the cross-ratio 
\cite{Alday:2009aq}. We compare this result to the analogous expansion obtained when gluing $T_2$ building blocks
%
%
\be \label{eq:SU2Nf4-from-T2}
 \mathcal{Z}_ {\text{hex}} 
 ({\bf Q},{\bf Q}', {\bf Q}_B;\ft,q) = \sum_{\lambda_1,\lambda_2} 
 (-{Q}_{B_1})^{|\lambda_1|} (-{Q}_{B_2})^{|\lambda_2|} 
 \mathcal{Z}^{T_2}_{\lambda_1,\lambda_2} ({\bf Q};\ft,q) 
 \mathcal{Z}^{T_2}_{\lambda_2^t,\lambda_1^t} ({\bf Q}';q,\ft) ~.
 \ee 
Setting here the K\"ahler parameters ${\bf Q}_B$ to be 
\be \label{eq:mapT2toStripGluing3}
Q_{B_1}= u Q_2 ~, \qquad Q_{B_2}= u Q_2' 
\ee 
similarly to equation \eqref{eq:mapT2toStripGluing2}, we find the instanton expansion from  
$\mathcal{Z}_{\text{hex}}  =  \mathcal{Z}^{\rm pert}_{\text{hex}}  \mathcal{Z}^{\rm inst}_{\text{hex}} $ is 
%
%
 \be \label{instexp-T2}
 \mathcal{Z}_{\text{hex}}^{{\rm inst}} ({\bf Q},{\bf Q}';\ft,q) =
 \frac{
 \mathcal{Z}_{\text{oct}}^{\text{inst}} ({\bf Q_m},{\bf Q_{m'}};\ft,q) 
 }{
\cM(Q_{B_1}')\cM(Q_{B_2}' \frac{t}{q})} ~,
 \ee
assuming a simple and natural dictionary between ${\bf Q},{\bf Q}'$ and ${\bf Q_m},{\bf Q_{m'}}$ and removing  
what is known as the non-full spin content $\cM(Q_{B_1}')\cM(Q_{B_2}' \frac{t}{q})$.  
The comparison between the partition functions obtained by gluing $T_2$ blocks and strip geometries can only be done 
order by order in the instanton expansion, unless one is able to perform the sum in equation \eqref{eq:T2-nu2} 
analytically and derive a product formula. It is a technically very difficult computation and thus we will present here a 
check up to second order in the instanton parameter. 


\subsection{$SU(2)$ four-flavour $\cN=1$ theory in 5D from gluing of strips}


It will help to start by reviewing the case of the strip first. 
We can explicitly compute the topological strings partition function for the octagon diagram in Figure \ref{fig:SU2Nf4-2T2}, 
which is obtained by gluing two strips, like in \cite{Mitev:2014jza, Coman:2018uwk} 
\bea \label{eq:octagonSU2Nf4}
&&\mathcal{Z}_{\text{oct}} ({\bf Q}_m,Q_F,{\bf Q}_B';t,q) = 
\frac{\cM(\frac{Q_F}{Q_{m_1} Q_{m_2}}) \cM(\frac{Q_F}{Q_{m_3} Q_{m_4}}) \cM(Q_F)\cM(Q_F\frac{t}{q})}{
\prod_{i=1}^4 \cM(Q_{m_i}\sqrt{\frac{t}{q}}) \cM(\frac{Q_F}{Q_{m_i}}\sqrt{\frac{t}{q}})} \\
&& \qquad\qquad\qquad\qquad\quad 
\sum_{\tau_1 , \tau_2}   (Q_{B_1}')^{|\tau_1 |}  (Q_{B_2}'\frac{t}{q})^{|\tau_2 |}  \,   q^{||\tau_1 ||^2}t^{||\tau_2^t||^2}  \,
\prod_{\ell=1}^2 \tilde{Z}_{\tau_\ell} (t,q) \tilde{Z}_{\tau_\ell^t} (q,t) 
\nn\\
&& \qquad\qquad \frac{\prod_{i=1,4} \cN_{\tau_1\emptyset}(Q_{m_i}\sqrt{\frac{t}{q}}) 
\cN_{\emptyset \tau_2}(\frac{Q_F}{Q_{m_i}}\sqrt{\frac{t}{q}})  
 \prod_{i=2,3} \cN_{\emptyset \tau_2}(Q_{m_j}\sqrt{\frac{t}{q}})  
   \cN_{\tau_1\emptyset}(\frac{Q_F}{Q_{m_j}}\sqrt{\frac{t}{q}})  
   }{\cN_{\tau_1 \tau_2}(Q_F) \cN_{\tau_1 \tau_2}(Q_F\frac{t}{q}) }  \, . \nn
\eea
The {\it base} K\"ahler parameters of the octagon must be identified like in \eqref{eq:mapQBstrip}, 
which is the same as equation (4.88) of \cite{Taki:2014pba}. 
Only after this step we can identify the perturbative and instanton parts for this 
partition function 
\be
\mathcal{Z}_\text{oct} = \mathcal{Z}^{\text{pert}}_\text{oct} \mathcal{Z}^{\text{inst}}_\text{oct}  \quad \mbox{with} \quad  \mathcal{Z}^{\text{pert}}_\text{oct} =  \frac{\cM(\frac{Q_F}{Q_{m_1} Q_{m_2}}) \cM(\frac{Q_F}{Q_{m_3} Q_{m_4}}) \cM(Q_F)\cM(Q_F\frac{t}{q})}{
\prod_{i=1}^4 \cM(Q_{m_i}\sqrt{\frac{t}{q}}) \cM(\frac{Q_F}{Q_{m_i}}\sqrt{\frac{t}{q}})}  \, .
\ee
With this factorization, the instanton sum 
\be \label{eq:instantonexpansion1-A}
\mathcal{Z}^{\text{inst}}_\text{oct} = 1+ u \mathcal{Z}^{1-\text{inst}}_\text{oct} + u^2 \mathcal{Z}^{2-\text{inst}}_\text{oct} 
+ \ldots 
\ee 
starts at zero order with unity,  where $\tau_1=\tau_2=\emptyset$. At the next order, the coefficient 
$\mathcal{Z}^{1-\text{inst}}_\text{oct}$ is given by
%
%
\bea \label{eq:octagonSU2Nf4-1inst}
&&\mathcal{Z}^{1-\text{inst}}_\text{oct} =  \frac{q ~ Q_F }{(1-q)(1-t)}  
\big( \frac{1}{Q_{m_1}Q_{m_4}} + \frac{1}{Q_{m_2}Q_{m_3}} \frac{\ft}{q} \big) + \nn\\
&&\frac{q}{\ft} 
\frac{(1+\frac{q}{\ft})\left(1+\sum_{\stackrel{i\neq j}{i,j=1}}^4 \frac{Q_F}{Q_{m_i}Q_{m_j}} +\frac{Q_F^2}{\prod_{k=1}^4 Q_{m_k}}
\right) - \sqrt{\frac{q}{\ft}} (1+Q_F)\sum_{i=1}^4 \left( \frac{1}{Q_{m_i}} + \frac{Q_F Q_{m_i}}{\prod_{j=1}^4 Q_{m_j}} \right)   
}{(1-q)(1-\ft^{-1})(1-Q_F q/\ft)(1-Q_F^{-1}q/\ft)} ~. \nn\\
&&
\eea
Removing from equation \eqref{eq:octagonSU2Nf4} what is referred to in literature as the {\it non-full spin content} 
\cite{Bao:2013pwa}, and which in this case is given by the product $\cM(Q_{B_1}')\cM(Q_{B_2}' \frac{t}{q})$, gives 
\be
 \mathcal{Z}_{\text{norm}}  =  
 \frac{\mathcal{Z}_{\text{oct}} ({\bf Q}_m,Q_F,{\bf Q}_B';t,q)}{
 \cM(Q_{B_1}')\cM(Q_{B_2}' \frac{t}{q})\cM({Q_F \over Q_{m_1} Q_{m_2}  })\cM({ Q_F  \over Q_{m_3} Q_{m_4}  } ) }  \, .
\ee
This is the partition function of the 5D $\cN=1$, $SU(2)$ four-flavour theory on $\mathbb{R}^4_{q,t} \times \mathbb{S}^1$ 
and the non-full spin content corresponds to (non 5D) 
degrees of freedom coming from the pairs of parallel external legs. The 
parameter identification \eqref{eq:mapQBstrip} then allows to split the partition function into a product of perturbative and 
instanton parts, where  
\be
\mathcal{Z}_{\text{norm}} 
= \mathcal{Z}_{\text{norm}}^{\text{pert}} 
\mathcal{Z}_{\text{norm}}^{\text{inst}} 
\quad \mbox{with} \quad  \mathcal{Z}_{\text{norm}}^{\text{pert}} 
=  \frac{\cM(Q_F)\cM(Q_F\frac{t}{q})}{\prod_{i=1}^4 \cM(Q_{m_i}\sqrt{\frac{t}{q}}) \cM(\frac{Q_F}{Q_{m_i}}\sqrt{\frac{t}{q}})}  
\, \nn
\ee
and   
\be \label{eq:octagonSU2Nf4Normal}
\mathcal{Z}_{\text{norm}}^{\text{inst}} 
({\bf Q}_m,Q_F;t,q) = 
\frac{\mathcal{Z}^\text{inst}_\text{oct} ({\bf Q}_m,Q_F;t,q)}{\cM(Q_{B_1}')\cM(Q_{B_2}' \frac{t}{q})} ~. 
\ee
The normalised partition function \eqref{eq:octagonSU2Nf4Normal} has an enhanced $E_5$ symmetry \cite{Taki:2014pba}, 
see also \cite{Mitev:2014jza}, and its $1-$instanton term is given by  
\bea  \label{eq:octagonSU2Nf4Normal-1inst}
&& \mathcal{Z}_{\text{norm}}^{1-\text{inst}} = \nn\\
&&\frac{q}{\ft} 
\frac{(1+\frac{q}{\ft})\left(1+\sum_{\stackrel{i\neq j}{i,j=1}}^4 \frac{Q_F}{Q_{m_i}Q_{m_j}} +\frac{Q_F^2}{\prod_{k=1}^4 Q_{m_k}}
\right) - \sqrt{\frac{q}{\ft}} (1+Q_F)\sum_{i=1}^4 \left( \frac{1}{Q_{m_i}} + \frac{Q_F Q_{m_i}}{\prod_{j=1}^4 Q_{m_j}} \right)   
}{(1-q)(1-\ft^{-1})(1-Q_F q/\ft)(1-Q_F^{-1}q/\ft)} ~. \nn\\
&&
\eea
Notice this expression is symmetric under the interchange of all $Q_{m_i}$ for $i=1,\ldots ,4$. 
We can also present the result compactly at second order in the case of the unrefined partition function, where $q=t$, 
if we further set $q=e^h$ and take the limit where $h\to 0$. Then to leading order in $h$, the $2-$instanton term is
\bea \label{eq:Zinst-u2-2}
&& (1-q)^4 \mathcal{Z}_{\text{norm}}^{2-\text{inst}} = \nn\\
&& \frac{1}{2} \Big[\frac{Q_F}{(1-Q_F)^2}  \Big(
2\big(1+\sum_{\stackrel{i\neq j}{i,j=1}}^4 \frac{Q_F}{Q_{m_i}Q_{m_j}} +\frac{Q_F^2}{\prod_{k=1}^4 Q_{m_k}}\big) 
- (1+Q_F)\sum_{i=1}^4 \big( \frac{1}{Q_{m_i}} + \frac{Q_F Q_{m_i}}{\prod_{j=1}^4 Q_{m_j}} \big) \Big)\Big]^2 .
\nn\\
&&
\eea

%
%

\subsection{$SU(2)$ four-flavour $\cN=1$ theory in 5D from gluing of $T_2$ blocks}

We now turn to the geometry depicted on the right side of this 
figure and the gluing of $T_2$ vertices. To look at this in a similar way, we first need the generalisation \eqref{eq:T2-nu2-6}  
of the $T_2$ partition function 
when two of the Young tableaux associated to external diagonal legs 
of the web are non-empty
\bea\label{eq:T2-nu2} 
\mathcal{Z}_{\vec{\lambda}}^{\text{top},2}(Q_1,Q_2,Q_3;t,q)&=& 
\ft^{\frac{||\lambda_1^t||^2+||\lambda_2^t||^2}{2}} \tilde{Z}_{\lambda_1^t}(q,\ft) \tilde{Z}_{\lambda_2^t}(q,\ft) 
\frac{\cM(Q_1 Q_2;\ft , q)}{\cM(Q_1\sqrt{\frac{\ft}{q}};\ft , q) \cM(Q_2\sqrt{\frac{\ft}{q}};\ft , q)}\nn\\
&~& \sum_\nu \left( Q_3 \sqrt{\frac{q}{\ft}} \right)^{|\nu|}
\frac{\cN_{\lambda_1 \nu}(Q_1\sqrt{\frac{\ft}{q}};\ft , q) \cN_{\nu \lambda_2}(Q_2\sqrt{\frac{\ft}{q}};\ft , q)}{
\cN_{\nu\nu}(1;\ft , q) \cN_{\lambda_1 \lambda_2}(Q_1 Q_2;\ft ,q)} ~.
\eea
%
%
%
%
%
%
Because we aim to use this as a building block when gluing, we keep here explicit the dependence on the $t,q$ deformation 
parameters. Recalling the discussion from Section \ref{sec:Resummation},  
when the external tableaux in \eqref{eq:T2-nu2} are empty the partition function reduces to 
\be \label{eq:T2localNonempty}
\mathcal{Z}^{\text{top}}_{T_2}(Q_1,Q_2,Q_3;\ft,q) = \frac{\cM(Q_1 Q_2;\ft,q)\cM(Q_2 Q_3;\ft,q)\cM(Q_1 Q_3\frac{\ft}{q};\ft,q)}{
\cM(Q_1 Q_2 Q_3 \sqrt{\frac{\ft}{q}};\ft,q)\prod_{\ell=1}^3 \cM(Q_\ell \sqrt{\frac{\ft}{q}} ;\ft,q)} ~.
\ee
It will be useful for us to write the topological strings partition function \eqref{eq:T2-nu2} for the $T_2$ web depicted in 
Figure \ref{fig:T2-nonempty} is the following way\footnote{ 
One may note here that  $\cN_{\lambda_1 \lambda_2}^{-1} (Q_1 Q_2;\ft ,q)$ is half of the contribution of a 5D vector multiplet 
.}, 
where it is normalized by $\mathcal{Z}^{\text{top}}_{T_2}$ 
\eqref{eq:T2localNonempty}
\be 
\mathcal{Z}^{\text{top},2}_{\lambda_1,\lambda_2} (Q_1,Q_2,Q_3;\ft,q) = \mathcal{Z}^{\text{top}}_{T_2}(Q_1,Q_2,Q_3;\ft,q)  
\prod_{i=1}^2 \ft^{\frac{||\lambda_i^t||^2}{2}} 
\tilde{Z}_{\lambda_i^t}(q,\ft) \frac{P_{\lambda_1,\lambda_2} (Q_1,Q_2,Q_3;\ft,q)}{\cN_{\lambda_1 \lambda_2}(Q_1 Q_2;\ft ,q)} 
 ~,
 \ee 
and we have defined the function 
\bea \label{eq:Plambda}
P_{\lambda_1,\lambda_2} (Q_1,Q_2,Q_3;\ft,q) &=& \frac{\cM(Q_1 Q_2 Q_3 \sqrt{\frac{\ft}{q}};\ft,q)
\cM(Q_3 \sqrt{\frac{\ft}{q}} ;\ft,q)}{
\cM(Q_2 Q_3;\ft,q)\cM(Q_1 Q_3\frac{\ft}{q};\ft,q)} \nn\\
&~& 
\sum_\nu \left( Q_3 \sqrt{\frac{q}{\ft}} \right)^{|\nu|}
\frac{\cN_{\lambda_1 \nu}(Q_1\sqrt{\frac{\ft}{q}};\ft , q) \cN_{\nu \lambda_2}(Q_2\sqrt{\frac{\ft}{q}};\ft , q)}{\cN_{\nu\nu}(1;\ft , q)}
\,.
\eea
Let ${\bf Q}=(Q_1,Q_2,Q_3)$ denote a vector of K\"ahler parameters. The topological strings partition function obtained 
by gluing $T_2$ blocks, as depicted on the left of Figure \ref{fig:SU2Nf4-2T2}, is 
 \be 
 \label{eq:SU2Nf4-from-T2-main}
 \mathcal{Z}_{\text{hex}} ({\bf Q},{\bf Q}';\ft,q) = \sum_{\lambda_1,\lambda_2} 
 (-Q_{B_1})^{|\lambda_1|} (-Q_{B_2})^{|\lambda_2|} 
 \mathcal{Z}^{\text{top},2}_{\lambda_1,\lambda_2} ({\bf Q};\ft,q) 
 \mathcal{Z}^{\text{top},2}_{\lambda_2^t,\lambda_1^t} ({\bf Q}';q,\ft) ~.
 \ee
To determine the first orders in the instanton expansion, we need to know explicitly the function 
$P_{\lambda_1,\lambda_2} (Q_1,Q_2,Q_3;\ft,q)$ defined in equation \eqref{eq:Plambda} when the Young diagrams 
$\lambda_1,\lambda_2$ have one or two boxes. In the case with only one box, we find 
\bea 
P_{\{1\},\emptyset} (Q_1,Q_2,Q_3;\ft,q) &=& 1 - Q_1 \left(1+Q_2 Q_3 \right)\sqrt{\frac{q}{\ft}} + Q_1 Q_3 ~, \\
P_{\emptyset,\{1\}} (Q_1,Q_2,Q_3;\ft,q) &=& 1 - Q_2 \left(1+Q_1 Q_3 \right)\sqrt{\frac{\ft}{q}} + Q_2 Q_3 ~. \nn
\eea
The first of these expressions has been previously computed by \cite{Taki:2014pba}. We have furthermore calculated 
\bea \label{eq:P2boxes}
P_{\{2\},\emptyset} ({\bf Q};\ft,q) &=& 1 + Q_1 Q_3 +(Q_1 Q_3)^2 + q Q_1 Q_3 (1 + Q_1 Q_2 \frac{1}{\ft}) 
+ Q_1^2 (1 + Q_2 Q_3 +(Q_2 Q_3)^2) \frac{q^2}{\ft}  \nn\\ 
&~& - Q_1 (1 + Q_1 Q_3) (1 + Q_2 Q_3) \sqrt{\frac{q}{\ft}}(1+q) \\
P_{\{1,1\},\emptyset} ({\bf Q};\ft,q) &=& 1 + Q_1 Q_3 (1+\frac{1}{\ft})  +(Q_1 Q_3)^2  + Q_1^2 \frac{q}{\ft} 
\left(Q_2 Q_3 +\frac{1}{\ft} (1 + Q_2 Q_3 +(Q_2 Q_3)^2) \right) \nn\\
&~& - \sqrt{\frac{q}{\ft}} Q_1 (1+\frac{1}{\ft})  (1 + Q_1 Q_3) (1 + Q_2 Q_3)      \nn\\
P_{\{1\},\{1\}} ({\bf Q};\ft,q) &=& \left( 1 - Q_1 \left(1+Q_2 Q_3 \right)\sqrt{\frac{q}{\ft}} + Q_1 Q_3  \right) 
\left(1 - Q_2 \left(1+Q_1 Q_3 \right)\sqrt{\frac{\ft}{q}} + Q_2 Q_3\right)  \nn\\
&~& + Q_1 Q_2 Q_3 (\sqrt{\frac{q}{\ft}} - \sqrt{q\ft} - \frac{1}{\sqrt{q\ft}} + \sqrt{\frac{\ft}{q}})\nn\\
P_{\emptyset ,\{1,1\}} ({\bf Q};\ft,q) &=& 1 + Q_2 Q_3 (1+\ft)  +(Q_2 Q_3)^2  + Q_2^2 \frac{\ft}{q} 
\left(Q_1 Q_3 +\ft (1 + Q_1 Q_3 +(Q_1 Q_3)^2) \right) \nn\\
&~& - \sqrt{\frac{\ft}{q}} Q_2 (1+\ft)  (1 + Q_1 Q_3) (1 + Q_2 Q_3)   \nn\\
P_{\emptyset ,\{2\}} ({\bf Q};\ft,q) &=& 1 + Q_2 Q_3 +(Q_2 Q_3)^2 + \frac{1}{q} Q_2 Q_3 (1 + Q_1 Q_2 \ft) 
+ Q_2^2 (1 + Q_1 Q_3 +(Q_1 Q_3)^2) \frac{\ft}{q^2}  \nn\\ 
&~& - Q_2 (1 + Q_1 Q_3) (1 + Q_2 Q_3) \sqrt{\frac{\ft}{q}}(1+\frac{1}{q}) ~. \nn
\eea
If we now apply the following mapping 
\be \label{eq:mapT2toStripGluing1}
Q_{m_3}=Q_1~,\quad Q_{m_1}=Q_1'~,\quad Q_{m_2}=\frac{1}{Q_3}~,\quad Q_{m_4}=\frac{1}{Q_3'}~,\quad 
Q_F= Q_1 Q_2 = Q_1' Q_2'
\ee
and set the base K\"ahler parameters to be 
\be \label{eq:mapT2toStripGluing2}
Q_{B_1} = u\frac{Q_F}{Q_{m_1}}~, \qquad Q_{B_2} = u\frac{Q_F}{Q_{m_3}} ~,
\ee
then the partition function \eqref{eq:SU2Nf4-from-T2} can be factored to 
\be \label{eq:instantonexpansion1}
\mathcal{Z}_{\text{hex}} = \mathcal{Z}^{\text{pert}}_{\text{hex}}\mathcal{Z}^{\text{inst}}_{\text{hex}}~, 
\quad \mbox{where} \quad
\mathcal{Z}_{\text{hex}}^{\text{pert}} ({\bf Q},{\bf Q}';\ft,q) = \mathcal{Z}^{\text{top}}_{T_2}({\bf Q};\ft,q) 
\mathcal{Z}^{\text{top}}_{T_2}({\bf Q}';q,\ft)
\ee
and the instanton part has an expansion in the instanton parameter $u$. 
The first order term in this expansion is 
\bea\label{eq:Zinst-u} 
u \mathcal{Z}_{\text{hex}}^{1-\text{inst}} &=& \sqrt{qt} \tilde{Z}_{\{1\}}(q,\ft)  \tilde{Z}_{\{1\}}(\ft,q)  
\left( -Q_{B_1}\frac{P_{\{1\},\emptyset} ({\bf Q};\ft,q) P_{\emptyset,\{1\}} ({\bf Q}';q,\ft) }{
\cN_{\{1\},\emptyset} (Q_1 Q_2;\ft,q) \cN_{\emptyset,\{1\}} (Q_1' Q_2';q,\ft)}   \right. \nn\\
&~& ~ \qquad \qquad \qquad \qquad \qquad \left.
-Q_{B_2} \frac{ P_{\emptyset,\{1\}}  ({\bf Q};\ft,q) P_{\{1\},\emptyset}({\bf Q}';q,\ft) }{
 \cN_{\emptyset,\{1\}}(Q_1 Q_2;\ft,q) \cN_{\{1\},\emptyset} (Q_1' Q_2';q,\ft)} \right)~  ~~\quad
\eea
and it becomes 
\bea \label{eq:Zinst-u-2} 
&&\mathcal{Z}_{\text{hex}}^{1-\text{inst}} = \nn\\
&&\frac{q}{\ft} 
\frac{(1+\frac{q}{\ft})\left(1+\sum_{\stackrel{i\neq j}{i,j=1}}^4 \frac{Q_F}{Q_{m_i}Q_{m_j}} +\frac{Q_F^2}{\prod_{k=1}^4 Q_{m_k}}
\right) - \sqrt{\frac{q}{\ft}} (1+Q_F)\sum_{i=1}^4 \left( \frac{1}{Q_{m_i}} + \frac{Q_F Q_{m_i}}{\prod_{j=1}^4 Q_{m_j}} \right)   
}{(1-q)(1-\ft^{-1})(1-Q_F q/\ft)(1-Q_F^{-1}q/\ft)} ~. \nn\\
&&
\eea
Notice this is identical to equation \eqref{eq:octagonSU2Nf4Normal-1inst}, expressing the $1-$instanton term in the 
normalised partition function derived by gluing horizontally strip geometries.
We can repeat this calculation at second order of the instanton expansion, 
where the pairs of Young tableaux $\lambda_1,\lambda_2$ which need to be considered are  
\be 
(\{2\},\emptyset )~,\qquad (\{1,1\},\emptyset )~,\qquad (\{1\},\{1\})~,\qquad (\emptyset, \{1,1\} )~,\qquad (\emptyset, \{2\} ) 
~. \nn
\ee
Taking the terms corresponding to these Young diagrams from equation \eqref{eq:SU2Nf4-from-T2} gives 
\bea \label{eq:Zinst-u2}
u^2 \mathcal{Z}_{\text{hex}}^{2-\text{inst}} &=& Q_{B_1} Q_{B_2} q\ft ~  \left(\tilde{Z}_{\{1\}}(q,\ft) \tilde{Z}_{\{1\}}(\ft , q) \right)^2 
\frac{P_{\{1\},\{1\}}  ({\bf Q};\ft,q) P_{\{1\},\{1\}}({\bf Q}';q,\ft) }{ \cN_{\{1\},\{1\}}  (Q_F;\ft,q) \cN_{\{1\},\{1\}} (Q_F;q,\ft) } \nn\\
&+& Q_{B_1}^2 \ft q^2 ~ \tilde{Z}_{\{1,1\}}(q,\ft) \tilde{Z}_{\{2\}}(\ft , q) 
\frac{P_{\{2\},\emptyset}({\bf Q};\ft,q) P_{\emptyset,\{1,1\}}({\bf Q}';q,\ft) }{ 
\cN_{\{2\},\emptyset} (Q_F;\ft,q) \cN_{\emptyset,\{1,1\}} (Q_F;q,\ft) } \nn\\
&+& Q_{B_1}^2 \ft^2 q ~ \tilde{Z}_{\{2\}}(q,\ft) \tilde{Z}_{\{1,1\}}(\ft , q) 
\frac{P_{\{1,1\},\emptyset}({\bf Q};\ft,q) P_{\emptyset,\{2\}}({\bf Q}';q,\ft) }{ 
\cN_{\{1,1\},\emptyset} (Q_F;\ft,q) \cN_{\emptyset,\{2\}} (Q_F;q,\ft) } \nn\\
&+& Q_{B_2}^2 \ft q^2 ~ \tilde{Z}_{\{1,1\}}(q,\ft) \tilde{Z}_{\{2\}}(\ft , q) 
\frac{P_{\emptyset,\{2\}}({\bf Q};\ft,q) P_{\{1,1\},\emptyset}({\bf Q}';q,\ft) }{ 
\cN_{\emptyset,\{2\}} (Q_F;\ft,q) \cN_{\{1,1\},\emptyset} (Q_F;q,\ft) } \nn\\
&+& Q_{B_2}^2 \ft^2 q ~ \tilde{Z}_{\{2\}}(q,\ft) \tilde{Z}_{\{1,1\}}(\ft , q) 
\frac{P_{\emptyset, \{1,1\}}({\bf Q};\ft,q) P_{\{2\},\emptyset}({\bf Q}';q,\ft) }{ 
\cN_{\emptyset,\{1,1\}} (Q_F;\ft,q) \cN_{\{2\},\emptyset} (Q_F;q,\ft) } ~.
\eea
Substituting appropriately the functions \eqref{eq:P2boxes}, going to the unrefined case where $\ft=q$, 
and further setting $q=e^h$ for simplicity, gives to leading order in $h$ in the limit $h\to 0$
\bea \label{eq:Zinst-u2-2}
&& (1-q)^4 \mathcal{Z}_{\text{hex}}^{2-\text{inst}} = \nn\\
&& \frac{1}{2} \Big[\frac{Q_F}{(1-Q_F)^2}  \Big(
2\big(1+\sum_{\stackrel{i\neq j}{i,j=1}}^4 \frac{Q_F}{Q_{m_i}Q_{m_j}} +\frac{Q_F^2}{\prod_{k=1}^4 Q_{m_k}}\big) 
- (1+Q_F)\sum_{i=1}^4 \big( \frac{1}{Q_{m_i}} + \frac{Q_F Q_{m_i}}{\prod_{j=1}^4 Q_{m_j}} \big) \Big)\Big]^2 .
\nn\\
&&
\eea
%
%
%
%
We therefore note that, like in the case of the $1-$instanton term \eqref{eq:Zinst-u-2}, the $2-$instanton coefficient 
is also identical to its counterpart from the previous subsection \eqref{eq:Zinst-u2-2} calculated from the 
horizontal gluing of strip geometries. Therefore at leading orders in the instanton expansion, we have identified 
%
%
\be \label{eq:identificationInstPFHexNorm}
\big[ \cM(Q_{B_1}')\cM(Q_{B_2}' t/q) \big]^{-1}
 \mathcal{Z}_{\text{oct}}^{\text{inst}}  = 
\mathcal{Z}_{\text{norm}}^{\text{inst}}=\mathcal{Z}_{\text{hex}}^{\text{inst}}
\ee 
from equations \eqref{eq:octagonSU2Nf4Normal} and respectively \eqref{eq:instantonexpansion1}. This result is highly 
non-trivial and supports the slicing invariance conjecture \cite{Iqbal:2007ii} whereby, under a correct identification of 
parameters, the instanton expansions computed for vertical and diagonal slicing of a web diagram are identical.

\subsection{Mapping to Virasoro four point conformal blocks}

It is known that the instanton expansion in equation \eqref{eq:instantonexpansion1-A} can be mapped to the expansion 
of Virasoro four point conformal blocks in powers of the conformal cross-ratio \cite{Alday:2009aq}. 
Therefore, via equations \eqref{eq:octagonSU2Nf4Normal} and \eqref{eq:identificationInstPFHexNorm}, one would 
naturally expect to be able to also map the coefficients of the instanton expansion \eqref{eq:instantonexpansion1} 
derived from gluing $T_2$ blocks to the same expansion on the side of conformal field theory.  
Following the analysis from Section \ref{sec:topol2matrix} which related the $T_2$ K\"ahler parameters to Liouville CFT 
momenta, we repeat the process for the two $T_2$ vertices glued to form the web on the right of Figure 
\ref{fig:SU2Nf4-2T2}. We thus identify the following dictionary 
\begin{align}
& Q_1 = q^{-\beta(\frac{Q}{2}+a_1-a_2+a)} ~, \qquad Q_2 = q^{-\beta(\frac{Q}{2}-a_1+a_2+a)} ~, 
\qquad Q_3 = q^{\beta(\frac{3Q}{2}+a_1+a_2+a)} ~, 
\end{align}
where $Q=\beta^{-1}-\beta$ and $a_1=a_2+a+\beta s$ and similarly 
\be 
 Q_1' = q^{-\beta(\frac{Q}{2}+a_3-a_4+a)} ~, \qquad Q_2' = q^{-\beta(\frac{Q}{2}-a_3+a_4+a)} ~, 
\qquad Q_3' = q^{\beta(\frac{3Q}{2}+a_3+a_4+a)} ~.
\ee
Taking this dictionary together with the map \eqref{eq:mapT2toStripGluing1} and the relation $q^{\beta}=e^R$, in the 
limit $R\to 0$, equation \eqref{eq:octagonSU2Nf4-1inst} becomes
\be \label{eq:Zinst-u-2-normalised-limit} 
Z^{1-\text{inst}}_n =  -\frac{\left(\Delta_a-\Delta_{a_1}+\Delta_{a_2} \right)
\left(\Delta_a-\Delta_{a_3}+\Delta_{a_4} \right)}{2\Delta_a} + 2a_1(a_3+Q) ~,
\ee
with $\Delta_a=a(a+Q)$. This can be compared to the instanton expansion in \cite{Alba:2010qc}, whereby the term 
$2a_1(a_3+Q)$ is identified with the $U(1)$ factor of \cite{Alday:2009aq} while the first term appears in the expansion 
of the four point conformal block of Liouville CFT.

\section{Conclusions and Outlook}

\subsection{Using the $T_2$-vertex in the AGT-correspondence}

We have presented direct evidence for the conjecture that
gluing  $T_2$-vertices yields partition functions related to instanton partition functions,
or equivalently Liouville conformal 
blocks, in the four-dimensional limit. This is not completely unexpected in view of the relations between the 
$T_2$ vertex and strip discussed in Section \ref{sec:CompareT2-Strip}, given that such results
are well-known in the case of the strip.
However, on first sight one finds results for the $T_2$ vertex and strip
differing by a non-trivial factor. One needs to take into account the fact that both 
partition functions represent piecewise analytic functions on their respective domains of 
definition in a common parameter space. Agreement is found when comparing the results
for the chambers in which both are defined. It should also be noted
that the strip partition functions have 
a smaller domain of definition than the $T_2$ partition functions, 
indicating that the relation between the two types of vertices is somewhat subtle.

It would be nice to find a mathematical proof of  the conjectured relation between gluing of $T_2$ 
vertices and instanton partition functions. One should also notice that the $T_2$ vertex offers
more options for gluing than the strip. This additional flexibility should be useful for the 
investigations of the so-called Sicilian theories \cite{Benini:2009mz}.

Our results indicate that the geometric engineering of $4d$ field theories 
determines the one-loop contributions to the instanton partition functions
up to leg factors. The result can easily be summarised as follows.
The arguments of the 
Barnes double Gamma functions appearing in the one-loop contributions should always be chosen 
in such a way that the partition functions are analytic within the whole region in the parameter space
under consideration. This requirement fixes the arguments of the 
Barnes double Gamma functions uniquely, reproducing the results we had carefully 
derived by various methods before.

One should note that the rule formulated above, natural as it may be, implies that the $T_2$ partition functions are 
only piecewise analytic over the parameter space. The loci where analyticity fails 
are related to the walls separating different K\"ahler cones related by flop transitions within the 
extended K\"ahler moduli space.  

A direct field theoretic explanation of the  piecewise analyticity does not seem to be available at the moment. 
One may note, however,  that the instanton partition functions admit a physical interpretation
as hemisphere partition functions \cite{Gava:2016oep,Dedushenko:2018tgx}. It seems quite possible that
the precise form of the one-loop contributions depends in a subtle way on the boundary 
conditions defining the hemisphere partition functions.
Another possible approach could be to follow the approach 
of \cite{Aganagic:2013tta,Aganagic:2014oia} relating the relevant $5d$ partition functions 
to partition functions of a three-dimensional gauge theory. The chamber dependence  discussed
in this paper is somewhat reminiscent of the Stokes phenomena studied in the context 
of $3d$ superconformal field theories on (deformed) hemispheres in 
\cite{Pasquetti:2011fj,Beem:2012mb}.  
It would be interesting if similar phenomena could explain our results more directly in field theoretical terms.

The one-loop contributions to the instanton partition functions are related
by  the AGT-correspondence to the choice of 
a basis in the one-dimensional space of conformal blocks on the three-punctures sphere. 
The geometric engineering of the $4d$ field theories of interest
therefore refines the previous results on the AGT-correspondence
in an interesting way by selecting preferred bases for the spaces of  three-point conformal blocks.
The relation between the geometric parameters of the topological string theory to the parameters
used in Liouville theory derived in Sections \ref{sec:topol2matrix} and \ref{sec:MM4dLim}
involved a slightly non-obvious sign which was explained by 
the subtleties in the relation between geometric transitions and matrix model representation for the 
partition functions discussed in Section \ref{geomtrans}.

\subsection{Higher rank}

As we have already discussed in the introduction, the problems we encounter in this current paper studying $T_2$, become much more difficult for the case of the $T_N$-vertex with $N>2$. In the case 
of the $W_N$-algebras with $N>2$ one finds
an infinite-dimensional space of chiral vertex operators in general. As discussed in 
\cite{Coman:2015lna,Coman:2017qgv}, one may expect that useful
bases for this space can be constructed with the help of the free field representation. 
The powers of the  screening charges 
provide labels for the elements of this basis. However, such bases are not unique, 
different choices of the contours used to define the 
screening charges, or equivalently, different choices of the ordering 
of these objects define different bases for the spaces of conformal blocks. 
As it is possible to consider continuous powers of the screening charges, as first 
observed in \cite{Teschner:2001rv} for $N=2$, one can define bases labelled by continuous parameters 
in this way\footnote{It is not easy to extend the well-known results for 
positive integer powers  of the screening 
charges to continuous values of these parameters.  
Reconstructing a meromorphic function from it's residues is a nontrivial problem 
as there is the freedom to multiply with an entire function, in general. 
Moreover, without  invoking the duality $b \leftrightarrow b^{-1}$ it is difficult to 
find the full set of poles and residues the final answer has.}. 

\begin{figure}[h!]
   \centering
      \includegraphics{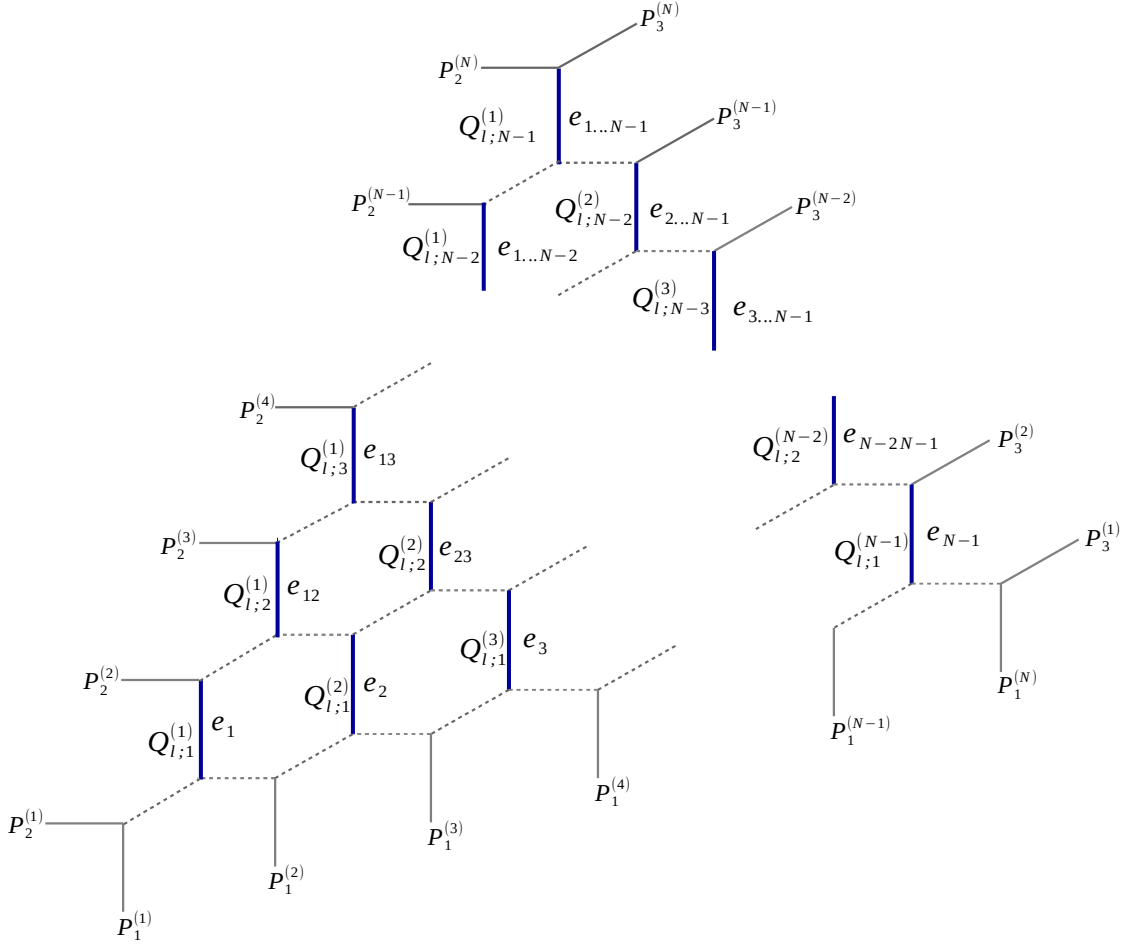} 
   \caption{\it $T_N$ web diagram with an assignment of K\"ahler parameters to the inner vertical segments, the positions 
   of the external semi-infinite branes, and the assignment among the K\"ahler parameters $Q_{l;q}^{(p)}$ and positive 
   roots $e_{ij}=\sum_{k=i}^j e_k$, for $e_k$ simple roots of the Lie algebra $sl_N$, $k=1,\ldots , N-1$.}
   \label{fig:TN}
\end{figure}

Considering $T_N$-vertices for $N>3$ one finds a qualitatively similar picture in the sense
that one can certainly define families of operators 
$\mathcal{V}_{\mathbf{u}_3,\mathbf{u}_1}^{\mathbf{u}_2,\mathbf{w}}(1)$ via  
(\ref{T2VO}) depending on the same number $\frac{1}{2}(N-2)(N-1)$ of parameters, 
now identified with the collection $\mathbf{w}$ of widths of the faces of the toric diagram in Figure \ref{fig:TN}.
One might 
hope that the operators defined from the $T_N$-partition functions via (\ref{T2VO}) 
indeed have a limit $q\rightarrow 1$ representing vertex operators of the  $W_N$-algebra. 

However, at the moment it is not even clear if this limit exists at all, in general. Even less clear is
if the limit, assuming it exists, is related to a vertex operator of the  $W_N$-algebra. One would need
to check that the Ward identities of the $W_N$ symmetry are satisfied. And it is furthermore 
unclear if the putative limit will be one of the bases 
proposed in \cite{Coman:2015lna,Coman:2017qgv}.

The free field representation simplifies considerably when the powers of the screening charges are
all positive integers. There is in fact a nice correspondence with a similar simplification in the 
topological strings partition function for the $T_N$-diagram, occurring when the K\"ahler parameters
satisfy certain integrality conditions, leading to integral formulae resembling the expressions 
from the free field representation of the $W_N$-algebra \cite{Aganagic:2014oia}. 
As a first step towards the identification with vertex operators of the $W_N$-algebra we 
have worked out the relation between the K\"ahler 
parameters and the numbers of screening operators in the resulting expressions.
The  map from the K\"ahler parameters associated to a $T_N$ web diagram to the momenta $a_1,~a_2,~a_3$ 
of Toda three point conformal blocks is as follows
\be 
\frac{P_1^{(a)}}{P_1^{(a+1)}}=\frac{q}{t}q^{\beta(a_1,e_a)} ~, \quad 
\frac{P_2^{(a)}}{P_2^{(a+1)}}=\frac{t}{q}q^{-\beta(a_2,e_a)} ~, \quad 
\frac{P_3^{(N-a)}}{P_3^{(N-a+1)}} = \frac{q}{t} q^{\beta(a_3,e_a)} ~.
\ee 
Here $e_a$ are the simple roots of the Lie algebra $sl_N$, with $a=1,\ldots , N-1$ and $(~,~)$ the inner product 
$(e_a,e_b)=\kappa_{ab}$, where $\kappa_{ab}$ is the Cartan matrix. The momenta are related as 
\be 
a_3= a_1+a_2+\beta \sum_{k=1}^{N-1} s_k e_k ~,
\ee
with screening numbers $s_k$ that are partitioned as $s_k = \sum_{i=1}^k \sum_{j=k}^{N-1} s_{ij}$.  The relation to the 
K\"ahler parameters in Figure \ref{fig:TN} is 
\be 
\label{eq:CompositeScreeningQ}
\beta s_{ij} = -\frac{1}{R} \mathrm{ln}\left(Q^{(i)}_{l;j-i+1}\right) + \frac{Q}{2} ~,
\ee
the derivation\footnote{
The relations between the K\"ahler parameters associated to internal edges of the $T_N$ diagram and the momenta 
parameters of Toda CFT are insensitive to the change in CFT conventions between those used in Section 
\ref{sec:topol2matrix} and the ones from \cite{Coman-Lohi:2018mgj}. The derivation of these relations in Chapter 19 of 
\cite{Coman-Lohi:2018mgj} therefore carries through.  
} of which can be found in \cite{Coman-Lohi:2018mgj}. 
Through \eqref{eq:CompositeScreeningQ} we discover the relation between the   $(N-1)(N-2)/2$ Coulomb moduli  of the $T_N$ theories and the  $(N-1)(N-2)/2$  composite screening charges of the free field representation which label a basis in the space of conformal blocks on $C_{0,3}$.

However, it is not clear to us, at the moment, how to take the limit 
$q\rightarrow 1$ using the contours of integration considered in \cite{Aganagic:2014oia}. Great care is needed to handle 
the possibility of contours getting pinched between poles of the 
integrand which collide in the limit $q\rightarrow 1$. We plan to return to this issue in future work.

\section*{Acknowledgements} 
The authors would like to thank Sara Pasquetti for useful discussions.

The work of I. Coman and J. Teschner is supported by the Deutsche Forschungsgemeinschaft (DFG) through the 
collaborative Research Centre SFB 676 ``Particles, Strings and the Early Universe'', project 
A10.
The work of I. Coman is also supported by the ERC starting grant H2020 ERC StG No.640159. 
E. Pomoni's work is supported by the German Research Foundation (DFG) via the Emmy
Noether program ``Exact results in Gauge theories''.

\appendix

\section{Special functions} \label{App:SpecialFunctions}

Here we collect definitions and identities for special functions. The quantum 
dilogarithm
\be \label{def:q-dilog}
\varphi(x) = (x;q) = \prod_{n=0}^\infty (1- q^n x)~, \qqq |q|<1
\ee
is a special case of shifted factorial function
\be
\label{eq:shiftedfactorials}
(x;q_1,\dots,q_r) = \prod_{n_1,\dots,n_r=0}^{\infty} 
\left( 1-x q_1^{n_1}\cdots q_r^{n_r} \right) \, , \quad \forall  \quad   |q_i|<1 ~.
\ee
A definition that overcomes the need to specify a regime 
of the $|q_i|$ and resolves issues of convergence is the 
polylogarithm function or the plethystic exponential
\be
(x;q_1,\dots,q_r) =  \exp \left( - \mbox{Li}_{r+1}(x) \right)  \, , \quad |x|<1
\ee
with the definition 
\be
\mbox{Li}_{r}(z) = \sum_{n=1}^\infty \frac{z^n}{n^r}   \, , \quad |z| < 1
\ee
for all complex arguments $z$ with $|z| < 1$.

The infinite product function $\calM(U) \equiv \calM(U;\ft, q)$ which appears often in the main text is also defined as a shifted factorial
\be
\label{eq:defcalMeverywhere}
\calM(U;\ft, q) = (U  q ;\ft, q)_{\infty}^{-1}=\left\{\begin{array}{ll}
\prod_{i,j=1}^{\infty}(1-U\ft^{i-1} q^j)^{-1} & \text{ for } |\ft|<1, | q|<1\\
\prod_{i,j=1}^{\infty}(1-U\ft^{i-1} q^{1-j}) & \text{ for } |\ft|<1, | q|>1\\
\prod_{i,j=1}^{\infty}(1-U\ft^{-i} q^j) & \text{ for } |\ft|>1, | q|<1\\
\prod_{i,j=1}^{\infty}(1-U\ft^{-i} q^{1-j})^{-1} & \text{ for } |\ft|>1, | q|>1\\ \end{array}\right.,
\ee
converging for all $U$. This function can be alternatively written as a plethystic exponential 
\be
\label{eq:defMpexp}
\calM(U;\ft, q)=\exp\left[\sum_{m=1}^{\infty} \frac{U^m}{m}\frac{ q^m}{(1-\ft^m)(1- q^m)}\right],
\ee
which converges for all $\ft$ and all $q$ provided that $|U|< q^{-1+\theta(| q|-1)}\ft^{\theta(|\ft|-1)}$, where only here $\theta(x)$ denotes the step function which is $\theta(x)=1$ if $x>0$ and  $\theta(x)=0$ if $x\leq 0$. From the analytic properties of the shifted factorials \eqref{eq:shiftedfactorials} we can derive the identities
\be
\label{eq:inversionidentities}
\calM(U;\ft^{-1}, q)=\frac{1}{\calM(U\ft;\ft, q)}, \qquad \calM(U;\ft, q^{-1})=\frac{1}{\calM(U q^{-1};\ft, q)},
\ee
as well as the following functional relations
\be
\label{eq:calMshift}
\calM(U\ft;\ft, q)=(U q; q)_{\infty} \calM(U;\ft,q),\qquad \calM(U q;\ft, q)=(U  q; \ft)_{\infty} \calM(U;\ft, q).
\ee
A further useful identity is 
\be \label{app:Mswap}
\cM (Q;q,t) = \cM (Q\frac{t}{q} ; t,q) ~.
\ee

\paragraph{Nekrasov partition function:} 
The quantum dilogarithm 
enters the definition of the Nekrasov partition function \cite{Awata:2008ed} through the functions 
%
%
%
%
%
%
$\cN_{RP}$
\be \label{def:Nek1}
\cN_{RP} (Q;t,q) = \prod_{i=1}^\infty \prod_{j=1}^\infty 
\frac{\varphi(Q q^{R_i-P_j +1} t^{j-i})}{\varphi(Q q^{R_i-P_j+1} t^{j-i-1})} 
\frac{\varphi(Q q t^{j-i-1})}{\varphi(Q q t^{j-i})} ~,
\ee
which is equivalent to 
\be \label{def:Nek2}
\cN_{RP} (Q;t,q) = \prod_{(i,j)\in R} \left( 
1 - Q q^{R_i - j + 1} t^{P_j^t -i } \right)  
\prod_{(i,j)\in P}  \left( 
1 - Q q^{-P_i + j } t^{-R_j^t + i - 1} \right) ~
\ee
and also to
\be 
\label{def:Nek3}
\cN_{RP} (Q;t,q) = \prod_{(i,j)\in P} \left( 
1 - Q q^{R_i - j + 1 } t^{P_j^t -i } \right)  
\prod_{(i,j)\in R}  \left( 
1 - Q q^{-P_i + j } t^{-R_j^t + i - 1 } \right) ~.
\ee
In the main text,  we label $\cN_{RP} (Q;t,q) \equiv \cN_{RP} (Q)$ in order not to make our formulas too baroque. 
The notation in these expressions is as follows: $R$ and $P$ are Young tableaux or 
partitons, $R_i$ represents the length of row $i$ and $R^t$ is the dual partition 
to $R$, with rows and column exchanged with respect to those of $R$. Furthermore, 
a box $s\in R$ on row $i$ and column $j$ of $R$ has leg-length $l_R (s) = R_j^t - i$. 
When one of the partitions in definition \eqref{def:Nek2} is empty, this becomes one of the following   
\be \label{def:Nek5}
\cN_{\emptyset R} (Q) = \prod_{(i,j)\in R} 
\left( 1 - Q q^{-R_i + j } t^{ i - 1 } \right) = 
 \prod_{i=1}^{N_R} \frac{\varphi(Q q^{-R_i + 1} t^{i - 1})}{\varphi(Q q t^{i-1} )}~ ,
\ee
\be 
\label{def:Nek6}
\cN_{R\emptyset } (Q) = \prod_{(i,j)\in R} 
\left( 1 - Q q^{R_i - j +1} t^{ -i } \right) = 
 \prod_{i=1}^{N_R} \frac{\varphi(Q qt^{-i} )}{\varphi(Q q^{R_i+1} t^{-i} )}~.
\ee
Note here for future reference that for $Q=v^{-2} t^{N}$, the partition function 
\be 
\cN_{R\emptyset}(Q) = 
\prod_{i=1}^{N_R}\frac{\varphi(Qqt^{-i})}{\varphi(Qq^{R_i+1}t^{-i})}
\ee
vanishes if $N_R>N$; the argument for the quantum 
dilogarithm in the numerator $\varphi(Qqt^{-i})$ becomes 
$1$ at $i=N+1$ and the numerator vanishes. Similarly, 
$\cN_{\emptyset R}(Q)$ vanishes if $Q=t^{-N}$ and $N_R>N$.  
This follows from the identity 
\be \label{def:NekID1}
\cN_{RP} (Qv^{-2}) = \cN_{PR} (Q^{-1}) (Qv^{-1})^{|R|+|P|} \frac{f_R}{f_P} ~,
\ee
where  
\be \label{def:factorNekID}
f_R = \prod_{(i,j)\in R} (-1) q^{R_i - j + 1/2} t^{-R_j^t + i - 1/2} = 
(-1)^{|R|} q^{||R||^2 / 2} t^{- ||R^t||^2 / 2} ~
\ee
for $|R|=\sum_i R_i$ and $||R||^2=\sum_i R_i^2$. 
Finally, note that when the partitions $R$ and $P$ are 
finite, whereby the Young tableau have a finite number of 
rows, equation \eqref{def:Nek1} becomes
\be \label{def:Nek4}
\cN_{RP} (Q) = \prod_{i=1}^{N_R} \prod_{j=1}^{N_P} 
\frac{\varphi(Q q^{R_i-P_j+1} t^{j-i})}{\varphi(Q q^{R_i-P_j+1} t^{j-i-1})} 
\frac{\varphi(Q q t^{j-i-1})}{\varphi(Qq t^{j-i})}  \cN_{R\emptyset } (t^{N_P} Q) \cN_{\emptyset P} (t^{-N_R} Q)   ~.
\ee
Similarly to \eqref{def:Nek5}-\eqref{def:Nek6}, the partition function \eqref{def:Nek4} also vanishes 
when $Q=v^{-2}t^N$ unless $N_R-N_P\leq N$ (and likewise when $Q= t^{-N}$). The function 
$\cN_{RP}$ further satisfies the identity 
\be \label{app:Nekswap} 
\cN_{RP}(Q;q,t) = \cN_{P^t R^t}(Q \frac{t}{q} ; t,q) ~.
\ee

%

\paragraph{Special functions for topological string amplitudes:} 
When writing topological string amplitudes, the following two functions are ubiquitous. Firstly, the the function
\be
\label{eq:deftildeZ}
\tilde{Z}_{\nu}(\ft, q) = \prod_{i=1}^{\ell(\nu)}\prod_{j=1}^{\nu_i}\left(1-\ft^{\nu^t_j-i+1} q^{\nu_i-j}\right)^{-1},\\
\ee
which up to an overall coefficient is the principal specialisation of the Macdonald function and secondly, the infinite product
\begin{equation}
\label{eq:defcalR}
\calR_{\lambda\mu}(Q;\ft, q) = \prod_{i,j=1}^{\infty}\left(1-Q \ft^{i-\frac{1}{2}-\lambda_j} q^{j-\frac{1}{2}-\mu_i}\right)=\calM(Q\sqrt{\frac{\ft}{ q}};\ft, q)^{-1}\calN_{\lambda^t\mu}(Q\sqrt{\frac{\ft}{ q}};\ft, q).
\end{equation}

\subsection{The limit $q\to1$}
 \label{App:phiMq21limit}
In this subsection we study how to take the $q\to 1$ of  $\varphi(z)$ and $\mathcal{M}(z)$. We begin with the quantum dilogarithm $\varphi(z)$ and its plethystic exponential expression
\be
\label{eq:diLogExpansion}
\varphi(z) = (z;q)_{\infty} =  \exp\left[- \sum_{m=1}^{\infty} \frac{z^m}{m}\frac{ 1}{(1- q^m)}\right] ~,
\ee 
using which we can  derive
\be 
\label{eq:qdilogDiv}
\log{\varphi(z)} = \frac{1}{\log q} \mbox{Li}_2 (z) + \frac{1}{2} \log (1-z)   +\mathcal{O}\left( \log q\right) \, .
\ee
This identity was also derived form Kirillov in \cite{Kirillov:1994en}.
Observe that by forming the ratio  $\frac{\varphi(q)}{ \varphi(z)}$ we eliminate the leading $\mbox{Li}_2$ divergence in the $q\rightarrow1$ limit and then by further dividing by the subleading $\log$ term we obtain a  function
\be
\label{eq:qGamma}
 \Gamma_q(z) =  \frac{ \varphi(q)}{ \varphi(q^z)}  (1-q)^{1-z} \, ,
\ee
 that is finite in the $q\rightarrow1$ limit
\be
\lim_{q\to 1}  \Gamma_q(z) = \exp \left( \zeta(0) (1-x) \left( \log q\right)  +\mathcal{O}\left( \log q\right)^2 \right)
\ee
with $\zeta(0)=-\frac{1}{2}$.
This is a $q$-deformed version of the usual Gamma  function studied by many authors and satisfies the functional relation 
\be
\Gamma_q(z+1) = \frac{1-q^z}{1-q} \Gamma_q(z) = [z]_q \Gamma_q(z) \, .
\ee
It is proven by Koornwinder in the Appendix B of \cite{KOORNWINDER199044} that 
\be
\label{eq:q21limqGamma}
\lim_{q \to 1} \Gamma_q(z) =  \Gamma(z) \, .
\ee
We may further compute the $q \to 1$ limit of 
\be 
\label{eq:diverencephitphiq}
\log{\left( \varphi(t) / \varphi(q) \right)} = - \left(  b^2+1 \right)\zeta(1) +\mathcal{O}\left( \log q\right) \, ,
\ee
using \eqref{eq:diLogExpansion},  although it is better to do so using
\be
\label{eq:GoodLimitphitphiq}
\lim_{q\to 1}{\left( \varphi(t) / \varphi(q) \right)}  =  \left( \Gamma(-b^2) \right)^{-1}  (1-q)^{1+b^2}  \,.
\ee
Finally, we can derive the following useful identity for the limit of the ratio of quantum dilogarithm functions
\bea
\label{eq:quantumdilogRatioLimit}
\lim_{q\to 1}\frac{\varphi(q^{\alpha_1} x)}{\varphi(q^{\alpha_2} x)} 
 = (1-x)^{\alpha_2-\alpha_1} ~,
\eea
using at an intermediate step the identity
\be 
\textrm{Li}_2 (x)=-\int_0^x \frac{dy}{y}\textrm{log}(1-y)~.
\ee
We can now proceed similarly with the $\mathcal{M}$ function using \eqref{eq:defMpexp}
\be
\mathcal{M}\big(z\big)  =   (zq ;t,q)_{\infty}^{-1} = \exp\left[\sum_{m=1}^{\infty} \frac{z^m}{m}\frac{ q^m}{(1-\ft^m)(1- q^m)}\right] \,,
\ee
and  in the $q\rightarrow1$ limit we obtain
\be
\label{eq:zqtexpansion}
b^2 \log (z;q,t)_{\infty} = - \frac{ \mbox{Li}_3 (z)}{(\log q )^2}  -  \frac{1}{2} \frac{\mbox{Li}_2 (z)}{\log q}  (b^2-1)  +  \frac{\mbox{Li}_1 (z)}{3!} \frac{b^4-3b^2+1}{2} 
+\mathcal{O}\left( \log q\right) \, . 
\\
\ee
We find that the function
\be
\label{eq:qdoublegamma}
 \Gamma_{q,t}(x)  = \frac{ \mathcal{M}(t/q) }{ \mathcal{M}(t^x/q) }\varphi(q)^{x-1}
 (1-q)^{\frac{1}{2}(x-1)(2-b^{-2}x) }
\ee
has a finite $q\to 1$ limit and it furthermore satisfies, using \eqref{eq:calMshift},  the functional relation
\be
\label{eq:FunctionalRelqDoubleGamma}
 \Gamma_{q,t}(x+1) =   \Gamma_{q}(x)   \Gamma_{q,t}(x) \, .
\ee

\section{Variants of Jackson integrals and the $q \to 1$ limit}
\label{ap:JacksonIntegrals}

Equation \eqref{contour-Jackson} is useful for studying the limit $q\rightarrow 1$. The Jackson integral 
\eqref{contour-Jackson}, which we reproduce here,
\be 
\cI'_q=\int_0^1d_q x \;x^{t-1}\frac{\varphi(qx)}{\varphi(q^s x)}
\ee
is known to reproduce the integral 
\be
\lim_{q \to 1} \cI'_q=\int_{0}^1dx \;x^{t-1}(1-x)^{s-1}   \, .
\ee
It remains to study the factor in front of $\cI'_q$  in \eqref{contour-Jackson}. We first note 
that $\vartheta_q(z)$ is closely related to the Jacobi theta function $\theta_1(x,\tau)$
defined as 

\be
\theta_1(x,\tau)=-e^{\frac{\pi\mathrm{i}}{4}\tau}\,2\sin(\pi x)
\prod_{n=1}(1-e^{2\pi\mathrm{i} n \tau}e^{2\pi \mathrm{i}x})(1-e^{2\pi\mathrm{i} n \tau}e^{-2\pi \mathrm{i}x})
(1-e^{2\pi\mathrm{i} n \tau})
\ee
Indeed, the relation between $\vartheta_q(z)$ and 
$\theta_1(x,\tau)$ is
\be
\vartheta_q(e^{2\pi \mathrm{i}x})=\mathrm{i}\,e^{\pi\mathrm{i} x}\,
e^{-\frac{\pi\mathrm{i}}{4}\tau}\,\theta_1(x,\tau),\qquad q=e^{2\pi\mathrm{i}\tau}.
\ee
In order to study the limit $q\rightarrow 1^-$, or equivalently $\tau\rightarrow 0$, $\mathrm{Im}(\tau)>0$, we may use the modular
transformation property of $\theta_1(x,\tau)$,

\be
\theta_1(x,\tau)=\mathrm{i}(-\mathrm{i}\tau)^{-\frac{1}{2}}e^{-\frac{\pi\mathrm{i}}{\tau}x^2}\theta_1(x/\tau,-1/\tau).
\ee
It follows that 
\be
\theta_1(s\tau,\tau)=\mathrm{i}\,(-\mathrm{i}\tau)^{-\frac{1}{2}}e^{-\pi\mathrm{i}{s^2}{\tau}}\theta_1(s,-1/\tau)
\sim - \mathrm{i}\,(-\mathrm{i}\tau)^{-\frac{1}{2}}\,e^{-\frac{\pi\mathrm{i}}{4\tau}}\,2\sin(\pi s),
\ee
leading to 
\be
\vartheta_q(q^s)\sim
(-\mathrm{i}\tau)^{-\frac{1}{2}}\,e^{-\frac{\pi\mathrm{i}}{4\tau}}\,2\sin(\pi s).
\ee
It remains to study the asymptotics of $(q,q)_\infty^3$. To this aim
we may use the relation between $(q,q)_\infty$ and the Dedekind eta-function, 
\be
(q,q)_\infty=e^{-\frac{\pi\mathrm{i}\tau}{12}}\eta(\tau)\,.
\ee
Using the modular transformation property 
$
\eta(\tau)=(-i\tau)^{-\frac{1}{2}}\eta(-1/\tau)
$, one finds
\be
(q,q)_\infty=e^{-\frac{\pi\mathrm{i}\tau}{12}}(-i\tau)^{-\frac{1}{2}}\eta(-1/\tau)=
e^{-\frac{\pi\mathrm{i}\tau}{12}}(-i\tau)^{-\frac{1}{2}}e^{-\frac{\pi\mathrm{i}}{12\tau}}\prod_{n=1}(1-e^{-2\pi\mathrm{i}/\tau})\,.
\ee
which implies
\be
(q,q)_\infty\sim e^{-\frac{\pi\mathrm{i}}{12\tau}}\,
(-i\tau)^{-\frac{1}{2}}\,.
\ee
Taken together this yields
\be
\frac{2\pi\mathrm{i}}{1-q}\,\frac{\vartheta_q(q^s)}{(q;q)_{\infty}^3}\sim
\frac{2\pi\mathrm{i}}{(-2\pi \mathrm{i}\tau)}
\frac{(-\mathrm{i}\tau)^{-\frac{1}{2}}}{(-i\tau)^{-\frac{3}{2}}} \,2\sin(\pi s)
\sim 2\mathrm{i}\sin(\pi s).
\ee

\section{Integral representation of the $T_N$ partition function}
\label{sec:TNFF}

For completeness, 
in this section we summarise the relation between the $T_N$ topological string partition function and a matrix integral 
that looks like the free field representation of $A_{N-1}$ Toda three point function. The derivation of the results  
can be found in the thesis \cite{Coman-Lohi:2018mgj}. Here we will not recall the form of $\mathcal{Z}_{T_N}^{\text{top}}$, we refer the interested reader to
 \cite{Bao:2013pwa,Mitev:2014isa,Coman-Lohi:2018mgj}.

The relation between the partition function and the integral formulation is 
after specialisation of parameters 
$v^{a} A_i^{(a)}/A_{i-1}^{(a)} = t^{s_{a,i}} P_2^{(a+i)} $ as depicted in Figure \ref{fig:TN} is\footnote{The parameters of type $A_m^{(n)}$ are associated 
to faces of the $T_N$ web diagram. See  \cite{Bao:2013pwa,Mitev:2014isa,Coman-Lohi:2018mgj} for definitions and convensions.}
\bea \label{secTN:eq:RelationI3Ztop}
\cI_{N} &= & 
\left(2\pi\ii \frac{\varphi(t)}{\varphi(q)}\right)^{\sum_{a=1}^{N-1} s_{a}} \vartheta_M ~ \cM(t/q)^{\frac{N(N-1)}{2}} 
 ~
\prod_{a=1}^{N-1} \prod_{I=1}^{s_a}\left( y^{(a)}_{I}\right)^{\zeta_{a}+1}_\emptyset 
  \mathcal{Z}_{T_N}^{\text{top}}
\eea
where the $( y^{(a)}_{I})_{Y_{a,i}}$ are the appropriate  generalisations of \eqref{eq:poles} for $T_N$ with $a=1,\dots, N-1$ and $i=1,\dots,N-a$. The  integral $\cI_{N}$ is similarly a  generalisation of $\cI_{2}$, defined in \eqref{q-MatrixModel},
\be \label{secTN:def:MatInt}
\cI_N = \int d'_\fq {\bf y} ~ \prod_{a=1}^{N-1} \prod_{I=1}^{s_a} 
(y^{(a)}_I)^{\zeta_{a}} ~ \cI_{m_a} (y)  \cI_{a,a} (y)  \prod_{a=1}^{N-2} \cI_{a,a+1} (y)  \, ,
\ee
where ${\bf y} = \{y^{(a)}_I\}$ contains the set of all integration variables,
\be \label{secTN:def:Ia}
\cI_{m_a} (y)=\prod_{I=1}^{s_a} \frac{\varphi( v^2 P_2^{(a)} / y_I^{(a)} )}{\varphi(P_2^{(a+1)} /y_I^{(a)})} ~,
\ee
\be \label{secTN:def:Iaap1-NOtilde}
\cI_{a,a} (y) = \prod_{J\neq I=1}^{s_a} \frac{\varphi( y^{(a)}_{I}/y^{(a)}_{J})}{\varphi(\ft y^{(a)}_{I}/y^{(a)}_{J})}~,
\quad
\cI_{a,a+1}(y) = \prod_{I=1}^{s_{a}} \prod_{J=1}^{s_{a+1}} 
\frac{\varphi(\ft y^{(a+1)}_{J}/y^{(a)}_{I})}{\varphi(y^{(a+1)}_{J}/y^{(a)}_{I})} ~,
\ee
and 
\be \label{background:def:zeta-o}
\zeta_{a} =  \beta (a_1+a_2 , e_a) - \beta^2 s_{a+1} + \beta^2 (s_a - 1) ~, \qqq s_N=0 ~,
\ee
with $s_a=\sum_{i=1}^{N-a} N_{a,i}$. 
The vectors $e_a$ are the simple roots of the algebra $\mathfrak{sl}_N$, the inner products are taken with 
respect to the Cartan matrix  $\kappa_{ab}= (e_a,e_b)$, $a_2=(a_2^{(1)},\ldots ,a_2^{(N-1)})$ and $(a_2,e_a)=a_2^{(a)}-a_2^{(a+1)}$. 
The  $(N-1)$ component vectors $a_i$ in equation \eqref{background:def:zeta-o}  
are related to the gauge theory parameters $P_1,P_2,P_3$ through 
\be \label{dictionary}
\frac{P_1^{(a)}}{P_1^{(a+1)}}=\frac{q}{t}q^{\beta(a_1,e_a)} ~, \quad 
\frac{P_2^{(a)}}{P_2^{(a+1)}}=\frac{t}{q}q^{-\beta(a_2,e_a)} ~, \quad 
\frac{P_3^{(N-a)}}{P_3^{(N-a+1)}} = \frac{q}{t} q^{\beta(a_3,e_a)} ~, \quad   t= q^{\beta^2}~,
%
%
\ee
where
\be 
%
%
a_3= a_1+a_2+\beta \sum_{k=1}^{N-1} s_k e_k ~.
\ee
Finally, the term $\vartheta_M$ in the prefactor is given by
%
%
\begin{align}  
\vartheta_M = \prod_{a=1}^{N-2} & \prod_{j>i=1}^{N-a}
\prod_{J=1}^{N_{a,j}} \left(\frac{P_2^{(a+i)}}{P_2^{(a+j)}} \ft^{-J}\right)^{-\beta^2 s_{a,i}}
\vartheta_q \left(\frac{P_2^{(a+i)}}{P_2^{(a+j)}} \ft^{-J}, q^{1+\beta^2 s_{a,i} } \right) \nn\\
&\prod_{j \geq i=1}^{N-a-1} 
\prod_{I=1}^{N_{a,i}}
\left(\frac{P_2^{(a+i)}}{P_2^{(a+1+j)}} \ft^{I-1}\right)^{\beta^2 N_{a+1,j}} 
\vartheta_q \left(\frac{P_2^{(a+i)}}{P_2^{(a+1+j)}} \ft^{I-1}, q^{1-\beta^2 N_{a+1,j} } \right) ~. \quad 
\end{align}

We conclude this section by observing that the $q\to 1$ limit of the integrand of
\eqref{secTN:def:MatInt}, up to prefactors that should  and will be workout and 
presented in future work, is proportional to the integrand of
Warnaar's $A_{N}$ Selberg integral \cite{2007arXiv0708.1193W,2009arXiv0901.4176W}. 
See also section 5 of \cite{Schiappa:2009cc} for a nice review.
\bea \label{SelbergAN}
 \cI_{A_{N-1}} 
&=&\int \prod_{a=1}^{N-1} \prod_{I=1}^{s_a} dy^{(a)}_I 
 \left(y^{(a)}_I\right)^{\beta (a_1,e_a)} (1-y^{(a)}_I)^{\beta (a_2,e_a)} \nn\\
 &~& \qquad\qquad\qquad
 \prod_{J>I} (y^{(a)}_J-y^{(a)}_I)^{\kappa_{aa} \beta^2} 
\prod_{b>a} \prod_{J=1}^{s_b} (y^{(a)}_I-y^{(b)}_J)^{\kappa_{ab} \beta^2} 
, ~ \qquad 
\eea
where $\kappa_{ab}$ is the Cartan matrix of the $\mathfrak{sl}_N$  algebra defined above.

\section{Summary of important formulae}\label{Summary}

We collect here some of the key formulae from our paper. 
The topological string partition function, as computed in \cite{Bao:2013pwa}, is a function 
$\mathcal{Z}_2^{\text{top}}(P_1,P_2,P_3;t,q)$ of three K\"ahler parameters $P_i$, with $i=1,2,3$, 
as well as the Omega deformation parameters $q,t$.
In Section \ref{sec:Resummation} we have shown that the infinite sum inside $\mathcal{Z}_2^{\text{top}}$ 
has a finite radius of convergence and that the result is indeed given by a 
factorised expression, 
as was conjectured in \cite{Kozcaz:2010af,Bao:2013pwa}. 

One of our goals has been to clarify the relation to Virasoro conformal blocks expected to emerge in the limit where the 
deformation parameters
$q=e^{-R\epsilon_1}$, $t=e^{R\epsilon_2}$ have $R\rightarrow 0$. This limit is nontrivial, as we have shown in Section 
\ref{sec:4dLim}, since the function 
representing the basic building block of the factorised expression $\mathcal{Z}_2^{\text{top}}$ 
diverges in this limit. In the case $|q|<1$, $|t|<1$ we have shown in Section \ref{sec:MM4dLim} that the limit 
\begin{equation}\label{limit}
\mathcal{Z}_2^{\text{4d}}(a_1,a_2,a_3;\beta)=
\lim_{R\rightarrow 0}
{(2\pi i)^s  \mathcal{M}(t/q) } \left( \frac{\varphi(t)}{\varphi(q)} \right)^{s} 
\mathcal{Z}_2^{\text{top}}(P_1,P_2,P_3;t,q)
\end{equation}
exists if $\beta^2=-\epsilon_2/\epsilon_1$, assuming that
the parameters $a_1,a_2,a_3$ and $P_1,P_2,P_3$ are related as 
\be
\label{dictionary}
P_1^2 = \frac{q}{t} q^{2 \beta a_1 }
\, , \quad
P_2^2 = \frac{t}{q} q^{-2 \beta a_2} \, , 
\quad P_3^2 = \frac{q}{t} q^{ 2 \beta a_3}
\, , \quad 
t= q^{\beta^2}
\, ,
\ee
where  $a_3=a_1+a_2 + s\beta$ with $s\in \mathbb{N}$, $\varphi (x)$ is the quantum dilogarithm function \eqref{def:q-dilog} 
and $\cM (x)$ has a definition as a plethystic exponential \eqref{eq:defMpexp}. 
What equation \eqref{limit} implies is that some divergent factors in the 
asymptotic behaviour of  $\mathcal{Z}_2^{\text{top}}$ for $R\rightarrow 0$ depend on $s$.

In order to establish the precise relation with Liouville conformal blocks we have used an alternative representation 
for $\mathcal{Z}_2^{\text{top}}$ in Section \ref{sec:topol2matrix}, available in the special cases where 
\begin{equation}
\label{integerQ}
\sqrt{\frac{q}{t}}  \frac{P_2 P_3}{P_1}=t^s, \qquad s\in\mathbb{N} ~.
\end{equation}
We have shown for this case that 
$\mathcal{Z}_2^{\text{top}}$ has an alternative representation often called $q$-deformed matrix integral, see also 
\cite{Aganagic:2013tta,Aganagic:2014oia}, in terms of a function $\mathcal{I}_2 (P_1, P_2, P_3;t,q)$ defined as
\be\label{q-matmod}
\mathcal{I}_2 = 
\int \frac{d'_qy_1}{y_1} \cdots \frac{d'_qy_s}{y_s} \,  
\prod_{i =1}^s y^{\zeta+1}_i  \,  \mathcal{I}_{1,1}(y)  \,  \mathcal{I}_{m}(y) \, ,
\qquad 
q^{\zeta+1} = \sqrt{\frac{t}{q} } \frac{P_1 P_3}{P_2} ~,
\ee
where
\be
\label{eq:integrants}
\mathcal{I}_{1,1}(y) = \prod_{i\neq j=1}^s \frac{\varphi(y_i/y_j)}{\varphi(ty_i/y_j)} 
\, , \quad
\mathcal{I}_{m}(y) = \prod_{i =1}^s \frac{\varphi(P_2^2 v^2/y_i )}{\varphi(1/y_i)} 
\, .
\ee
The integrals $\int d'_qy_1 \cdots d'_qy_s \prod_{i =1}^s y^{-1}_i $ are variants of the Jackson 
integral\footnote{The integration measure appears as such in order to satisfy the identification 
with the summation \eqref{eq:VariantJacksonDefinition}. This has been further discussed in 
Section \ref{sec:topol2matrix}.} 
defined for meromorphic functions $M(y)$ of $s$ variables $y=(y_1,\dots,y_s)$ as 
\begin{equation}
\label{eq:VariantJacksonDefinition}
\int \frac{d'_qy_1}{y_1} \cdots \frac{d'_qy_s}{y_s} \; M(y):=(2\pi \mathrm{i})^s \sum_{\substack{R_1,\dots,R_s\in\mathbb{N}\\R_1>R_2>\dots R_s}}
\mathop{\rm Res}_{y=y_R} M(y)~,
\end{equation}
where $y_R=(t^{s-1}q^{R_1},t^{s-2}q^{R_2},\dots,q^{R_s})$. The precise relation between $\mathcal{Z}_2^{\text{top}}$
and $\mathcal{I}_2 $ has been shown to be
\be
\label{Ztop-MM}
\mathcal{Z}_2^{\text{top}}(P_1, P_2, P_3;t,q)=\frac{t^{-\frac{1}{2}s (s-1) (\zeta+1)}} 
{(2\pi i)^s  \mathcal{M}(t/q) } \left( \frac{\varphi(q)}{\varphi(t)} \right)^{s} 
\mathcal{I}_2 (P_1, P_2, P_3;t,q)~.
\ee
Representing  $\mathcal{Z}_2^{\text{top}}$ via equation \eqref{Ztop-MM} offers another 
way to study the limit \eqref{limit}. Despite the fact that all the ingredients in the definition 
of  
$\mathcal{I}_2 (P_1, P_2, P_3;t,q)$ diverge in this limit, we have found that a finite limit  exists for this function 
proportional to the Selberg integral, giving an independent confirmation for  \eqref{limit}.

The function $\mathcal{Z}_2^{\text{4d}}(a_1,a_2,a_3)$ has thereby been found to be
\begin{align}\label{GMM-N}
\mathcal{Z}_2^{\text{4d}}(a_1,a_2,a_3;\beta)=
&\left(\frac{\beta^{1-\beta^2}}{2\pi\mathrm{i}}\Gamma(\beta^2)\right)^{-s}
\frac{\Gamma_\beta(\beta
)}{\Gamma_\beta((s+1)\beta
)}\\
&\frac{\Gamma_\beta(\beta^{-1}+2a_1)}{\Gamma_\beta(\beta^{-1}+2a_1+s\beta)}
\frac{\Gamma_\beta(\beta - 2a_2)}{\Gamma_\beta( - 2a_2 + (1-s)\beta )}
\frac{\Gamma_\beta(2\beta^{-1}-\beta+2a_3)}{\Gamma_\beta(2\beta^{-1}-(s+1)\beta+2a_3)} \, ,
\notag\end{align}
the three point conformal block of the Virasoro algebra with  $c=1-6(\beta-\beta^{-1})^2<1$  central charge.  
It is interesting to note that a similar limit exists for $|t|>1$, $|q|<1$
giving Liouville conformal blocks. 
Relations of the form \eqref{limit} have previously been proposed in  \cite{Kozcaz:2010af,Bao:2013pwa}. 
However, the references above have neither identified a renormalisation prescription giving a finite limit such as 
\eqref{limit}, nor correctly identified the precise relation \eqref{dictionary} between the parameters.

To understand the $T_2$ vertex as a building block, we have compared this in Section \ref{sec:CompareT2-Strip} 
to a much more intensively studied counterpart in the relevant literature, the strip geometry. The main insight 
gained here has been that in various chambers of the K\"ahler moduli space, the partition functions agree up to leg 
factors for the corresponding web diagrams. 
However, not \emph{all} of the chambers are covered by the strip diagrams.  
We have then checked in Section \ref{sec:gluing} that gluing $T_2$ vertices gives the correct conformal blocks of higher 
point functions, explicitly analysing how to glue two such vertices to obtain the partition function of the 5D theory 
with $SU(2)$ gauge group and $N_f =4$ fundamental hypermultiplets.
We have first reviewed how the partition function of the $SU(2)$, $N_f =4$ theory can be 
obtained by gluing two strip functions $ \mathcal{Z}^{\rm strip}_{\tau_1,\tau_2} ({\bf Q_m};\ft,q)$  
%
%
\be \label{stripglue-a}
 \mathcal{Z}_{\rm oct} ({\bf Q_m},{\bf Q_{m'}};\ft,q) = \sum_{\tau_1,\tau_2} 
 (-Q'_{B_1})^{|\tau_1|} (-Q'_{B_2})^{|\tau_2|} 
 \mathcal{Z}^{\rm strip}_{\tau_1,\tau_2} ({\bf Q_m};\ft,q) 
 \mathcal{Z}^{\rm strip}_{\tau_2^t,\tau_1^t} ({\bf Q_{m'}};q,\ft) ~.
 \ee
The instanton piece in 
$\mathcal{Z}_{\rm oct}  =  \mathcal{Z}^{\rm pert}_{\rm oct}  \mathcal{Z}^{\rm inst}_{\rm oct} $ determined by specialising the 
K\"ahler parameters $Q'_{B_1}=uQ''_{B_1}$ and $Q'_{B_2}=uQ''_{B_2}$
can be recast as a sum over orders in the instanton expansion  
\be \label{instexp-strip-a}
 \mathcal{Z}^{\rm inst}_{\rm oct} ({\bf Q_m},{\bf Q_{m'}};\ft,q) = \sum_{k=0}^{\infty} \,
 u^k \, \mathcal{Z}_{k}^{\rm strip} ({\bf Q_m},{\bf Q_{m'}};\ft,q)~
 \ee
and reproduces the expansion of the corresponding Virasoro 
conformal block in powers of the cross-ratio.

Gluing $T_2$ building blocks requires decorating external legs of the corresponding web diagrams by 
Young tableaux and we have shown how this yields Virasoro conformal blocks. We have found that this procedure 
produces a partition function of the form 
\be \label{eq:SU2Nf4-from-T2-a}
 \mathcal{Z}_\text{hex}
 ({\bf Q},{\bf Q}';\ft,q) = \sum_{\lambda_1,\lambda_2} 
 (-{Q}_{B_1})^{|\lambda_1|} (-{Q}_{B_2})^{|\lambda_2|} 
 \mathcal{Z}^{\rm top,2}_{\lambda_1,\lambda_2} ({\bf Q};\ft,q) 
 \mathcal{Z}^{\rm top,2}_{\lambda_2^t,\lambda_1^t} ({\bf Q}';q,\ft) ~
 \ee 
depending on the K\"ahler parameters ${\bf Q,Q}'$. 
By specialising the K\"ahler parameters $Q_{B_i}$ in terms of the instanton expansion parameter $u$, this 
expression can be recast in a form 
$\mathcal{Z}_{\text{hex}}  =  \mathcal{Z}^{\rm pert}_{\text{hex}}  \mathcal{Z}^{\rm inst}_{\text{hex}} $ where the instanton 
piece is related to $\mathcal{Z}^{\rm inst}_{\rm oct}$ through 
%
%
%
 \be \label{instexp-T2-Intro}
%
 \mathcal{Z}_{\text{hex}}^{{\rm inst}} ({\bf Q},{\bf Q}';\ft,q) =
 \frac{
 \mathcal{Z}_{\text{oct}}^{\text{inst}} ({\bf Q_m},{\bf Q_{m'}};\ft,q) 
 }{
\cM(Q_{B_1}')\cM(Q_{B_2}' \frac{t}{q})} 
 \ee
at the first orders in $u$. 

\bibliographystyle{JHEP}
\bibliography{Biblio}

\end{document}